\documentclass[preprint,aps,showpacs]{revtex4-1}
\usepackage{geometry}                
\usepackage{amsmath,amssymb,amsfonts,dcolumn,color,graphicx,graphics,latexsym,placeins,epsfig,tikz}
\usetikzlibrary{arrows,shapes}
\usepackage{rotating,bm,mathrsfs}
\pdfoutput=1
\usepackage{comment}
\usepackage[T1]{fontenc}
\usepackage{multirow}
\usepackage{array}
\usepackage{siunitx}
\usepackage{float}
\usepackage{xcolor}
\usepackage{svg}
\usepackage{xcolor}
\usepackage[bottom]{footmisc}
\usepackage{comment}
\usepackage{microtype}
\usepackage{subcaption}

\usepackage{graphicx} 
\usepackage{hyperref} 

\begin{document}
\title{\boldmath Relativistic Tidal Disruption in Black Hole and Wormhole Backgrounds}

\author{Pritam Banerjee}
\email{bpritamphy@gmail.com}
\affiliation{Indian Institute of Astrophysics, Koramangala, Bangalore 560034, India}

\author{Kowsona Chakraborty}
\email{kowsonac24@iitk.ac.in}

\author{Niles Mondal}
\email{nilesm24@iitk.ac.in}

\author{Tapobrata Sarkar}
\email{tapo@iitk.ac.in}

\affiliation{Department of Physics, Indian Institute of Technology Kanpur,
	Kanpur 208016, India}

\begin{abstract}
Black holes (BHs) and wormholes (WHs) are characterized by distinct spacetime geometries, whose differences become pronounced close to the central objects. A useful way to probe such differences is via the dynamics of stellar tidal disruption events in the regime of strong gravity. Here, using a general relativistic smoothed particle hydrodynamics code inspired from an algorithm developed by Liptai and Price, we perform a suite of numerical simulations of solar mass polytropic stars in the background of supermassive Schwarzschild BHs and similar mass exponential WHs. Important differences between the two geometries near the BH event horizon or the WH throat is provided by the distinct outcomes of such events. For a given impact parameter, BH backgrounds lead to greater tidal stripping compared to WHs ones and further, the critical impact parameter, beyond which the star undergoes full tidal disruption is higher for WH backgrounds compared to BHs. We further study the differences in observable peak fallback rates in the two backgrounds. We also provide a quantitative explanation for the tendency of stars in partial tidal disruptions to retain larger cores around more massive centers, by computing tidal stresses in a Fermi normal coordinate system and introducing an appropriate measure of stellar compactness. Finally, we suggest a way to observationally distinguish BH and WH backgrounds, based on the properties of different observables.
\end{abstract}

\maketitle

\section{Introduction} \label{sec:intro}
\subsection{Motivation and summary}\label{motivation}
The study of massive compact astrophysical objects with or without an event horizon remains one of the most
active areas of research in theories of gravity. The tidal forces they produce can fully or partially destroy a smaller object, 
such as a star, that comes sufficiently close. Such events, called tidal disruption events (TDEs)
\cite{Rees, Phinney, LattimerSchramm, EvansKochanek, Syer, MagorrianTremaine, Piran1} provide a
powerful observational probe into the strong gravitational regime of massive compact objects. In this context,
it is common to model the massive object as a Black Hole (BH), and indeed there is abundant literature by now
on TDEs involving stellar objects and Schwarzschild or Kerr BHs (for recent reviews see the collection of articles in \cite{Jonker}), inspired by the fact that
supermassive BHs with masses $M_\bullet \sim 10^6~-~10^{10}M_{\odot}$ are by now believed to exist at the center of most galaxies
(see, e.g., the reviews \cite{Genzel, Kormendy, Alexander}). In the last few decades, TDEs have been extensively
studied both analytically and numerically, and is an ideal theoretical laboratory to understand the physics of luminous electromagnetic
transients that result from them. Further, TDEs are also probes of theories of gravity via accretion physics and stellar
dynamics. In a TDE, a star approaching a BH is disrupted near the tidal radius, and produces a stream of debris that might
have complicated dynamics due to various competing forces acting on it. Depending on the nature of the TDE, a part of the
debris falls back into the BH  over time, and this generates a characteristic light curve whose behaviour encodes both the stellar
structure and the nature of the BH (for a small sampling of the literature see
\cite{Guillochon2013, 2019GReGr..51...30S, Lodato2009, Lodato2, Stone1, Stone2, Kesden1}).

While TDEs by BHs are believed to be a common phenomenon near galactic centers, the literature on tidal disruption by
horizonless compact objects is relatively rarer. This fact assumes importance, as currently there is no direct empirical observation
of the event horizon, neither is there an accepted proof of the cosmic censorship conjecture.
An example of such an exotic object is the wormhole (WH), a geometry
which contains a throat that connects two different universes or two distant regions of a single universe.
WHs continue to remain a subject of interest long after they were first proposed as a manifestation of a nontrivial change in the topology of spacetime \cite{MW}, and early efforts to explain their possible emergence through cosmological phase transitions were undertaken in
\cite{Sato1,Sato2,Sato3}. A WH is said to be traversable if physical observers or signals can pass through its throat.
The authors of \cite{FullerWheeler} showed that the Schwarzschild WH (called the Einstein–Rosen bridge)
does not satisfy this criterion. A traversable WH geometry was obtained in \cite{MorrisThorne}, and this work was soon
followed by proposals for constructing a “time machine” based on such geometries \cite{MTY,Novikov}. A comprehensive discussion of these developments can be found in the monograph by \cite{VisserBook} (see also \cite{Lemos} for additional related work). Also, it is well established that the matter sources required to sustain WH spacetimes generally violate the standard energy
conditions of general relativity \cite{MorrisThorne}, the so called ``exotic matter.'' Nonetheless, a variety of approaches have been explored to avoid these violations, in particular, time-dependent WH solutions \cite{Sayan1} and modified gravity theories \cite{Lobo1} are known to provide frameworks in which WH configurations may be realised without violating energy conditions.

Apart from purely theoretical reasons, our main interest in WH spacetimes is due to the fact that these are BH mimickers. For example, \cite{DamourSolodukhin} pointed out that several features of BH
spacetimes can be closely resembled by WHs as well, such as quasi-normal modes, accretion properties,
no hair theorems, lensing features, etc. (see, e.g., \cite{Cardoso1, KZ}). This
fact assumes importance in the context of the previous discussion as far as astrophysical observations
involving TDEs are concerned. Namely, we ask if (and to what extent) TDEs in BH and WH backgrounds are
similar, and if there is a way to observationally differentiate between these two spacetimes.
In a previous paper \cite{Pritam1}, we explored this question via some ``static'' properties of TDEs in
the two different backgrounds. Namely, we showed that although at large distances both BHs and WHs
are expected to behave like Newtonian objects, as one goes closer to the event horizon of the BH or the WH throat,
different tidal effects on stars can be seen, due to the different geometries.
There we quantified this difference by a numerical procedure by transforming to a local Fermi-normal (FN)
tetrad basis and solving the stellar hydrostatic equations in this basis following the earlier work
of \cite{Ishii} (see also \cite{GReffects1,GReffects2,GReffects3}). This analysis focused on the computation of a minimum central density below which a star would
be tidally disrupted, from which a critical mass could be obtained.
Now, the tidal radius computed by equating the tidal force on a star due to
a massive object with mass $M_\bullet$ with its self gravity at the surface gives an estimate of the closest
radial distance that the star can come near the massive object without getting tidally disrupted, and
is given by
\begin{equation}
r_t = R_*\left(M_\bullet/M_*\right)^{1/3}~,
\label{tidalradius}
\end{equation}
with $M_*$ and $R_*$ denoting the mass and radius of the star, respectively. In \cite{Pritam1},
the critical mass mentioned above was used to cast this relation in the form
$r_t/R_* = \alpha \left(M_\bullet/M_*\right)^\delta$, with $\alpha$ and $\delta$ characterising the variations
in BH and WH backgrounds.

Here, we go beyond this method to compute the time dependent dynamics of TDEs in BH and WH backgrounds. This involves numerical simulations of stars in these backgrounds in global coordinates, focusing on observables like fallback rates and gravitational waves. For the numerical analysis, we use a general relativistic smoothed particle hydrodynamics (GRSPH) code developed by us that extends our previous code, whose details
appear in \cite{Garain1}, and we closely
follow the algorithm 
as described by Liptai and Price \cite{LiptaiPrice} (see also \cite{Liptai2019}). While we will present the results of the simulations in the main \textcolor{black}{text} of the paper, the necessary numerical tests establishing the robustness of the code is detailed in Appendix \ref{Test Results}. 

\subsection{The black hole and wormhole gravity backgrounds}

To distinguish between BHs and WHs, we will study ultra deep tidal
encounters in the regime where General Relativity (GR) plays a
major role. To keep the analysis simple, we work with the Schwarzschild BH and Morris-Thorne type exponential WH \cite{MorrisThorne}, and compare TDEs in these backgrounds.
For both of the cases, the masses are $\sim {\mathcal O}(10^6-10^7 M_{\odot})$
and the stellar mass \textcolor{black}{and radius} is chosen as $1~M_{\odot}$ \textcolor{black}{and 1$R_\odot$ with polytropic equation of state having $\gamma =5/3$.} , so that back-reaction effects can be safely ignored and the BH and WH provide a fixed
background geometry.
The BH is quantified by the metric
\begin{equation}
ds^2 = -\left(1 - \frac{2GM_\bullet }{c^2 r}\right) c^2 dt^2 + \frac{1}{1 - \frac{2GM_\bullet}{c^2 r}} dr^2 + r^2 d\theta^2 + r^2 \sin^2 \theta d\phi^2~,
\end{equation}
with $M_\bullet$ being its ADM mass and $G$ and $c$ denote the Newton's gravitational constant and the speed of light, respectively.
On the other hand, the quintessential example of a static traversable WH of the Morris-Thorne type is provided by the metric written
in the same coordinates,
\begin{equation}
ds^2 = -e^{2\Phi(r)} c^2 dt^2 + \frac{1}{1 - \frac{b(r)}{r}} dr^2 + r^2(d\theta^2 + \sin^2 \theta d\phi^2)~.
\end{equation}
The example that we consider here is the exponential WH with
\begin{equation}
e^{2\Phi(r)} = e^{-\frac{2GM_\bullet}{c^2 r}}, \quad b(r) = \frac{2GM_\bullet}{c^2},
\end{equation}
where $M_\bullet$ is the ADM mass of the WH. The  WH throat  is defined by  $r_{th}=\frac{2GM_\bullet}{c^{2}}$. 
From a Newtonian perspective, an object that is captured by a BH\textcolor{black}{, instead} tunnels through the throat of a WH of the same mass. The choice of a constant shape function $b(r)$ results in zero energy density, and the WH spacetime is supported only by tangential and radial pressures. Consequently, there is no matter interaction with the star, making this a somewhat convenient case to compare with vacuum BH.
We consider parabolic stellar orbits (with eccentricity $e=1$) in the background of the BH and WH spacetimes introduced above. The interaction of the star with the background geometry is parametrized by the
\textcolor{black}{impact parameter} $\beta$. Defining the orbital periapsis distance as $r_p$,
this is given by $\beta \equiv r_t / r_p$,
with $r_t$ defined in Eq. (\ref{tidalradius}). For a fixed stellar structure, increasing $\beta$ corresponds to deeper encounters, bringing the star closer to the event horizon $r_s=\frac{2GM_\bullet}{c^2}$ in the BH spacetime, or to the throat $r_{\rm th}$ in the WH spacetime, thereby enhancing GR effects.

We define the critical impact parameter $\beta_c$ as the threshold separating encounters that leave behind a self-bound stellar remnant (partial disruptions) from those resulting in complete disruptions. For polytropic stars with adiabatic index $\gamma=5/3$ that we consider here, hydrodynamical simulations have long established that this transition is sharp. Using SPH simulations, \cite{Guillochon2013} first demonstrated that the transition occurs at $\beta_c \simeq 0.9$, a result subsequently confirmed and refined by \cite{Mainetti2017}, who obtained $\beta_c = 0.92 \pm 0.02$. An analytic prescription calibrated against these simulations by \cite{Coughlin2022} yields a slightly higher effective threshold, $\beta_c \simeq 0.95$--$0.96$, once differences in the definition of the tidal radius are taken into account. Relativistic studies incorporating BH spacetime curvature show that relativistic tidal enhancement produces only modest shifts in $\beta_c$ for encounters occurring outside the innermost stable circular orbit \cite{Gafton2015}.

In order to study the differences in BH and WH TDEs, we first fix the regime of interest. Note that from the definition of the impact parameter and using Eq. (\ref{tidalradius}), in units of the gravitational radius $r_g = GM/c^2$, the periapsis distance scales as $\frac{r_p}{r_g}\sim M_\bullet^{-2/3}$. With our choice of stellar parameters, if for example $M_\bullet\sim 10^{6}M_{\odot}$, then with $\beta\sim 1$, the periapsis is found to be $r_p\sim 50 r_g$. At such a large distance, both the BH and the WH backgrounds are effectively Newtonian and thus cannot possibly be distinguished by TDEs. With this value of the central mass, we find that $\beta_c \simeq 0.93$ for the BH spacetime (consistent with previous relativistic studies) while the WH spacetime exhibits a slightly higher threshold, $\beta_c \simeq 0.95$, and the difference is thus minor.

For examining genuine relativistic effects due to the two backgrounds, we thus need to consider higher values of the central mass \textcolor{black}{leading to lower $r_p(r_g)$}. In this paper, we have chosen to perform numerical simulations with the values of the BH and WH masses ranging from $6\times 10^6 - 1.2\times 10^7 M_{\odot}$, for which $r_p$ varies from $\sim 7 - 15 r_g$ and therefore our simulations capture genuine relativisic effects that should be useful in differentiating the nature of tidal dynamics in BH and WH backgrounds.

{\textcolor{black}{\section {Tidal tensor in the relativistic regime}}}
\label{FNC}

Since WHs are known to be BH mimickers, a natural question in the light of the above discussion would be to 
ask if TDEs from WHs look like ones from BHs with a different set of parameters. To this end, it is useful to examine two aspects of TDEs that
are most easily studied in the FN co-moving frame of the star, namely the tidal tensor and the spread of energy at the periapsis. 
To keep the analysis simple, we work in the so called ``tilde frame'' that is co-rotating with the star \cite{Ishii}. 
This will be enough to discuss the necessary physics.

We start with the tidal tensor defined as $C_{ij} = R_{\mu\nu\rho\lambda}{\hat e}^{\mu}_0{\hat e}^{\nu}_i{\hat e}^{\rho}_0{\hat e}^{\lambda}_j=R_{0i0j}$. Here,
$R_{\mu\nu\rho\lambda}$ is the Riemann tensor and ${\hat e}^{\alpha}_k$ denote the usual FN tetrads which
are written for parabolic orbits in \cite{Pritam1}. Using these, a simple
computation yields for our BH and WH backgrounds, the diagonal matrices with components
(evaluated at $r_p$; in this section we set $G=c=1$ to avoid cluttering of notation)
\begin{eqnarray}
C_{ij}^{\rm BH}&=& \left\{-\frac{2M_\bullet(M_\bullet+r_p)}{r^3(r_p-2 M_\bullet)}~ ,\frac{M_\bullet(4M_\bullet+r_p)}{r_p^3(r_p-2M_\bullet)}~,\frac{M_\bullet}{r_p^3}\right\}~,~~\nonumber\\
~C_{ij}^{\rm WH}&=&\left\{-\frac{M_\bullet}{r_p^5} \left(e^{\frac{2 M_\bullet}{r_p}} \left(2 M_\bullet^2-6 M_\bullet r_p+3
   r_p^2\right)+r_p^2\right), \frac{M_\bullet}{r_p^4} \left(e^{\frac{2 M_\bullet}{r_p}} (3 r_p-2M_\bullet)-2
   r_p\right), \frac{M_\bullet}{r_p^4}\left(r_p-2M_\bullet\right)\right\}~~~~,
\end{eqnarray}
where $M_\bullet$ is the mass of the central object (the BH result appeared previously in \cite{ChengEvans}).
These expressions become identical in the Newtonian limit $r_p \gg M_\bullet$, but clearly there is no simple change of parameters
that makes them equivalent at smaller distances. The above expressions show that whereas some of the tidal tensor components
diverge near the BH event horizon, they are finite valued near the WH throat. Hence, in this latter region, WHs are not
efficient BH mimickers as far as \textcolor{black}{deep tidal encounters} are concerned.

This can be further substantiated by the energy spread at the periapsis, that
captures the difference in specific energy between the center of mass of the star and a point on its surface, assuming that the
star retains its spherical shape at the periapsis. The expression for the energy spread was written in \cite{Kesden1} and
reads
\begin{equation}
\Delta E = -g_{\mu\nu}{\hat e}^{\mu}_0{\hat e}^{\rho}_i \Gamma^{\nu}_{\rho 0}X^i~,
\end{equation}
where $\Gamma$ is the Christoffel symbol and $X^i$ denotes the spatial distance measured in the FN coordinates.
A straightforward computation yields that at the periapsis,
\begin{equation}
\Delta E^{\rm BH} = \frac{M_\bullet}{\sqrt{r_p^3(r_p - 2M_\bullet)}}R_\ast~,~~
\Delta E^{\rm WH} = M_\bullet\sqrt{\frac{r_p-2M_\bullet}{r_p^5}}R_\ast~.
\label{spreadperiapsis}
\end{equation}
These are identical in the limit $r_p \gg M_\bullet$, but we again see that there is no simple relation that would effectively mimic the energy
spread of the WH with that of a BH at smaller values of $r_p$,
namely, $\Delta E$ diverges near the BH event horizon, but vanishes near the WH throat.

Now, a Keplarian approximation \cite{Lodato2} to the motion of stellar debris after a full TDE is robust even for close encounters as the debris spends
most of the time far from the central mass before it returns (approximately) to its initial position. With such an approximation, the
time of return of the most bound debris is known to be given by $T=\frac{\pi M_\bullet  }{\sqrt{2}}(\Delta E)^{-3/2}$, and hence we have, assuming
TDEs at the same $r_p$ and for the same central mass,
\begin{equation}
\frac{T^{\rm BH}}{T^{\rm WH}} =\left(1-\frac{2M_\bullet}{r_p}\right)^{3/2}~.
\end{equation}
Thus, for similar central masses and periapsis distances, $T^{\rm WH} \gg T^{\rm BH}$ if $r_p$ is close to $2M_\bullet$. From this we can
also conclude that since the peak of the mass fallback rate of the stellar debris into the BH horizon or WH throat can be
approximated to be ${\dot M}_{\rm peak} \sim 1/T$, then we have
\begin{equation}
\frac{{\dot M}_{\rm peak}^{\rm BH}}{{\dot M}_{\rm peak}^{\rm WH}} =\left(1-\frac{2M_\bullet}{r_p}\right)^{-3/2}~,
\end{equation}
Implying that the peak fallback rate of debris into the BH will be much higher than that into the WH, for full TDEs with similar parameters.
We point out here that the above analysis pertains to full TDEs. For partial TDEs, a FN frame analysis becomes more
challenged due to the competing gravity of the central mass and the remnant core, and a more refined approach is usually needed
that incorporates stellar hydrodynamics.

From the above discussion, it should be clear that when $r_p\approx {\rm a ~ few}~ r_g$, BHs and WHs \textcolor{black}{should} quantitatively differ as far as TDEs are concerned. Such
events at such small periapsis distances naturally call for \textcolor{black}{hydrodynamical} simulations and we now present our results from these.

\textcolor{black}{\section{Simulation results and observables}}

In this section, we present the results of our numerical simulations. These will include both theoretical outcomes of the TDEs as well as observables.

\textcolor{black}{\subsection{Visual differences of stellar density distribution}}

\begin{figure}[H]
    \centering

    \begin{subfigure}{0.48\textwidth}
        \centering
        \includegraphics[width=\linewidth]{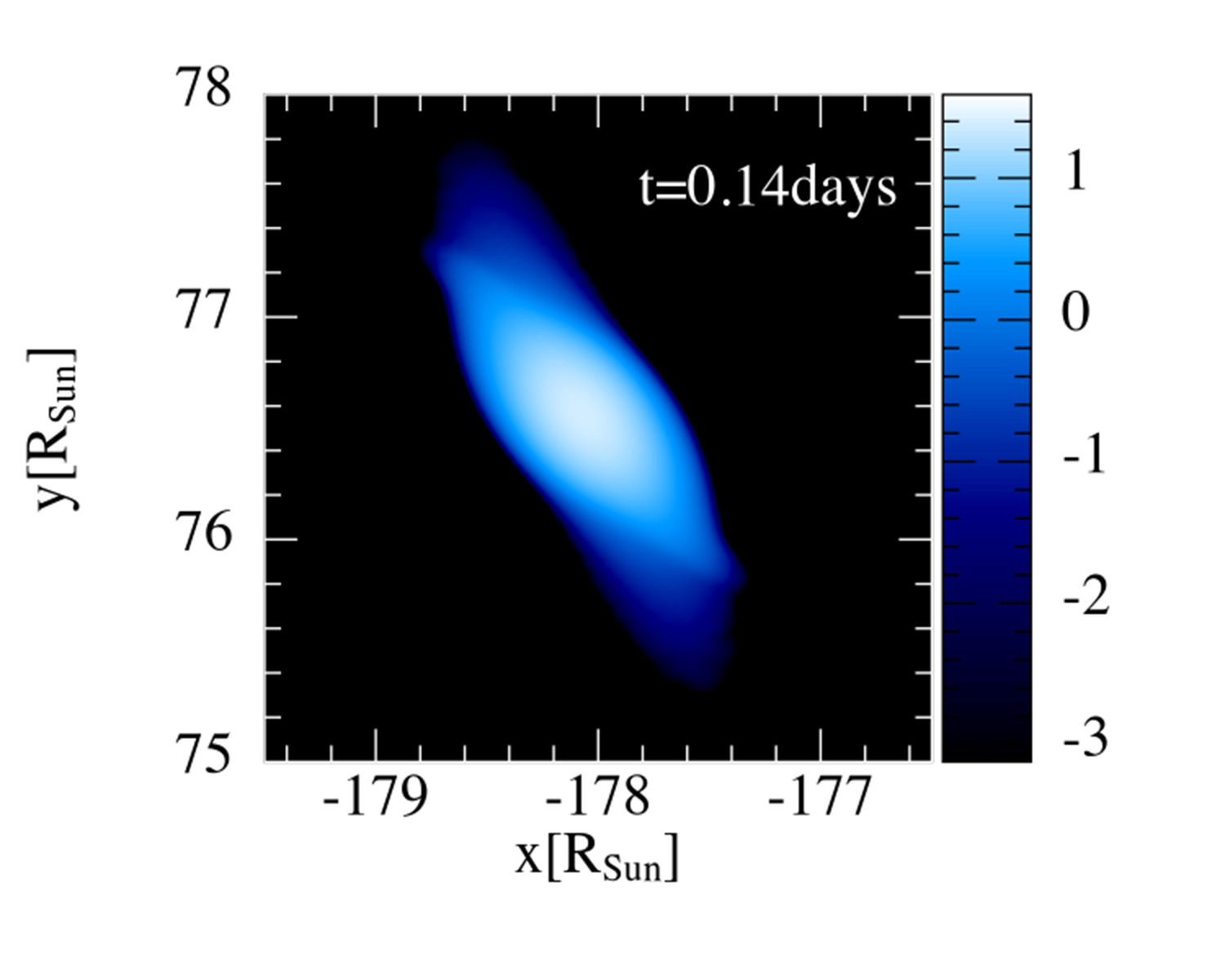}
        \caption{BH: Periapsis}
    \end{subfigure}
    \hfill
    \begin{subfigure}{0.48\textwidth}
        \centering
        \includegraphics[width=\linewidth]{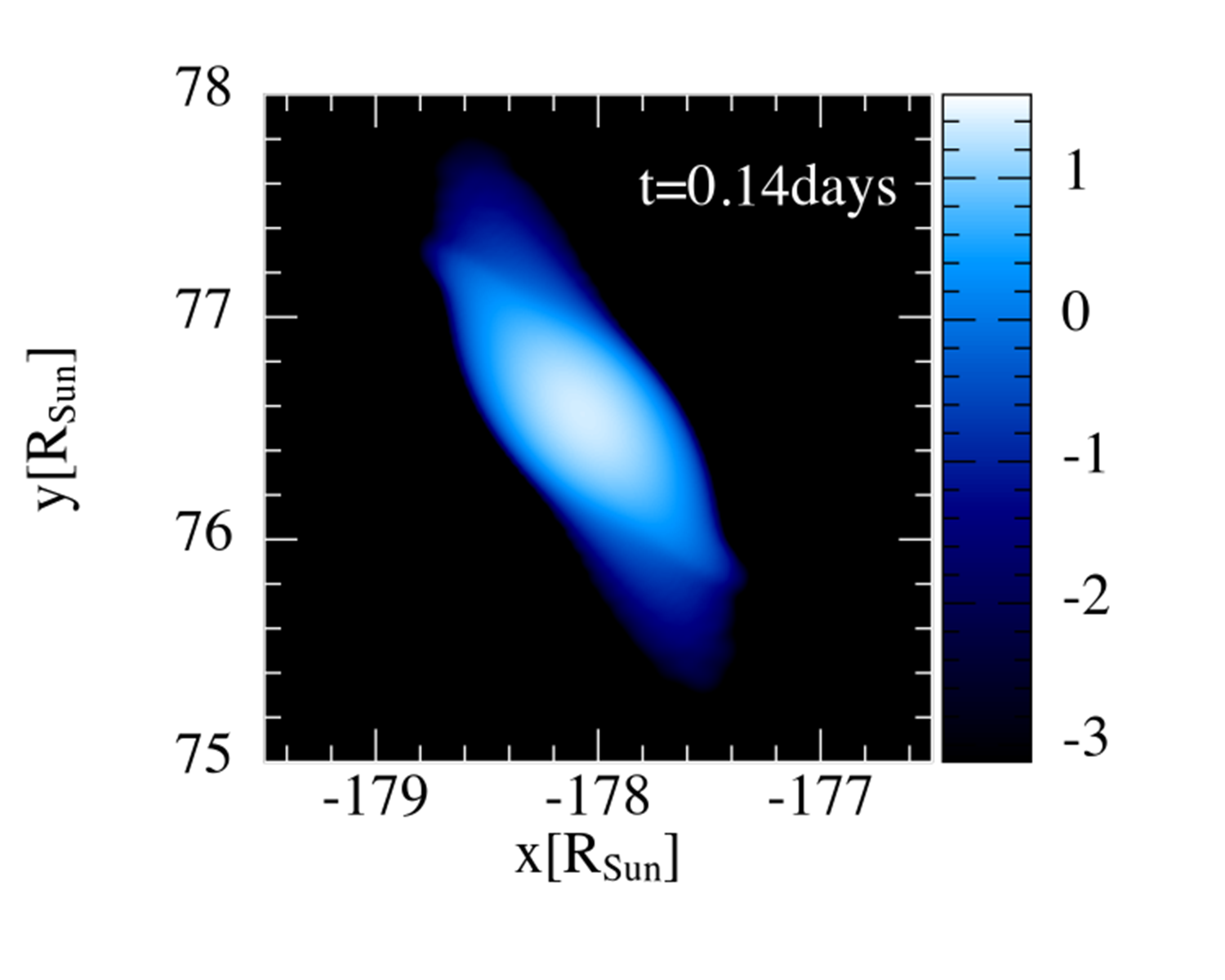}
        \caption{WH: Periapsis}
    \end{subfigure}

    \vspace{-0.20cm}

    \begin{subfigure}{0.48\textwidth}
        \centering
        \includegraphics[width=\linewidth]{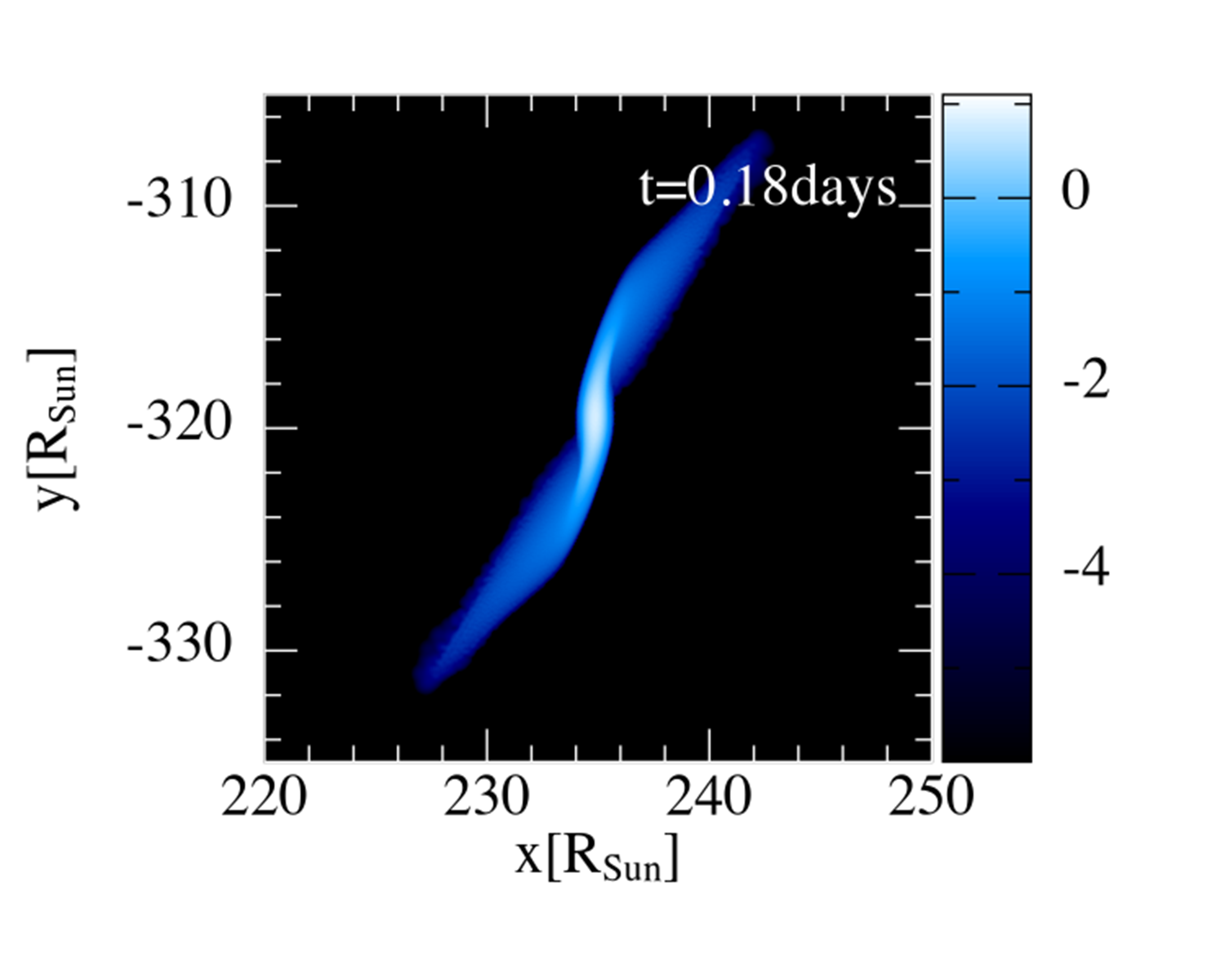}
        \caption{BH: $t = 0.18$ day}
    \end{subfigure}
    \hfill
    \begin{subfigure}{0.48\textwidth}
        \centering
        \includegraphics[width=\linewidth]{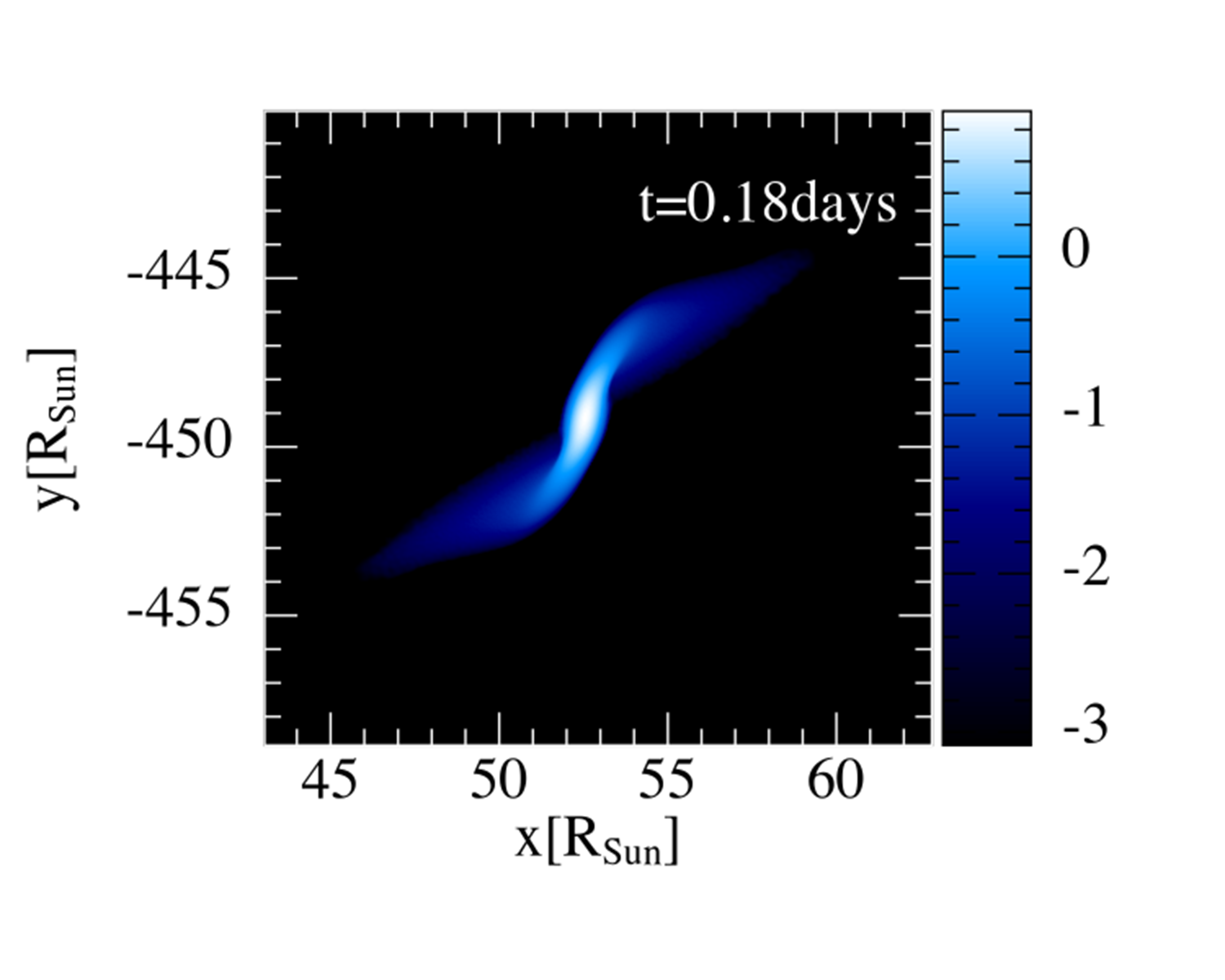}
        \caption{WH: $t = 0.18$ day}
    \end{subfigure}

    \vspace{-0.20cm}

    \begin{subfigure}{0.48\textwidth}
        \centering
        \includegraphics[width=\linewidth]{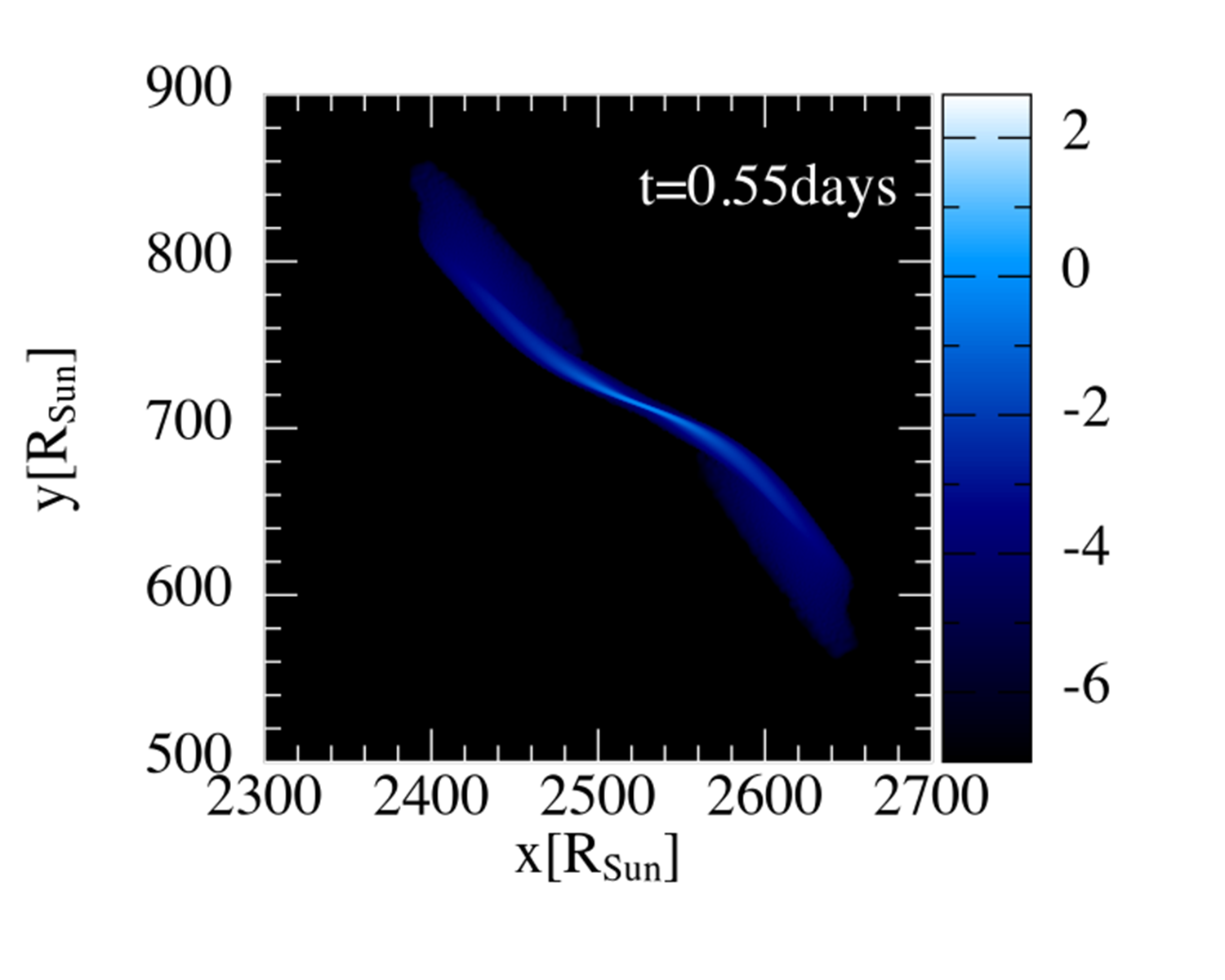}
        \caption{BH: $t = 0.55$ day}
    \end{subfigure}
    \hfill
    \begin{subfigure}{0.48\textwidth}
        \centering
        \includegraphics[width=\linewidth]{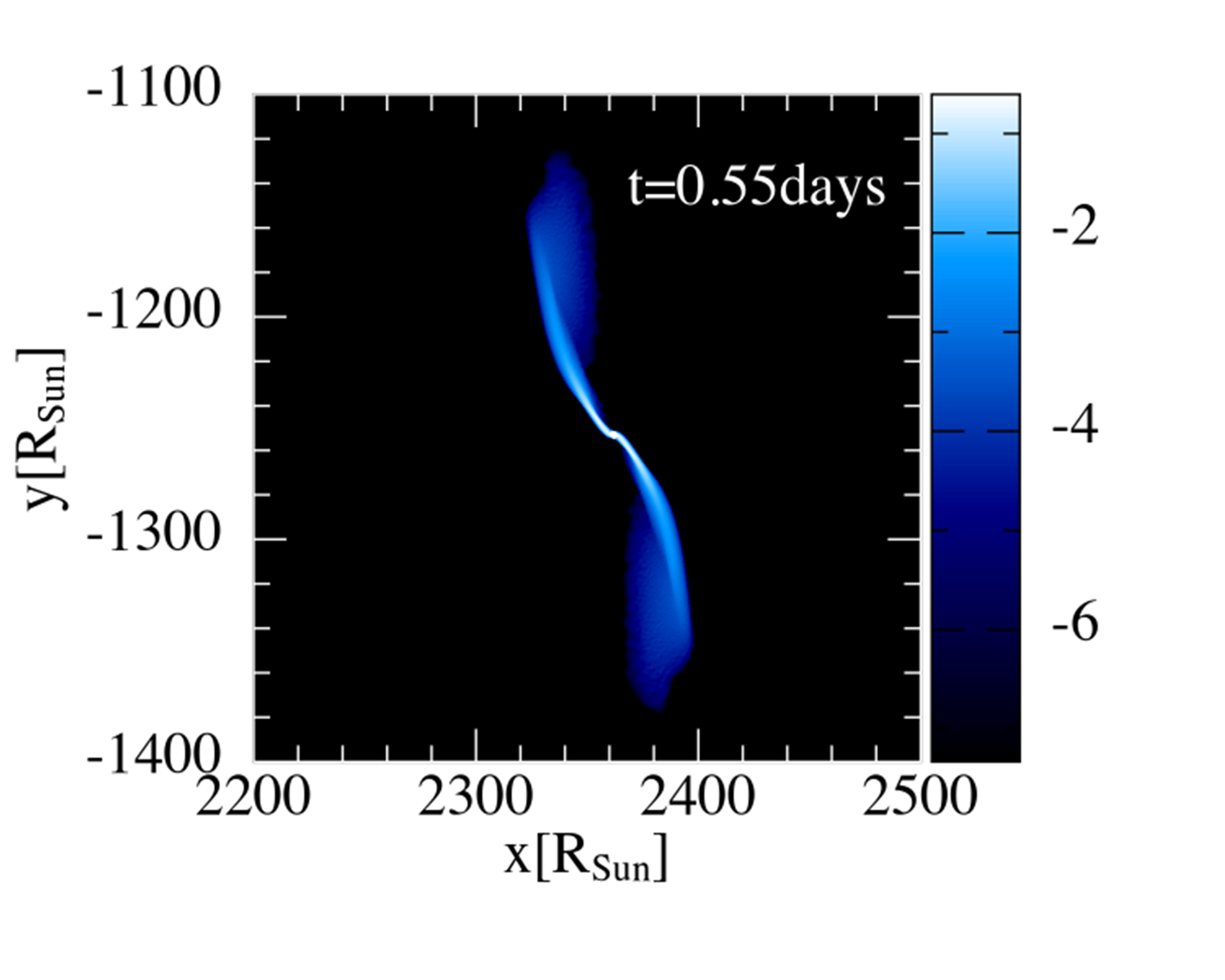}
        \caption{WH: $t = 0.55$ day}
    \end{subfigure}

    \vspace{-0.2cm}

    \caption{Evolution of debris around BH (left column) and WH (right column) of
    $1.2 \times 10^7 M_\odot$, $\beta = 1.3$ ($r_p \approx 6.9 r_g$).
    Log density (in $g\,cm^{-2}$) is shown via the color bar. A remnant
    self-bound core is visible at $t = 0.55$ days in the WH case (bottom right),
    indicating partial disruption.}
    
    \label{fig:bh_wh_combined}
\end{figure}

To begin with, it is useful to provide a \textcolor{black}{comparison} of the two scenarios.
To this end, we show aspects of the remnant core survival with fixed initial parameters.
We compare tidal disruption events produced by a BH and an WH, evolved from identical initial stellar and orbital parameters, with a central mass $M_\bullet = 1.2\times 10^7M_{\odot}$, and impact parameter $\beta = 1.3$. Both spacetimes exhibit comparable tidal deformation near periapsis ($t/t_p \simeq 1$). \textcolor{black}{Subsequently, at }$0.18$ days and $0.55$ days, the differences in the two cases are clearly \textcolor{black}{visible}. While the BH encounter results in complete destruction of the stellar core (left column of Figure~\ref{fig:bh_wh_combined}), the WH retains a compact, gravitationally bound core (right column of Figure~\ref{fig:bh_wh_combined}).\\

\subsection{Nature of \textcolor{black}{the} bound debris in partial disruptions}

\begin{figure}[h!]
    \centering
    \setlength{\tabcolsep}{1pt}   
    \renewcommand{\arraystretch}{0.01} 
    \begin{tabular}{cc}
        \includegraphics[width=0.5\textwidth]{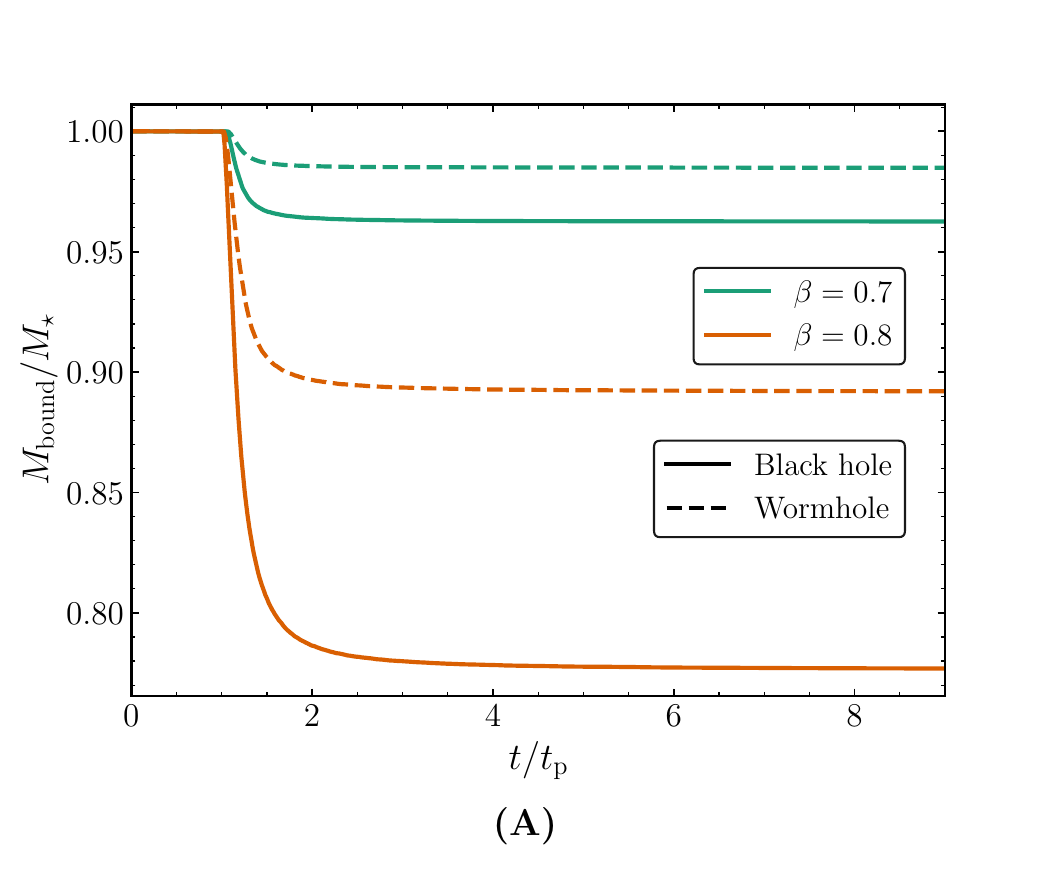} &
        \includegraphics[width=0.5\textwidth]{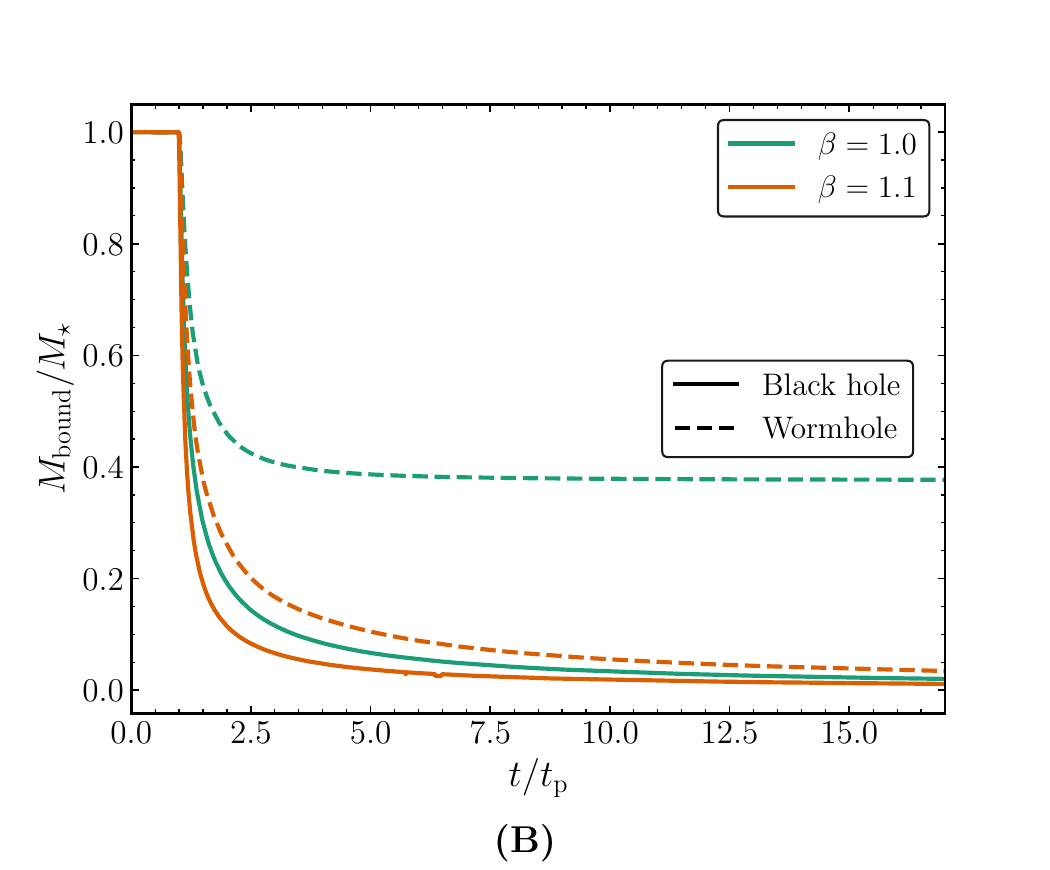} \\
        \includegraphics[width=0.5\textwidth]{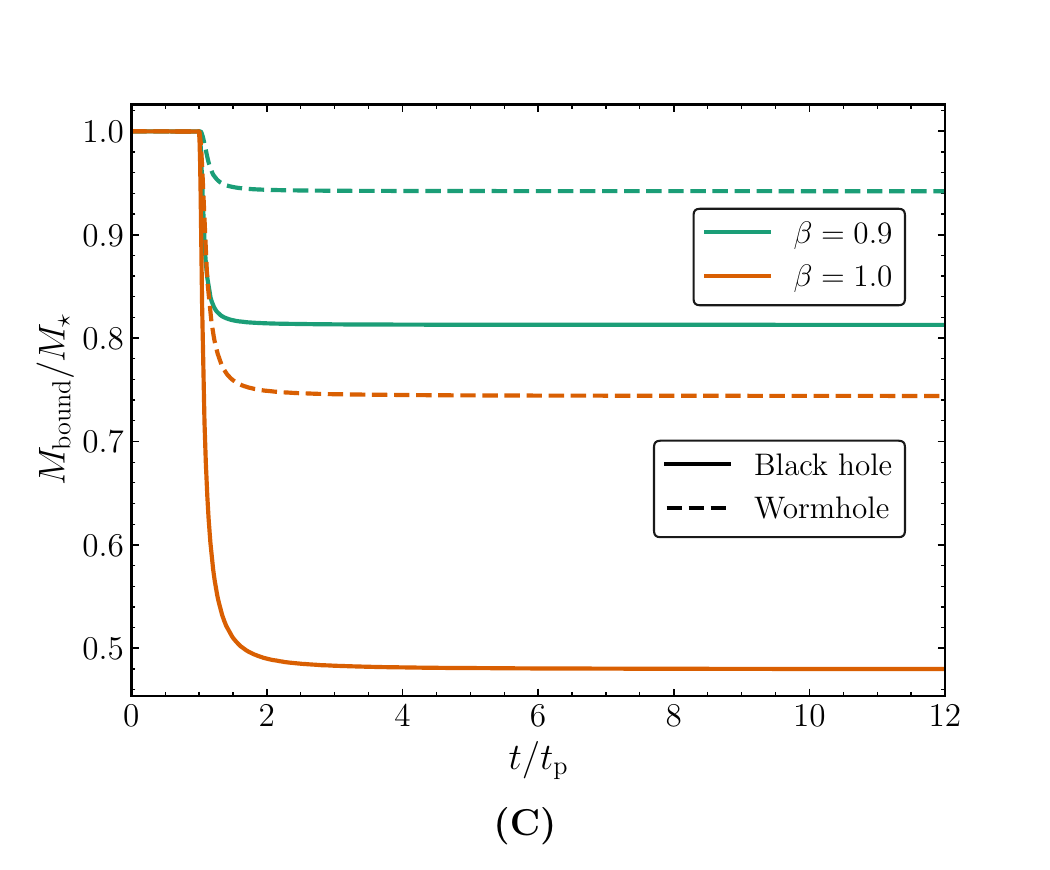} &
        \includegraphics[width=0.5\textwidth]{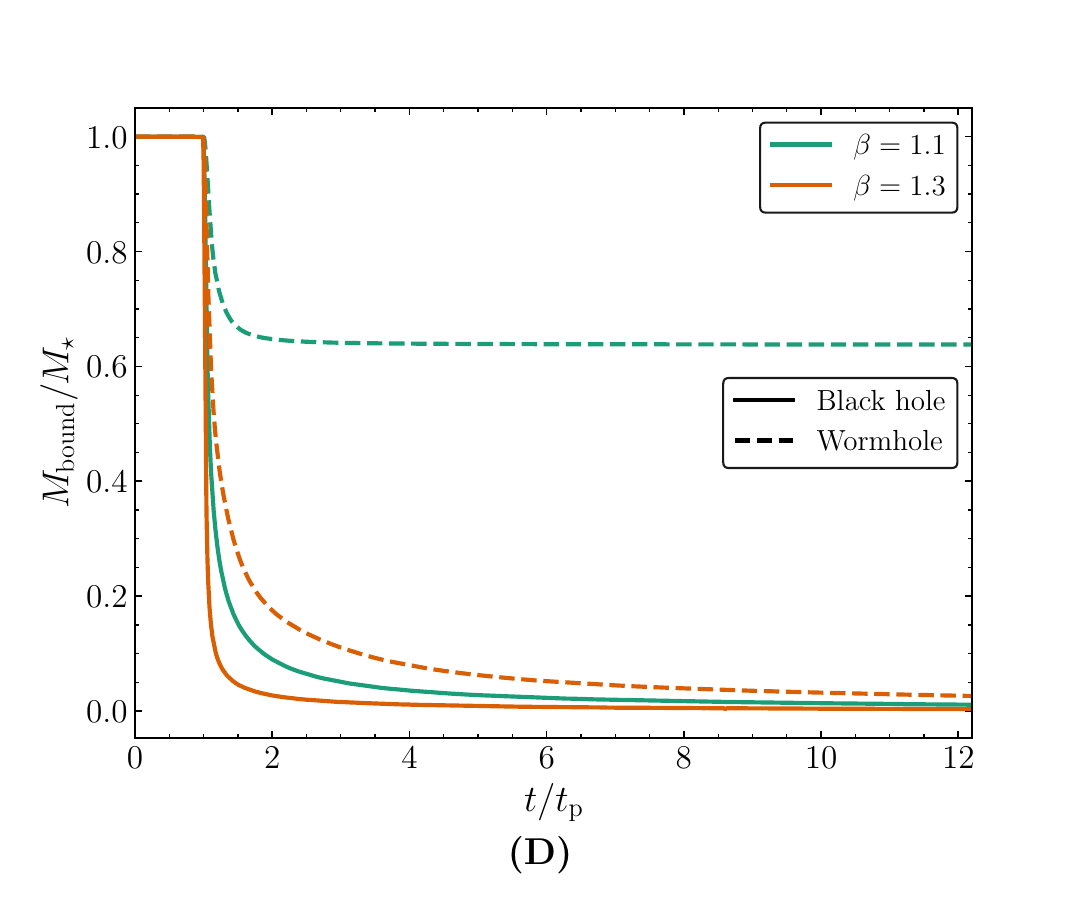}
    \end{tabular}
    \caption{
Bound core mass fraction, $M_{\rm bound}/M_\ast$, as a function of periapsis-normalized time, $t/t_p$, for BH (solid lines) and WH (dashed lines) encounters. 
Panels (A) and (B) correspond to   $M_\bullet = 6\times 10^6\,M_{\odot}$, while panels (C) and (D) correspond to $M_\bullet = 1.1\times 10^7\,M_{\odot}$. 
Each panel shows two values of the impact parameter $\beta$, as indicated in the legends. 
In all cases the WH consistently retaining a larger bound core compared to the BH for the same $\beta$.
}
    \label{fig:four_compact}
\end{figure}

For relatively smaller values of $\beta$ than those considered above, there is a self-bound core that is left behind after the tidal encounter. The mass of this core is different in BH and WH backgrounds. In  Figure~\ref{fig:four_compact}, we show the behaviour of the self-bound core for two different central masses $6\times 10^6 M_{\odot}$ (Figure~\ref{fig:four_compact}: A and B) and $1.1\times 10^7 M_{\odot}$ (Figure~\ref{fig:four_compact}: C and D) for a few different values of $\beta$. It is clearly seen that for partial TDEs, WHs retain significantly higher core mass compared to BHs of the same mass, for a given value of $\beta$. On the other hand, \textcolor{black}{in higher $\beta$ values}, BHs produce full disruptions while WHs still preserve cores. This brings us to the issue of the critical impact parameter beyond which a star is disrupted.

\subsection{Critical Impact Parameter}

\begin{figure}[h!]
    \centering
    \setlength{\tabcolsep}{1pt} 
    \begin{tabular}{cc}
        \includegraphics[width=0.49\textwidth]{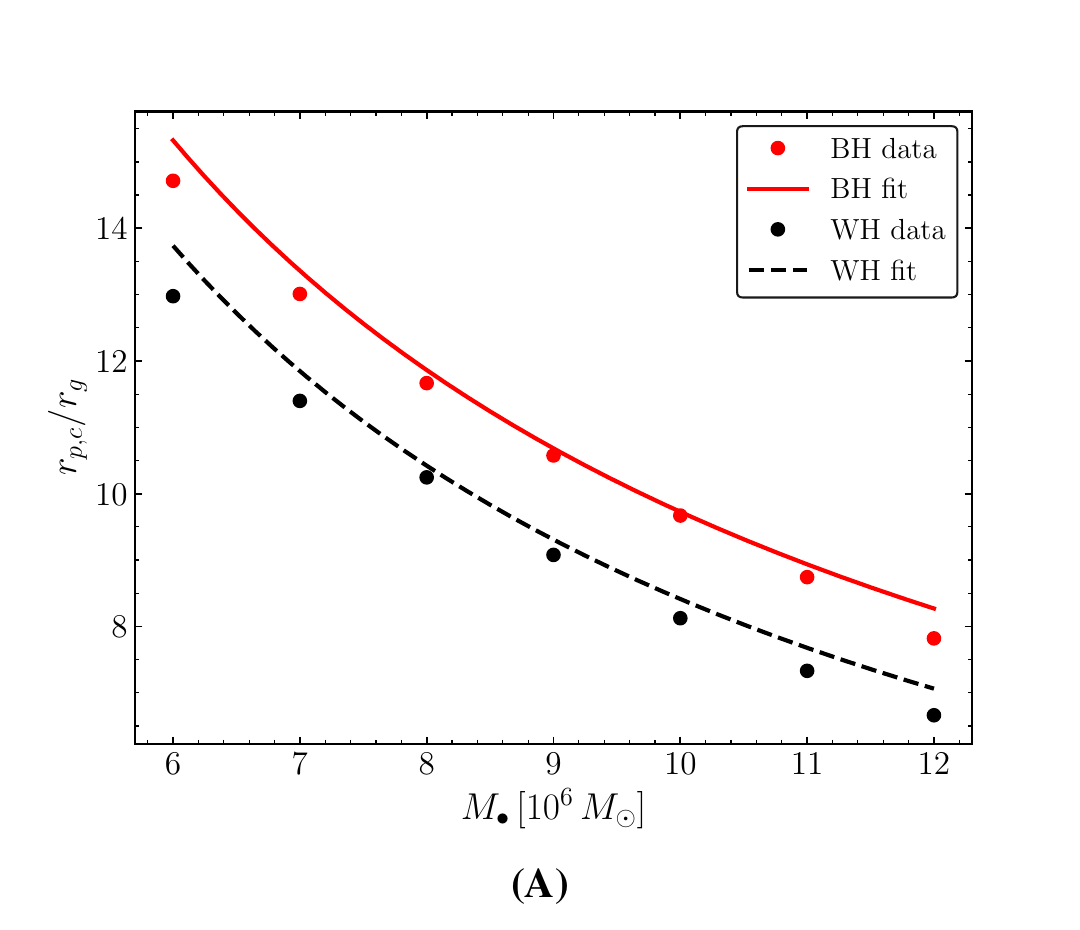} &
        \includegraphics[width=0.49\textwidth]{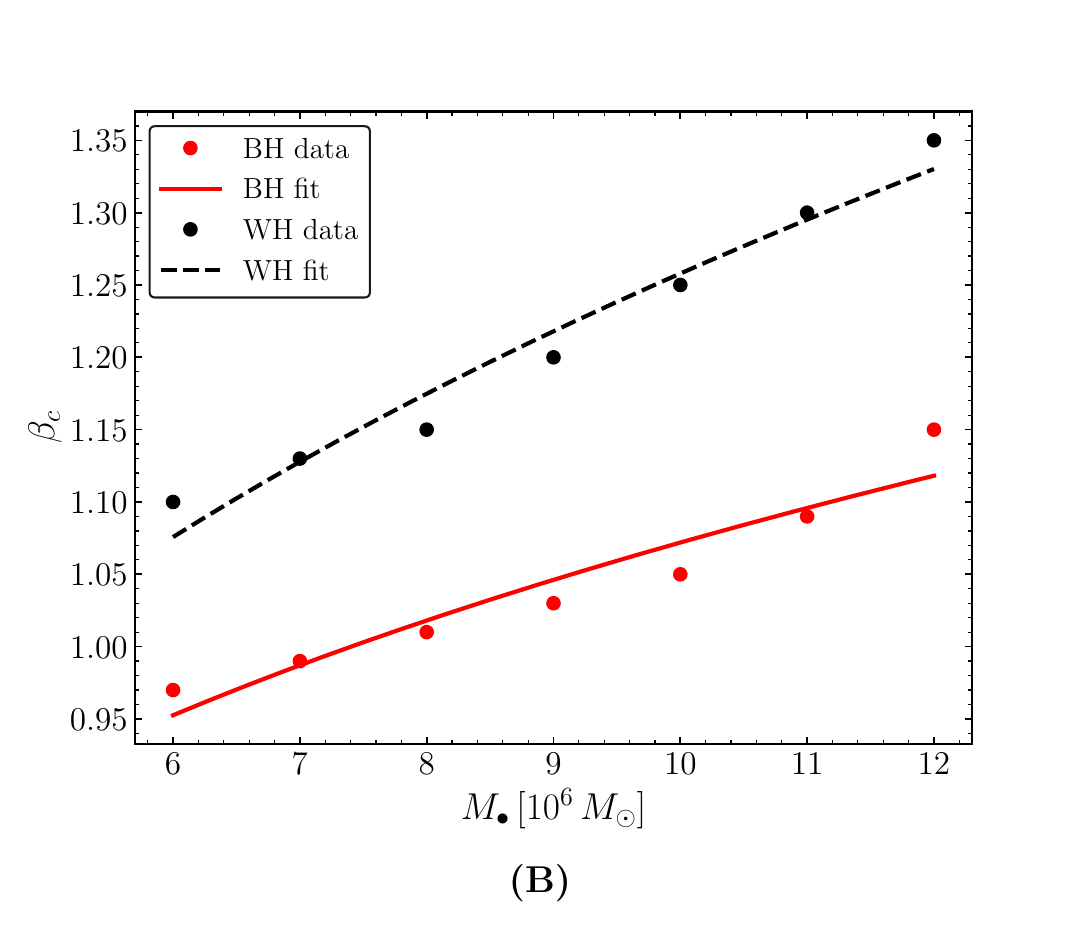}
    \end{tabular}
    \caption{
Critical periapsis distance and critical impact parameter as functions of the BH and WH mass, $M_\bullet$, shown in units of $10^6\,M_\odot$. 
Panel (A) shows the critical periapsis distance, $r_{p,c}/r_g$, and panel (B) shows the critical impact parameter, $\beta_c$, for the BH and WH cases. 
The red points and solid curves correspond to the BH data and fit, while the black points and dashed curves correspond to the WH data and fit, respectively. 
}
    \label{fig:beta_c}
\end{figure}

We characterize the transition between partial and full disruption using the critical periapsis distance $r_{p,c}$ (smallest periapsis for which a self-bound core survives) or, equivalently the critical impact parameter $\beta_c \equiv r_t / r_{p,c}$. Figure~\ref{fig:beta_c} shows that in both spacetimes, $r_{p,c}$ decreases with increasing mass, but WHs consistently exhibit smaller $r_{p,c}$ than BHs, indicating stars can penetrate deeper before complete disruption. Correspondingly, WHs exhibit systematically larger $\beta_c$ at all masses. We perform power-law fits for both the critical pericenter and the critical impact parameter. For the critical pericenter, we obtain $r_{p,\mathrm{c}}/r_g = 75.50\,\times (M_\bullet/10^6\,M_\odot)^{-0.89}$ for the BH, and $r_{p,\mathrm{c}}/r_g = 76.72\,\times (M_\bullet/10^6\,M_\odot)^{-0.96}$ for the other case. Similarly, for the critical impact parameter, we find $\beta_{\mathrm{c}} = 0.63\,\times (M_\bullet/10^6\,M_\odot)^{0.23}$ for the BH, and $\beta_{\mathrm{c}} = 0.62\,\times (M_\bullet/10^6\,M_\odot)^{0.30}$ for the WH case.


\subsection{Fallback rates}

\begin{figure}[h!]
    \centering
    \setlength{\tabcolsep}{1pt} 
    \begin{tabular}{cc}
        \includegraphics[width=0.49\textwidth]{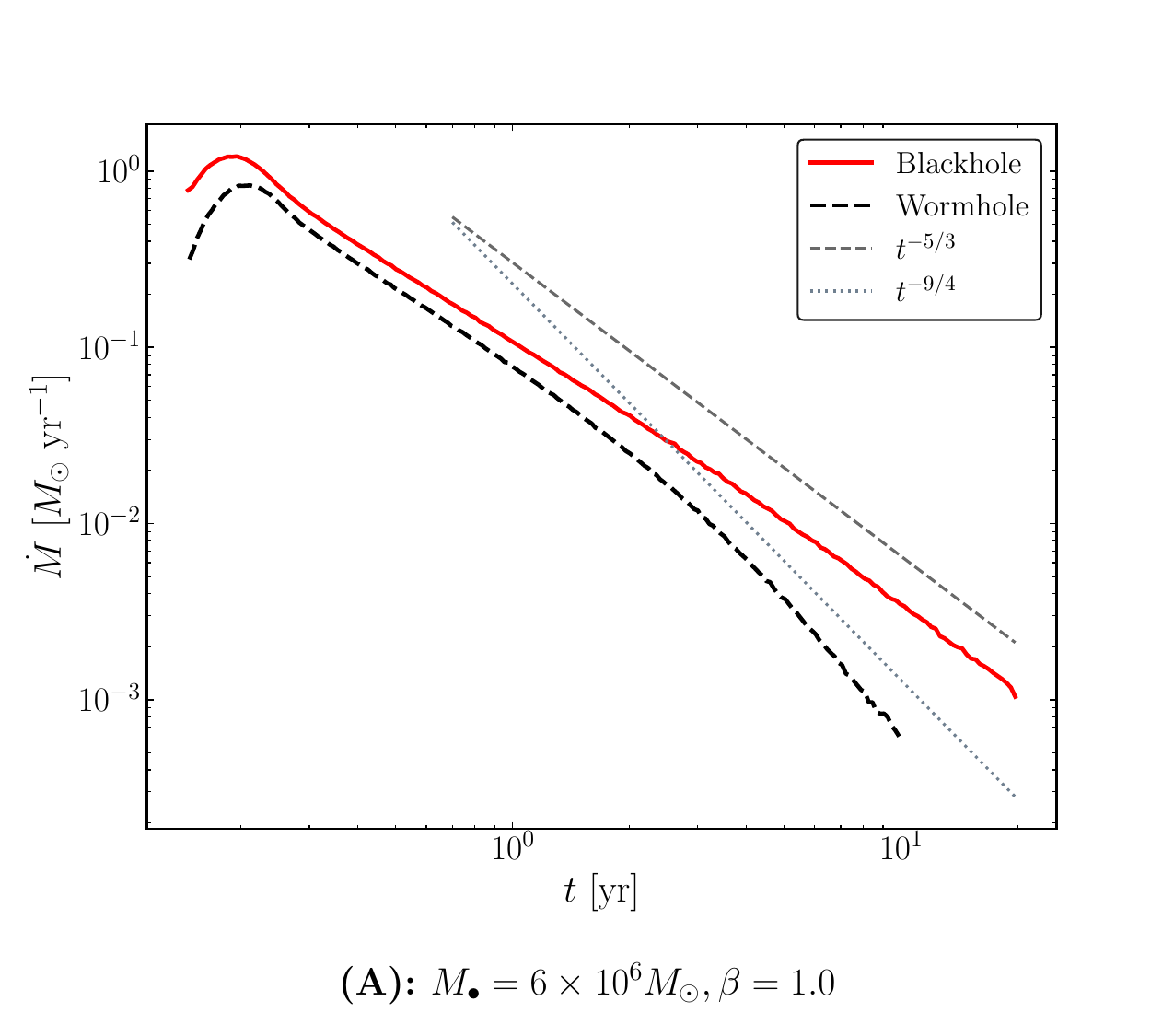} &
        \includegraphics[width=0.49\textwidth]{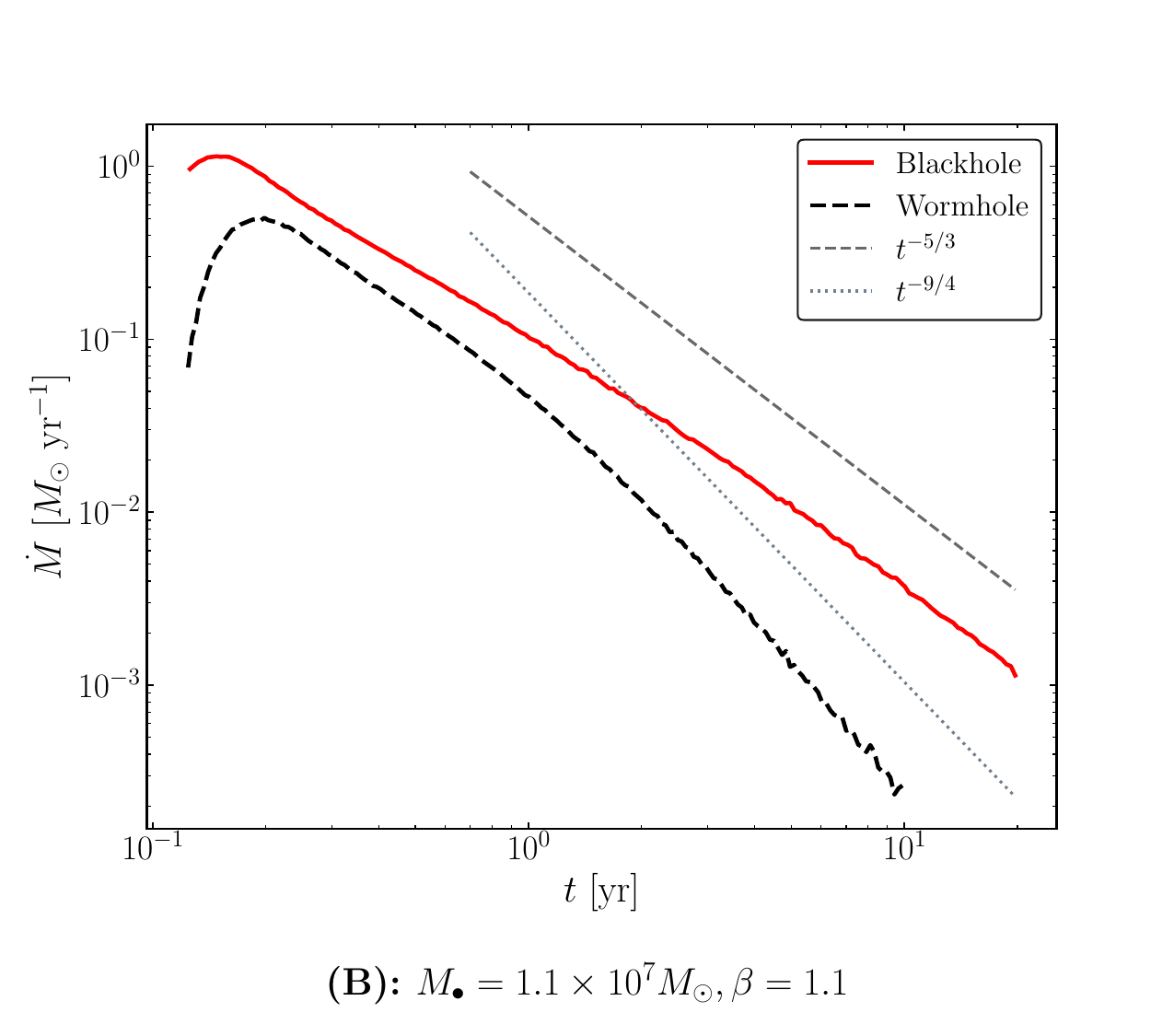}
    \end{tabular}
   \caption{
Fallback rate, $\dot{M}$, as a function of time, $t$, for BH (solid red) and WH (dashed black) cases. 
Panel (A) corresponds to $M_\bullet = 6\times 10^6\,M_\odot$, $\beta = 1.0$, and panel (B) to $M_\bullet = 1.1\times 10^7\,M_\odot$, $\beta = 1.1$, with reference slopes $t^{-5/3}$ and $t^{-9/4}$ shown. 
The WH case exhibits a steeper late-time decay compared to the BH.
}
    \label{fig:fallback}
\end{figure}

We now study aspects of the fallback rates of debris into the BH or the WH. These are expected to mimic observed light curves from TDEs.
Figure~\ref{fig:fallback} shows the fallback rate in our system by  compact object of masses $M_\bullet = 6 \times 10^6 M_\odot$  and
$M_\bullet = 1.1 \times10^7 M_\odot$ with impact parameters $\beta = 1.0$ ($r_p \approx 14.2 r_g$) and $\beta = 1.1$ ($r_p \approx 9.2 r_g$) respectively.
The results presented here are obtained within the frozen-in approximation. In this \textcolor{black}{approximation}, the debris energy distribution \textcolor{black}{once established,} is assumed to remain
unchanged thereafter. We have checked in a few cases that fallback rate obtained from the frozen-in approximation
is in excellent agreement with those from \textcolor{black}{hydrodynamical} simulations, although the latter often involves substantial computation costs.

For the BH case, the disruption is \textcolor{black}{full}, i.e., no bound stellar core survives the encounter.
In contrast, the WH case results in partial disruption,
with approximately $52\%$ of the stellar core remaining gravitationally bound.
This qualitative difference in the disruption outcome is directly reflected
in the structure of the fallback curves.
Note that the BH case exhibits a higher peak fallback rate compared
to the WH case. The peak is also marginally earlier in time.
At late times, the BH fallback curve follows the usual canonical
$t^{-5/3}$ power-law decline expected for a full disruption\cite{Lodato2009},
consistent with the standard analytical prediction
By contrast, the WH curve exhibits a steeper decay, closer to
$t^{-9/4}$ as indicated in the figure which was predicted by  \cite{Miles2020} in partial disruption cases.
This comparison demonstrates that, even for identical
$M_\bullet$ and $\beta$, the spacetime geometry strongly influences
the disruption outcome and the subsequent fallback evolution.

The behaviour of the fallback rate is explained by differential mass distribution of stellar debris. 
\begin{figure}[h!]
    \centering
    \setlength{\tabcolsep}{1pt} 
    \begin{tabular}{cc}
        \includegraphics[width=0.49\textwidth]{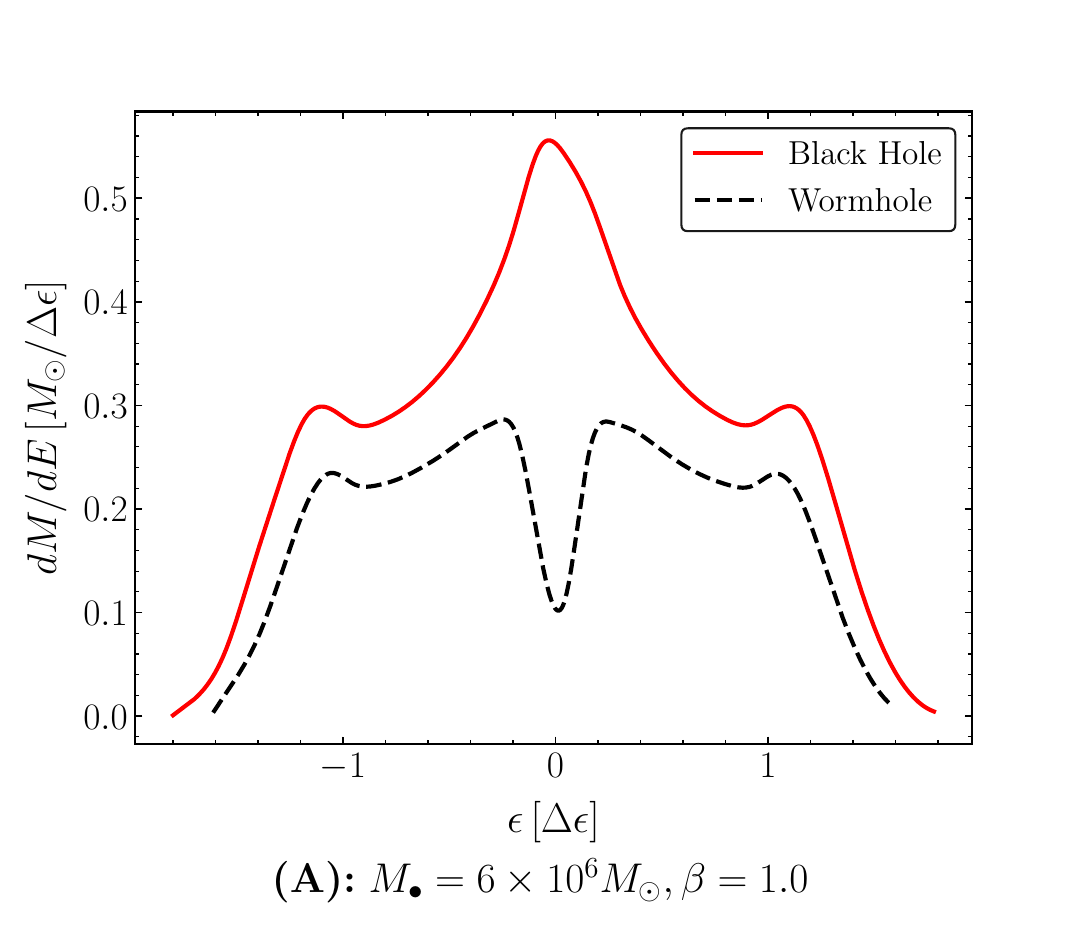} &
        \includegraphics[width=0.49\textwidth]{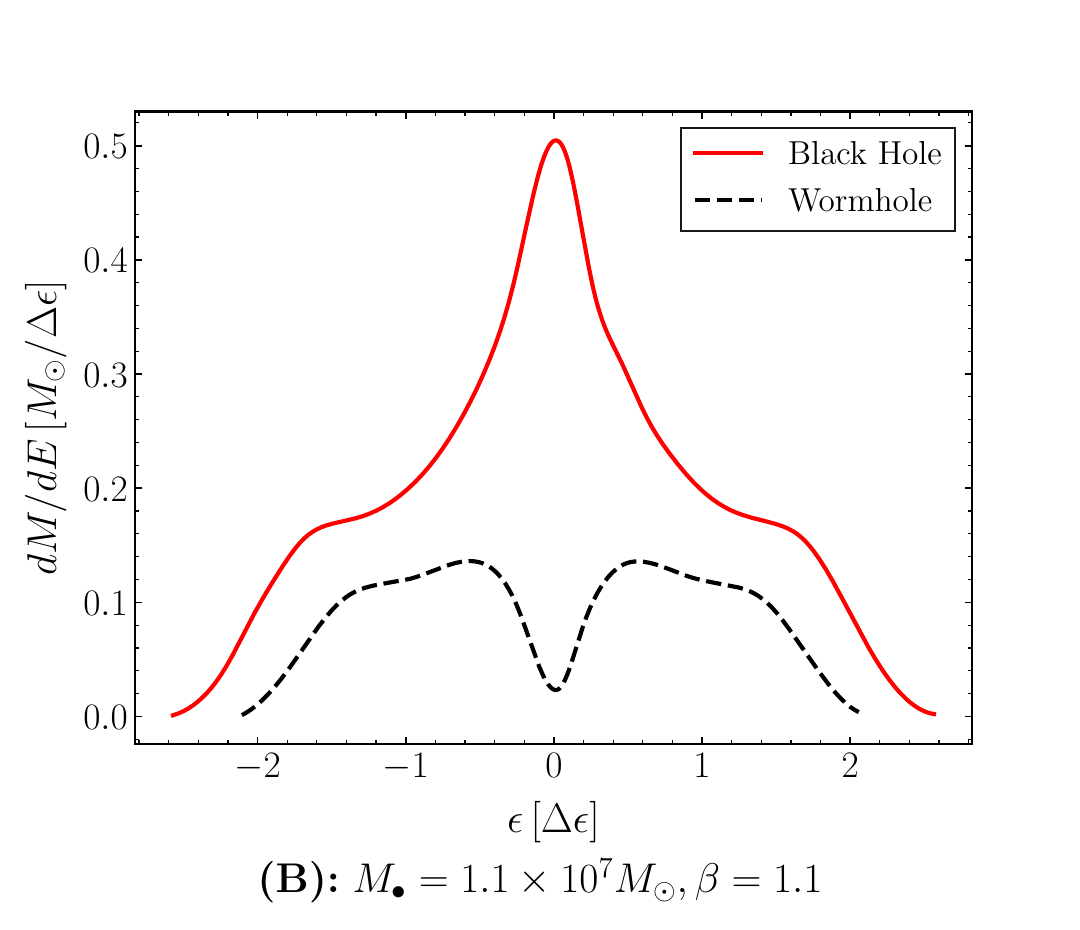}
    \end{tabular}
   \caption{
Energy distribution, $dM/d\epsilon$, as a function of normalized energy, $\epsilon/\Delta\epsilon$, for the BH (solid red) and WH (dashed black) cases. 
Panel (A) corresponds to $M_\bullet = 6\times 10^6\,M_\odot$, $\beta = 1.0$, and panel (B) to $M_\bullet = 1.1\times 10^7\,M_\odot$, $\beta = 1.1$. 
The WH case shows a suppressed and split distribution near $\epsilon \sim 0$, while the BH case remains more centrally peaked.
}
    \label{fig:two_compact}
\end{figure}
In Figure~\ref{fig:two_compact}, the differential mass distribution $dM/d\epsilon$ is plotted as a function of the specific energy in units of $\Delta\varepsilon = \frac{G M_\bullet R_\ast}{r_t^{2}}$, for the same values of parameters used in Figure~\ref{fig:fallback}. From this figure, we see that the WH exhibits a clear dip at $\varepsilon\simeq 0$ in contrast to the BH case. Since the core particles have been removed prior to constructing the histogram, any deficit near zero energy implies that a substantial fraction of mass remains concentrated in a bound, compact core that does not contribute to the debris distribution. The dip therefore signals a partial disruption, where a significant remnant core survives the tidal encounter.

In contrast, the BH case does not exhibit such a dip, since this corresponds to a full TDE, i.e., the entire stellar mass is dispersed into tidal debris.
An important feature of Figure \ref{fig:two_compact} is the difference in the overall width of the energy distribution between the BH and WH cases, and this directly relates to our discussion in section \ref{FNC}, see Eq. (\ref{spreadperiapsis}). As discussed there, the BH produces a noticeably broader $dM/d\epsilon$ indicating a larger spread in specific orbital energies imparted to the stellar debris during the disruption. In contrast, the WH distribution is narrower, implying that the debris occupies a more restricted range of energies. A broader energy distribution, as seen in the BH cases, implies the presence of more tightly bound material. Consequently, the most-bound debris in the BH spacetime returns in shorter timescales, leading to an earlier and higher peak in the fallback rate. In contrast, the narrower energy spread in the WH case indicates that fewer debris elements achieve large binding energies. The most-bound material is therefore less tightly bound than in the BH case, resulting in a longer return time for the earliest fallback and a reduced peak accretion rate.

\textcolor{black}{\subsection{Gravitational wave emission from tidal encounters}}
\label{GW}

We now compare Gravitational wave (GW) emission between BH and WH backgrounds. GW emission from TDEs arises from the time-varying quadrupole moment during \textcolor{black}{the star's }periapsis passage \cite{Peters1963}. The characteristic GW frequency associated with a TDE is set by the dynamical timescale at periapsis,
\begin{equation}
f_{\rm GW} \sim \left( \frac{G M_\bullet}{r_t^3} \right)^{1/2} \sim 6 \times 10^{-4}\,{\rm Hz}\,\beta^{3/2}
\left( \frac{M_\ast}{M_\odot} \right)^{1/2}
\left( \frac{R_\ast}{R_\odot} \right)^{-3/2},
\label{eq:fgw_scaling}
\end{equation}
while the strain amplitude at luminosity distance $d_L$ is:
\begin{equation}
h_{\rm GW} \sim 2 \times 10^{-22}\,
\beta
\left( \frac{M_\bullet}{10^6\,M_\odot} \right)^{2/3}
\left( \frac{M_\ast}{M_\odot} \right)^{4/3}
\left( \frac{R_\ast}{R_\odot} \right)^{-1}
\left( \frac{d_L}{10\,{\rm Mpc}} \right)^{-1}.
\label{eq:hgw_scaling}
\end{equation}
This estimate confirms that TDE GW signals are intrinsically weak but increase in strength for deep encounters and nearby sources.

\begin{figure}[h!]
    \centering
    \begin{subfigure}{\textwidth}
        \centering
        \includegraphics[width=\linewidth]{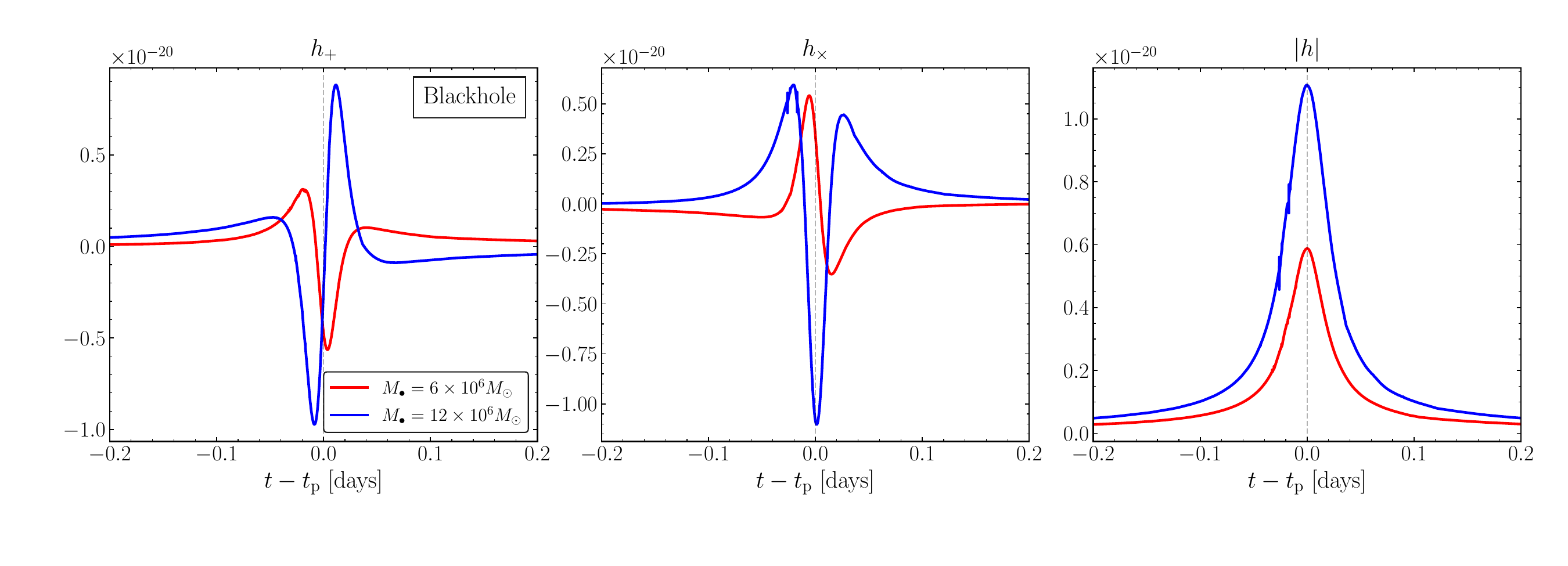}
        \caption{}
        \label{fig:BH_GW_faceon}
    \end{subfigure}
    \vspace{0.5em}
    \begin{subfigure}{\textwidth}
        \centering
        \includegraphics[width=\linewidth]{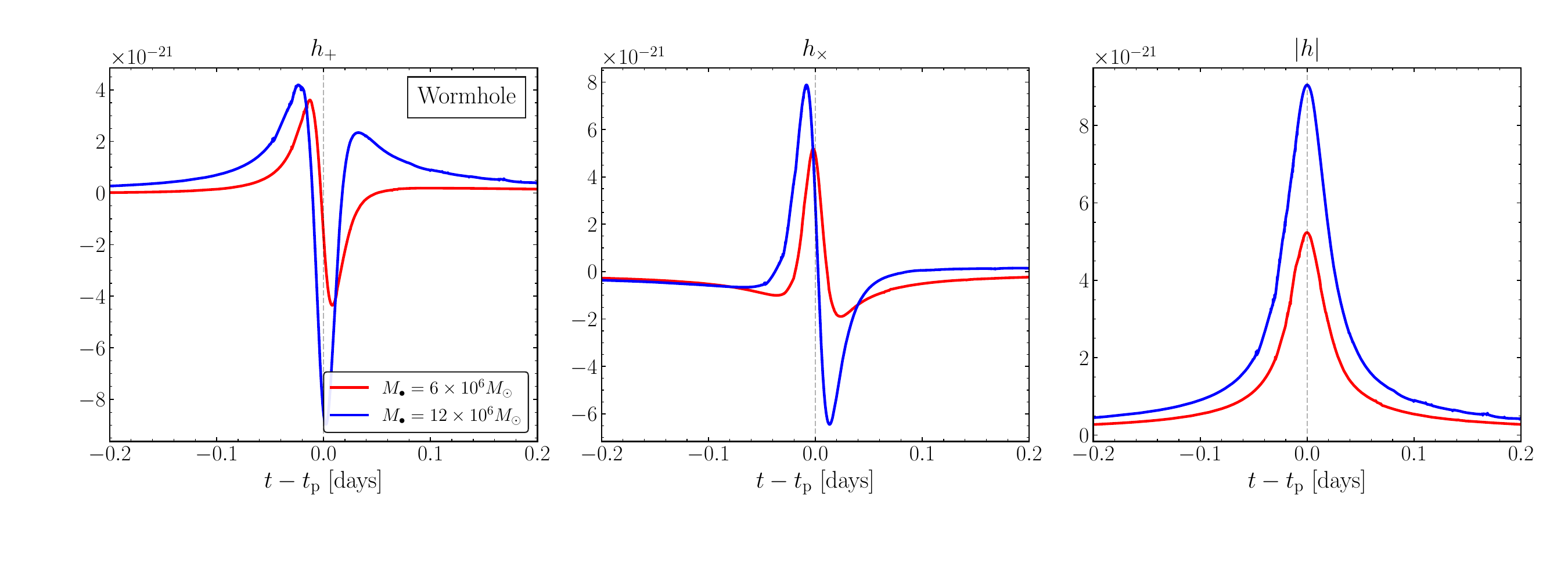}
        \caption{}
        \label{fig:WH_GW_faceon}
    \end{subfigure}
    \caption{Gravitational wave strain components, $h_+$ and $h_\times$, and total strain amplitude $|h|$, as functions of $(t - t_p)$ for face-on inclination, obtained from our simulations at $\beta = 1.1$.
    Each row displays three panels corresponding to $h_+$ (left), $h_\times$ (middle), and $|h|$ (right).
    In both cases the red and blue curves correspond to $M_\bullet = 6.0\times10^6\,M_\odot$ and $M_\bullet = 1.2\times10^7\,M_\odot$, respectively.
    (a) BH case: The peak amplitude occurs near periapsis, with higher mass producing a stronger and sharper GW signal.
    (b) WH case: Similar to the BH case, the signal peaks near periapsis, but the WH shows slightly lower amplitude.}
    \label{fig:GW_faceon_combined}
\end{figure}

For deep encounters with $\beta \gtrsim 1$ and periapsis distances approaching the gravitational radius, relativistic effects become increasingly important. In the post-Newtonian (PN) framework developed by \cite{BlanchetDamour1986}, the waveform is expressed in terms of radiative mass and current multipole moments, which differ from their Newtonian counterparts due to relativistic corrections. In our analysis, these effects are incorporated by consistently including 1PN corrections to the orbital motion when estimating the GW strain and frequency, ensuring validity for moderately relativistic TDE encounters.
At leading PN accuracy, the dominant contribution to GW emission arises from the radiative mass quadrupole moment
which is, following Eq.~(1.3b) of \cite{BlanchetDamour1986}, given by
\begin{align}
I^{\mathrm{rad}}_{ij}(t) ={}&
\int d^3\mathbf{x}\, \hat{x}_{ij}\, c^{-2}
\left[T^{00}(\mathbf{x},t) + T^{ss}(\mathbf{x},t)\right]
\nonumber\\
&+ \frac{1}{14 c^2}\frac{d^2}{dt^2}
\int d^3\mathbf{x}\, \hat{x}_{ij}\, \mathbf{x}^2\, c^{-2}
\left[T^{00}(\mathbf{x},t) + T^{ss}(\mathbf{x},t)\right]
\nonumber\\
&- \frac{20}{21 c^2}\frac{d}{dt}
\int d^3\mathbf{x}\, \hat{x}_{k\langle i} x_{j\rangle}\,
T^{0k}(\mathbf{x},t)
+ \mathcal{O}\!\left(c^{-4}\right),
\label{eq:BD_rad_quad}
\end{align}
where $T^{\mu\nu}$ is the stress--energy tensor of the material source expressed in harmonic coordinates, angular brackets denote the symmetric trace-free (STF) projection, and $\hat{x}_{ij}$ denotes the STF part of $x_i x_j$. For tidal disruption events, the stress--energy tensor corresponds to the evolving stellar debris, whose rapid deformation near periapsis leads to a strongly time-dependent radiative quadrupole moment. While the Newtonian contribution dominates for shallow encounters, the 1PN corrections become increasingly relevant for deep encounters with $\beta \gtrsim 1$, where relativistic velocities are attained.

For an observer located along the $z$-axis, the two GW polarizations are given by
\begin{align}
h_+(t) = \frac{G}{c^4 D}
\left[
\ddot{I}^{\mathrm{rad}}_{xx}(t)
- \ddot{I}^{\mathrm{rad}}_{yy}(t)
\right]~,~
h_\times(t) = \frac{2G}{c^4 D}
\ddot{I}^{\mathrm{rad}}_{xy}(t).
\end{align}
These expressions remain valid to 1PN accuracy provided the radiative quadrupole moment includes the PN corrections given in Eq.~\eqref{eq:BD_rad_quad}.

In Figure \ref{fig:BH_GW_faceon}, we show the GW polarizations $h_+$ and $h_{\times}$ as well as the amplitude $|h|=\sqrt{h_+^2+h_{\times}^2}$ for the $6\times 10^6~M_{\odot}$ and $1.2\times 10^7~M_{\odot}$ BHs for the face-on inclination. Figure \ref{fig:WH_GW_faceon} shows these for the corresponding same-mass WHs in the same inclination. All the figures are plotted for $\beta = 1.1$, for which the background with mass $6\times 10^6~M_{\odot}$ results in full disruption in both cases whereas both backgrounds result in partial disruption for mass $1.2\times 10^7~M_{\odot}$. From the figures, we see that the behaviour of $|h|$ is qualitatively similar for both the BH and WH backgrounds, with only a change in amplitude observed at the same order. 

\textcolor{black}{\section{Comparison with Previous Studies}}\label{comparison}

Prior to our work, the authors of \cite{Ryu2020}  performed a suite of fully relativistic hydrodynamic simulations of tidal disruptions of main sequence stars by SMBHs, in a Schwarzschild spacetime, using a different numerical recipe than ours. In their setup, the BH spacetime and tidal field are treated relativistically, while the star's self gravity is computed via a Newtonian Poisson solver in a comoving local tetrad frame, where the metric is approximately Minkowski. The stellar potential is then added to $g_{tt}$ in that local frame as a post-Newtonian correction. Further, they define a ``physical tidal radius'' $\mathcal{R}_{t}$, as the maximum periapsis distance at which full disruption occurs, and show that this is enhanced relative to the Newtonian tidal radius $r_{t}$ (of Eq. \ref{tidalradius}).
For partial disruptions, they find that the remnant mass fraction $M_{\rm rem}/M_\ast$ at a given normalized periapsis $r_{p}/\mathcal{R}_{t}$ increases with $M_\bullet$. 

In our work,  we have also considered back-reaction due to star's gravity as a small linear perturbation to the background metric, following \cite{LiptaiThesis2020}. By solving the relativistic fluid equations, we get a velocity--dependent self--gravity correction of the form
\begin{equation}
    \mathbf{a}_{\rm sg} \;=\; -\,U^{0}\,\left(1 - \frac{\delta_{lm} v^{l}v^{m}}{c^{2}}\right)\,\nabla\Phi_{\rm sg}~,
\label{sg}
\end{equation}
where $U^{0}$ is the gravitational time-dilation factor, $v$ is the local three-velocity in the fluid frame, and $\Phi_{\rm sg}$ is the self gravitational potential of the star. 

In the deep relativistic regime ($\beta \gtrsim 0.95$, large $M_\bullet$), \cite{Ryu2020} finds that for a fixed normalized periapsis $r_{p}/\mathcal{R}_{t}$, the remnant mass fraction increases with $M_\bullet$: stars hold on to more of their mass when the encounter is more relativistic. They attribute this to relativistic modifications of the tidal tensor and debris dynamics, not to any change in self--gravity. In our framework, as $M_\bullet$ grows and $\beta$ increases, the periapsis velocity approaches $c$ and the factor $(1-v^{2}/c^{2})$ becomes  small,  weakening self-gravity. Naively, this might suggest even \textcolor{black}{a} stronger disruption. However, our simulations show that at sufficiently deep encounters the qualitative trend remains the same, i.e., for fixed $r_{p}/r_{p,c}$, the surviving core mass fraction increases with $M_\bullet$, and the critical impact parameter $\beta_{c}$ for full disruption rises.

The difference of our results from treating hydrodynamics in the Newtonian regime can be explained from the fact that geodesic congruence effects in relativistic tidal fields compete with, and eventually dominate over, the weakening of self gravity. Nearby fluid elements in the star follow a bundle of timelike geodesics whose separation is governed by the geodesic deviation equation.  This geodesic convergence leads to stronger tangential compression than in the Newtonian case and naturally produces a very dense core that is effectively decoupled from the large scale tidal field. This is in close analogy with the core survival found in \cite{Seoane2023} for underluminous TDEs.

At the same time, gravitational time dilation reduces the proper frame duration of the periapsis passage. The proper time interval experienced by the star during the strongest part of the encounter is shorter by a factor $\sqrt{1 - r_s/r}$ than the coordinate time. Our explicit orbit integrations show that, for the deep relativistic encounters we consider, the proper crossing time at periapsis becomes comparable to, or smaller than, the stellar sound crossing time. Pressure gradients and tidal counterforces therefore have very little proper time to deform or unbind the innermost region. The inner core is compressed and then exits periapsis essentially ballistically, with its structure largely preserved. 

To better understand the reason behind stars retaining a larger bound core mass for higher BH mass, we calculate the tidal potential as experienced by the star at the periapsis. We consider the local FN basis inside which the star co-rotates with its orbital motion (the ``tilde frame'' in which the computations of section \ref{FNC} were also done). While this is an assumption that does not hold true throughout the parabolic geodesic of the stellar core, the star is mostly tidally locked at the periapsis. As a result, the co-rotating frame is a good assumption there. We calculate the tidal potential in the FN frame, which increases with distance from the centre of mass (COM) of the star (the origin of the FN frame). We track the magnitude of the tidal potential along the BH direction ($\phi_{x}$) and along the two other perpendicular directions ($\phi_y$ and $\phi_z$). While the tidal force field acts towards expanding the star towards the BH or WH (and away from it), in the other two perpendicular directions, the tidal field compresses the stellar body. We calculate the magnitude of the tidal potential at a distance $1 R_\odot$ from the COM. We repeat this task for higher ADM masses. We find that the magnitude of tidal potential increases with ADM mass. 

\begin{figure}[h!]
    \centering
    \setlength{\tabcolsep}{1pt} 
    \begin{tabular}{cc}
        \includegraphics[width=0.49\textwidth]{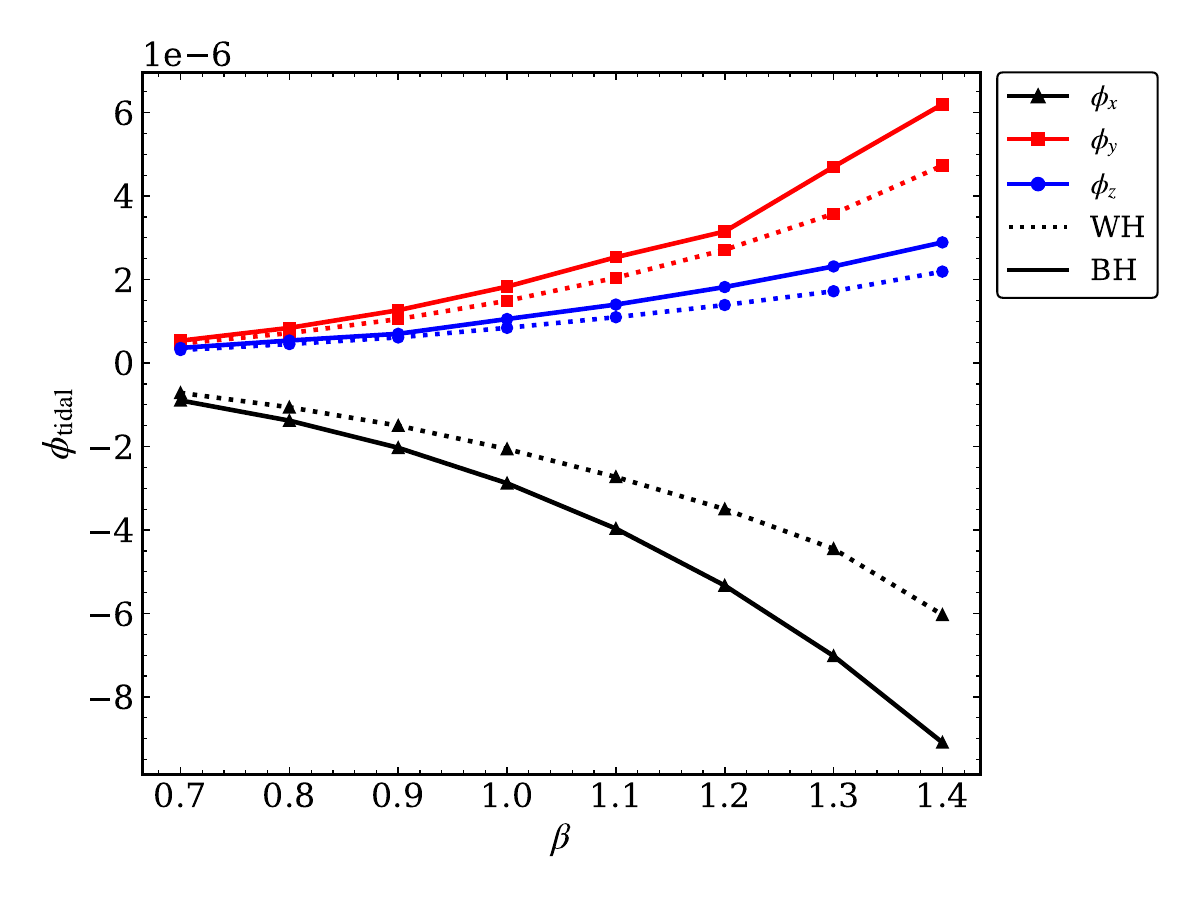} &
        \includegraphics[width=0.49\textwidth]{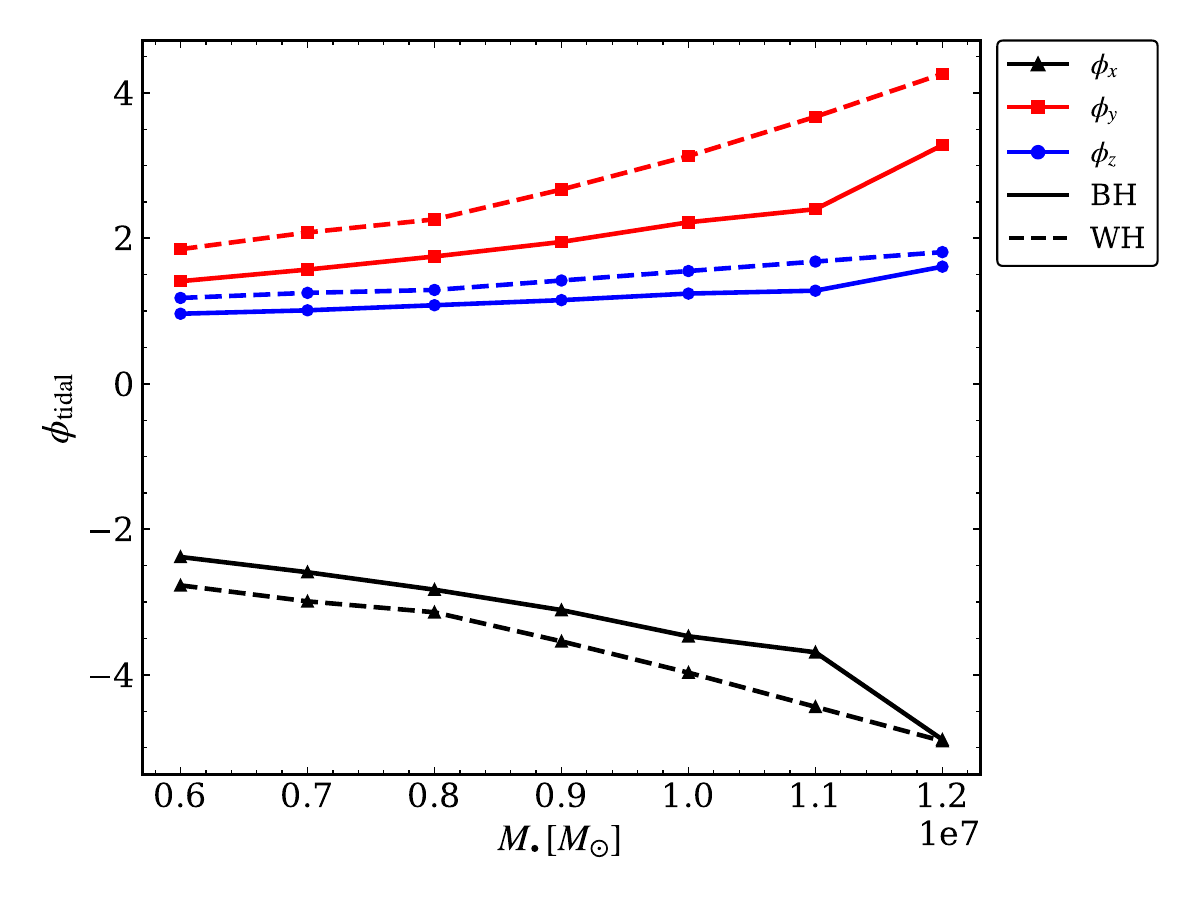}
    \end{tabular}
    \caption{
Tidal potential components, $\phi_x$, $\phi_y$, and $\phi_z$, for BH (solid) and WH (dashed) cases. 
The left panel shows $\phi_{\rm tidal}$ as a function of the impact parameter $\beta$ for $M_\bullet = 10^7\,M_\odot$, while the right panel shows $\phi_{\rm tidal}$ at $r_{p,c}$ as a function of $M_\bullet$. 
The WH case exhibits systematically weaker tidal compression compared to the BH, particularly in the $\phi_x$ component.
}
    \label{fig:phi_tidal}
\end{figure}

In Figure~\ref{fig:phi_tidal}, we show change of tidal potential in the FN set up described above with increasing ADM mass.  Since the tidal potential is zero at the COM of the star, and its value at the surface of the star ($1 R_\odot$) increases with BH mass, it is evident that the corresponding tidal force (compressing force) also increases with ADM mass. Therefore, we expect that the star will be able to retain more core mass in the case of a higher ADM mass in TDEs that show partial disruptions.
From the same figure we also plot tidal potential  at $r_{p,c}$ with increasing ADM mass. It supports our earlier observation that post TDE remnant core retains more mass for WH spacetime than that of BH spacetime.
In this sense, our study can be viewed as an extension of the  work of \cite{Ryu2020} that incorporates a physically motivated modification of stellar self--gravity. From both of hydrodynamic simulation and FNC calculation  we quantify how the disruption efficiency and critical impact parameter $\beta_{c}(M)$ change. The critical periapsis distance normalized by the gravitational radius of the BH decreases approximately twice as rapidly in our case compared to that reported in  \cite{Ryu2020}. This indicates that a proper relativistic treatment requires a deeper encounter to achieve the same degree of tidal disruption.\\

\section{Compactness as a measure of Tidal Disruption in the relativistic regime}

To quantify the outcome of close stellar encounters, we construct a compactness parameter from the tidal force evaluated at periapsis in the FN frame.
The radial tidal field $\phi_x$ represents the relativistic tidal stretching responsible for unbinding stellar material, while $\phi_y$ and $\phi_z$ encode stabilizing transverse compression that supports core survival. A natural geometric measure of the competition between these effects is provided by the anisotropy ratio
\begin{equation}
\mathcal{A}
=
\frac{\sqrt{\phi_y^2+\phi_z^2}}{|\phi_x|},
\end{equation}
which depends only on the relative magnitudes of the tidal eigenvalues and therefore captures relativistic effects arising from the underlying spacetime geometry.

Tidal disruption, however, is not determined by anisotropy alone. The strength of the encounter is controlled by the impact parameter $\beta$. In the Newtonian limit, the magnitude of the tidal field at periapsis scales as
\begin{equation}
|\lambda_x|
\sim
\frac{G M_\bullet R_\ast}{r_p^3}
=
\beta^3 \frac{G M_*}{R_*^2}.
\end{equation}
 The factor $\beta^3$ thus represents the Newtonian amplification of tidal strength associated with deeper encounters and is independent of the local anisotropy of the tidal tensor.

Motivated by these considerations, we define the compactness parameter as
\begin{equation}
\label{eq:compactness}
\mathcal{C}
=
\frac{\sqrt{\phi_y^2+\phi_z^2}}{|\phi_x|\,\beta^3}.
\end{equation}
This definition cleanly separates the physical origins of the two contributions: the ratio of tidal eigenvalues encodes relativistic tidal anisotropy through the 
FN frame, while the explicit $\beta^3$ factor incorporates the Newtonian scaling of tidal strength with periapsis distance.

\begin{figure}[t]
    \centering
    \setlength{\tabcolsep}{1pt} 
    \begin{tabular}{cc}
        \includegraphics[width=0.49\textwidth]{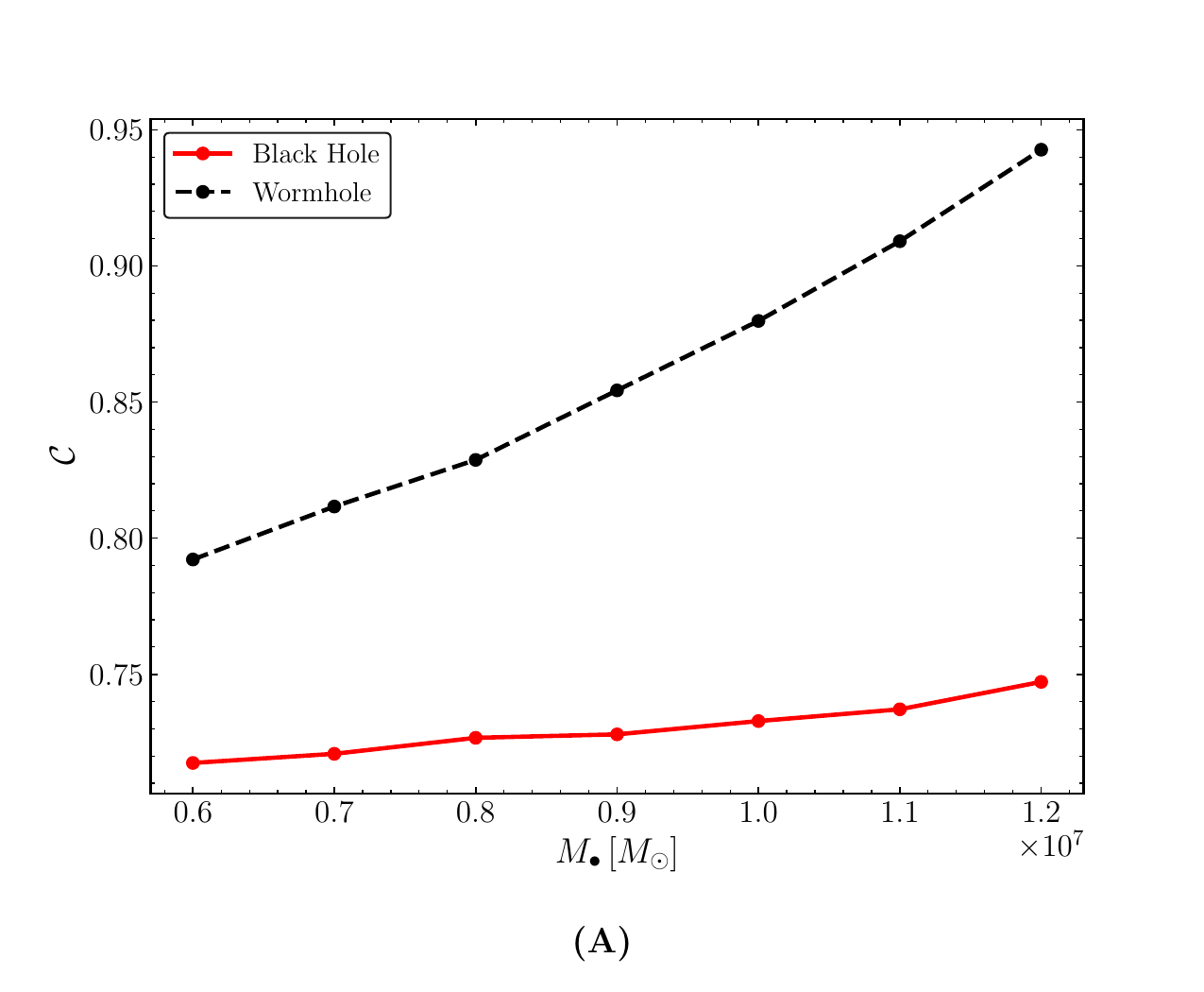} &
        \includegraphics[width=0.49\textwidth]{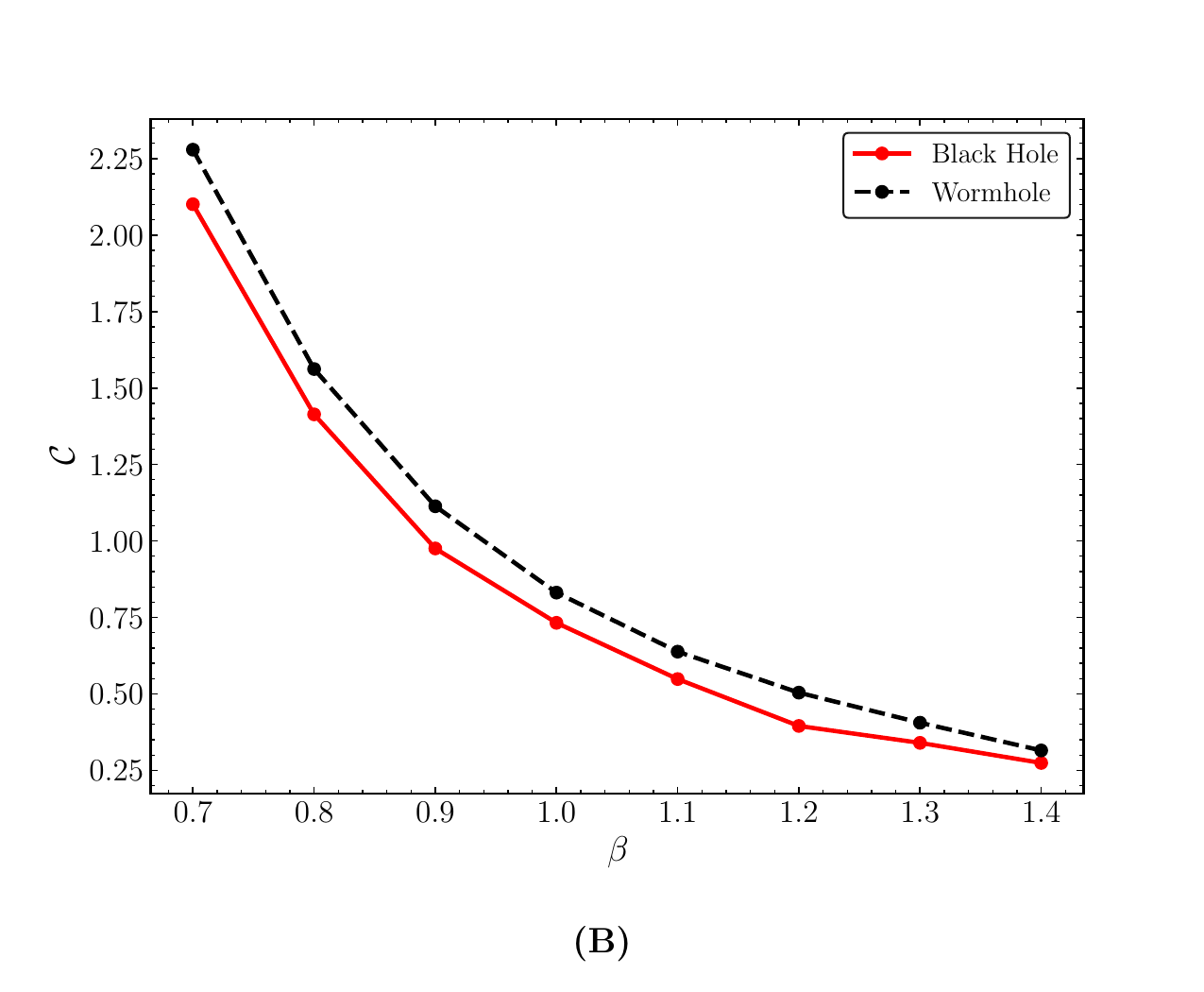}
    \end{tabular}
   \caption{
Compactness, $C$, of the star for BH (solid red) and WH (dashed black) cases. 
The left panel shows $C$ as a function of $M_\bullet$ at fixed $\beta = 1$, while the right panel shows $C$ as a function of $\beta$ for $M_\bullet = 10^7\,M_\odot$. 
In both cases, the WH yields a higher compactness than the BH, with $C$ decreasing as $\beta$ increases.
}
    \label{fig:compactness}
\end{figure}

For a fixed impact parameter ($\beta=1$), the compactness reduces to $\mathcal{C}=\mathcal{A}$ and depends solely on the geometry of the tidal field. In this regime, increasing the mass of the central object leads to a more isotropic tidal configuration, enhancing transverse compression relative to radial stretching and thereby increasing the compactness Figure~\ref{fig:compactness}(A). This behaviour explains why higher-mass central objects tend to retain a larger fraction of stellar material for encounters of identical depth.
For fixed central mass and varying impact parameter, the $\beta^3$ dependence dominates the behaviour of $\mathcal{C}$. As $\beta$ increases, the rapid growth of tidal strength overwhelms the stabilizing transverse compression, leading to a sharp decrease in compactness and enhanced disruption Figure~\ref{fig:compactness}(B). This trend reflects the increasing dominance of tidal work over stellar self-gravity in deep encounters.

A comparison between BH and WH spacetimes shows that WHs systematically exhibit larger compactness values across both parameter scans, indicating weaker disruption efficiency. This difference originates from relativistic modifications of the tidal eigenvalues $\phi_i$ in the WH geometry, which enhance transverse compression relative to radial stretching. The compactness defined in Eq.~\ref{eq:compactness} therefore provides a unified and physically motivated diagnostic for comparing tidal disruption outcomes across different spacetimes and encounter regimes.

In summary, the results presented in Figure~\ref{fig:compactness}(A) demonstrate that, for a fixed impact parameter $\beta_c$, the stellar compactness increases with increasing mass of the central compact object. This qualitative trend is consistent with our finding that the critical impact parameter $\beta_{\rm c}$ increases with BH or WH mass. In other words, more massive compact objects require deeper encounters (larger $\beta$) to achieve complete disruption.  The Figure~\ref{fig:compactness}(B) further shows that, for a fixed central mass, the stellar compactness decreases rapidly with increasing $\beta$, as expected. Deeper encounters enhance tidal deformation and mass stripping, thereby reducing the bound core and lowering the remnant compactness. Finally, we find that, for identical values of mass and $\beta$, WH spacetimes consistently produce higher post-encounter stellar compactness compared to BHs. This indicates that tidal interactions in WH geometries are comparatively less efficient at fully disrupting the star, leading to a more tightly bound remnant under otherwise identical orbital parameters.

\section{Discussions and conclusions} \label{sec:conclusion}

It is well known by now that WHs might be horizonless BH mimickers. This makes it important and interesting to study dynamics in these
backgrounds, and astrophysical tidal disruption events are very well suited to carry out this analysis. Although these have been studied
before using static methods where one transforms to a locally inertial FN frame, a comparative analysis disruption dynamics
using numerical simulations is not available in the literature. In this paper we fill this gap and present a number of novel results that
arise from simulations of solar mass stars in supermassive BH or WH backgrounds. In particular, using general relativistic smoothed particle hydrodynamics,
we probe deep encounters, i.e., strong gravity regimes in these. This is an important feature of our analysis, as similar mass BHs and WHs essentially
behave like similar Newtonian masses for mild encounters.

To begin with, we have shown that due to the inherently different spacetime geometry, WHs retain substantially more self-bound mass
than BHs in partial disruptions. This led us to a discussion on the critical impact parameter $\beta_c$ that separates the region of full disruption from
partial ones. We have quantified how this is consistently higher for the WH, as compared to the BH, due to the weaker gravity of the former
even in deep regimes. Further, we have shown that $\beta_c$ increases with BH or WH mass, and explained this by a FN frame computation.
We have also commented upon the distribution of energy in stellar debris and discussed GWs in both backgrounds.

Before we end this paper, we make a few important comments. Firstly, it might be prudent to ask if one can, in some sense, “distinguish between” BHs and their mimickers from observational data. To this end, note that the gravitational wave signal from a TDE encodes information about the binary interaction ($M$, $M_\ast$, $R_\ast$, $\beta$) through the characteristic frequencies, strain amplitudes and their temporal evolution. See, for example, Eqs ~\eqref{eq:fgw_scaling}, ~\eqref{eq:hgw_scaling}, and the expressions for the frequency and amplitude of GWs generated via strong compression near periapsis, given by 
\cite{2019GReGr..51...30S} :
\begin{equation}
f_{\rm comp} \simeq 4 \times 10^{-5}\,
\beta^{4}
\left( \frac{M_\ast}{M_\odot} \right)^{1/2}
\left( \frac{R_\ast}{R_\odot} \right)^{-3/2}
\,\mathrm{Hz}
\label{f_comp}
\end{equation}
\begin{equation}
h_{\rm comp} \simeq 10^{-27}\,
\beta^{2}
\left( \frac{M_\ast}{M_\odot} \right)^{2}
\left( \frac{R_\ast}{R_\odot} \right)^{-1}
\left( \frac{d_L}{20\,\mathrm{Mpc}} \right)^{-1}~.
\label{h_comp}
\end{equation}
We also assume that we have a given mass radius relation of the stellar object, and that the observer distance $d_L$ is known (see below). 
In an ideal scenario, if one estimates the $4$ parameters from the above constraints, \textcolor{black}{the TDE observables, when compared with the observed data, directly indicate whether the data prefer a BH model or a WH model. For example, having the estimated values of $M$, $M_\ast$, $R_\ast$, $\beta$ from GW, the detection of a late-time fallback rate scaling $\sim\dot{M}\propto t^{-9/4}$, rather than the canonical $t^{-5/3}$ expected for a full tidal disruption, may indicate a deviation from the standard Schwarzschild BH scenario and is potentially consistent with a static exponential WH. However, given the fact that $f_{\rm comp}$ and $h_{\rm comp}$ in Eqs~(\ref{f_comp}) and (\ref{h_comp}) are orders of magnitude less than that of Eqs.~\eqref{eq:fgw_scaling} and \eqref{eq:hgw_scaling}, they might turn out to be insufficient to estimate all the 4 parameters. In that case, one might have to utilise both GW and TDE constraints simultaneously for a Bayesian model selection. For the latter case, we need the analytical prescriptions for the peak fall-back rate $\dot{M}_{\rm peak}$ which occurs at time $t_{\rm peak}$, and the fitting expression for the fall-back rate as a function of time $\dot{M}_{\rm fit}(t)$ (which includes the late-time behaviour) for both partial and full TDEs for both BHs and WHs and for a given star and its evolutionary stage. One such approach to obtain the analytical formulas is given in \cite{Coughlin2022} and \cite{Bandopadhyay2026}. A relativistic extension of their study is required in order to implement the analytical models into the statistical analysis for our case.}

In order to substantiate our argument regarding the mass-radius relation and the observer distance $d_L$, we further note that the spectroscopic classification of TDEs provides direct evidence for determining whether the disrupted star was on the main sequence or in an evolved stage. TDE-H events, which exhibit strong hydrogen Balmer emission lines, indicate disruption of main-sequence stars that retain their hydrogen-rich envelopes \cite{Gezari2012, Arcavi2014}. In contrast, TDE-He events dominated by helium lines with minimal hydrogen signatures suggest the disruption of evolved stars that have shed their hydrogen envelopes, such as helium cores or stripped stars \cite{Gezari2012}. One might use spectroscopic data from TDEs to determine whether the disrupted star is a main-sequence star and, hence, use the appropriate mass-radius relation. Also, from the EM wave redshift and GW redshift, we can estimate $d_L$. 

\textcolor{black}{We also note that the fallback rate is used as an observable in our study. However, the real observable is the luminosity of a TDE. The fallback rate determines the rate at which debris falls towards the BH or WH, creating an accretion disk. The accretion process leads to the observed luminosity. The accretion efficiency depends on the viscous timescale, which is expected to be much shorter than the fallback timescale in the early stages of fallback (days to a few months) \cite{Mockler2019, Nicholl2022}. Therefore, the peak fallback is expected to lead the TDE light curve. The late-time behaviour, on the other hand, is primarily determined by the viscous timescale, making it challenging to follow the late-time slope of the fallback. Therefore, we emphasize that the entire fallback analytical model has to be utilized in the Bayesian analysis. Secondly, the $t_{\rm peak}$ used in our analysis measures the peak fallback time from the start of the simulation. In reality, the peak time can only be measured from the detection of the first light, which is the time $t_{\rm fd}$ taken by the most bound debris to return to the pericenter and drive the accretion-induced emission. Therefore, we need to use $t_{\rm fd}$ as an additional parameter in the Bayesian analysis.}  


Finally, we emphasise here that we have taken a particular example of a horizonless compact object, namely the exponential WH.
It is therefore natural to ask if our results are generic. In this context, we recall that this particular example was
chosen in order to avoid interactions between `exotic' matter due to the WH and stellar debris, 
that would make an SPH analysis more challenging. If we do not consider such interactions and assume
only gravitational interactions of stellar debris, then our results should remain robust for other WH backgrounds as
well, as borne out by a local FN frame analysis.

\section*{Acknowledgments}

We acknowledge the support and resources provided by PARAM Sanganak under the National Supercomputing
Mission, Government of India, at the Indian Institute of Technology Kanpur.

\section*{Data Availability Statement}
The data underlying this article will be shared upon reasonable request to the corresponding author.

\appendix

\section{Numerical tests}
\label{Test Results}
Here, we present the standard test results achieved by our general relativistic smoothed particle hydrodynamics code. We arrange the numerical tests in two parts: one for special relativity and another for general relativity.
\subsection{Tests of Special Relativity}
In this section, we show the results obtained using our code for the following tests: 3D relativistic shock tube, 1D blast wave, 1D and 3D sinusoidal shock tube, transverse shock tube and radial oscillation of a polytropic star. All tests have been performed in $c = G = 1$ units, with artificial viscosity parameter $\alpha_{\rm AV}=1$ and artificial conductivity $\alpha_{\rm u}=0.1$ unless otherwise stated. All figures in this subsection are generated using SPLASH. 

\subsubsection{3D shock tube test}
We perform this test for a fluid with 72012 particles with the following initial condition:
\begin{equation}
    [\rho,P] = \begin{cases} [10, 13.33333] &\mbox{for } x < 0 \\
[1.125,0.000001] & \mbox{for } x > 0 \\
\end{cases}
\end{equation}
The maximum Lorentz factor achieved in this test is $\Gamma \sim 1.4$. 

\begin{figure}[H]
    \centering
    \setlength{\tabcolsep}{2pt} 

    \begin{tabular}{cc}
        \includegraphics[width=0.40\textwidth]{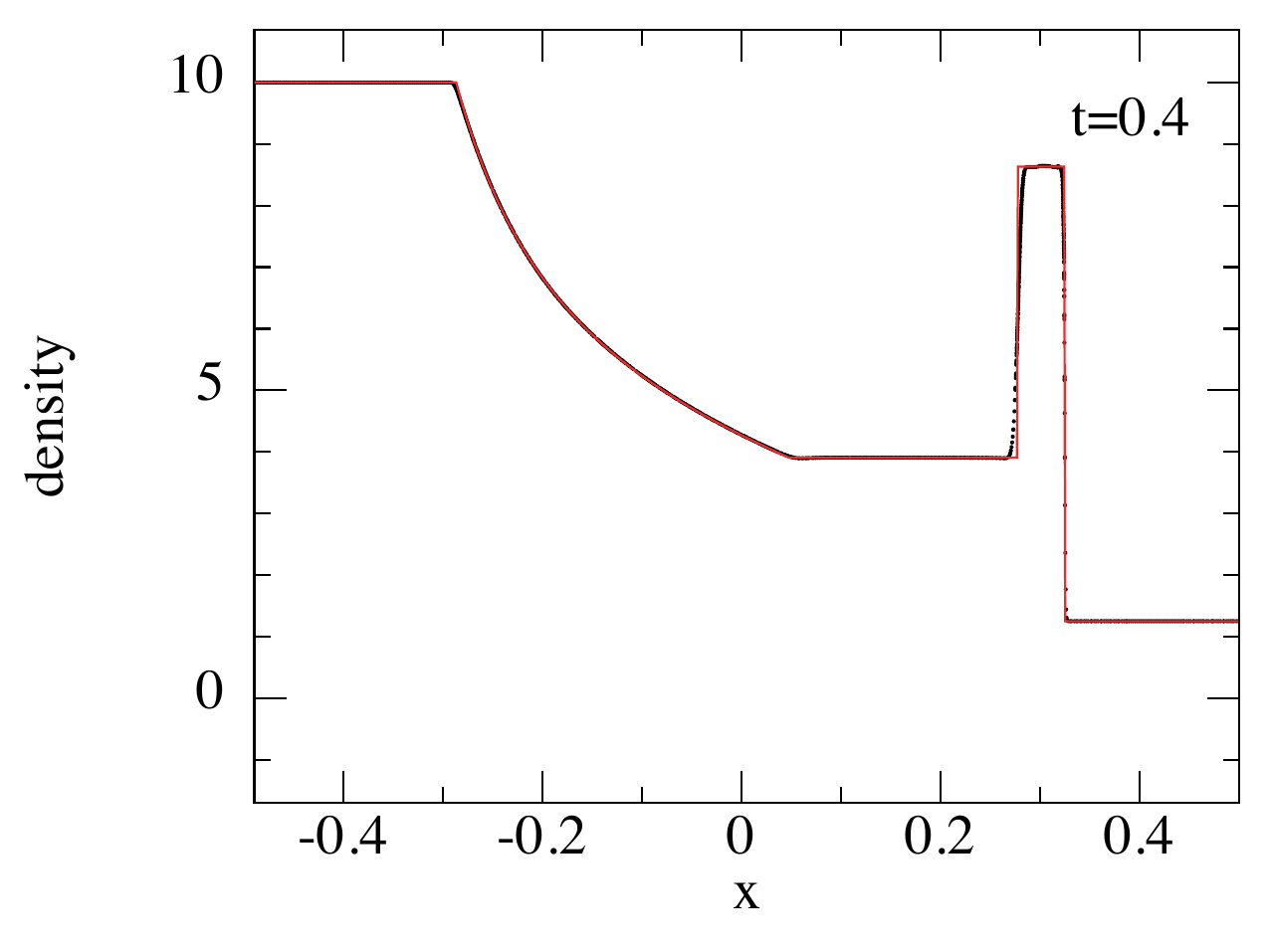} &
        \includegraphics[width=0.40\textwidth]{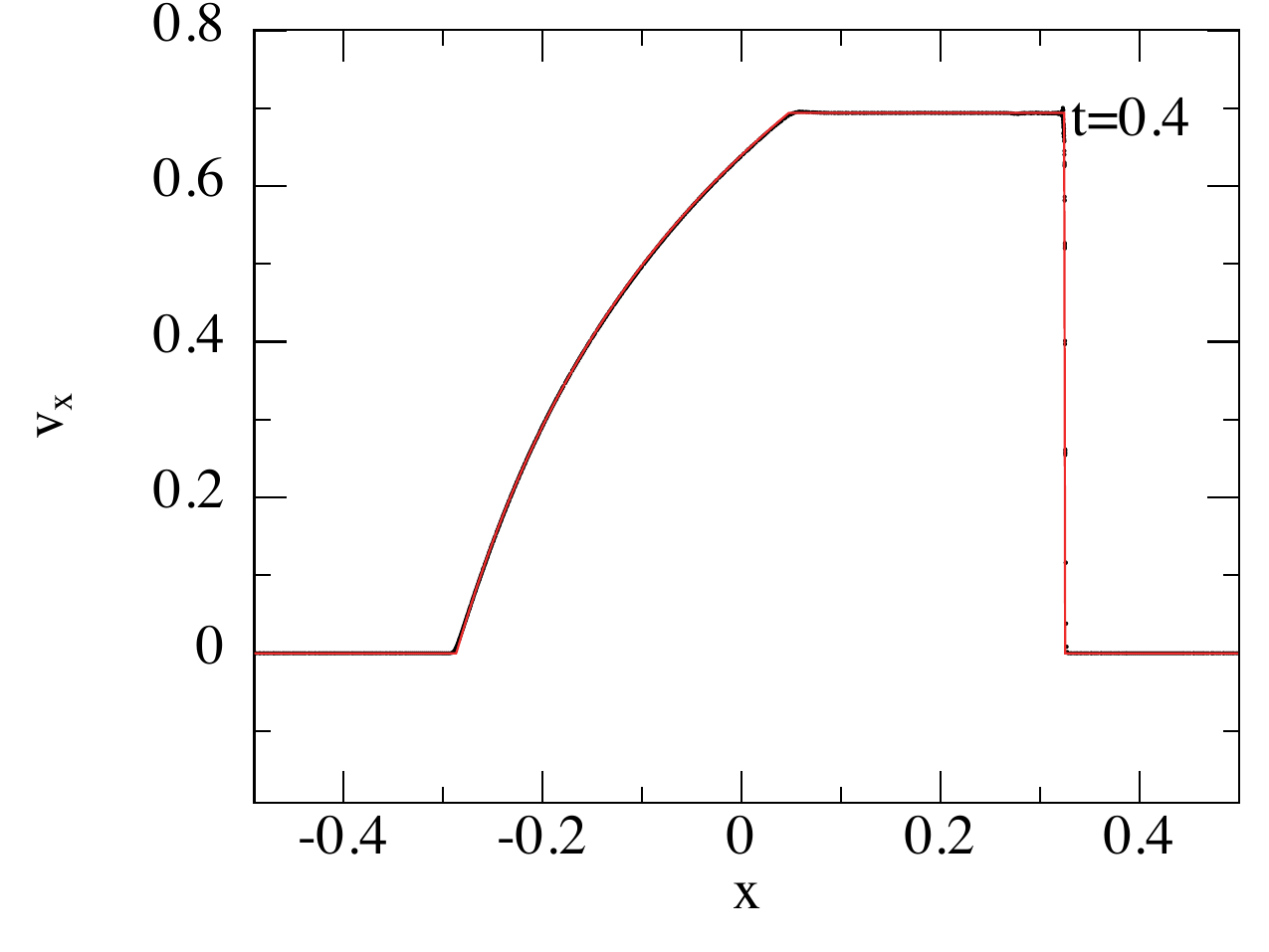} \\

        \includegraphics[width=0.40\textwidth]{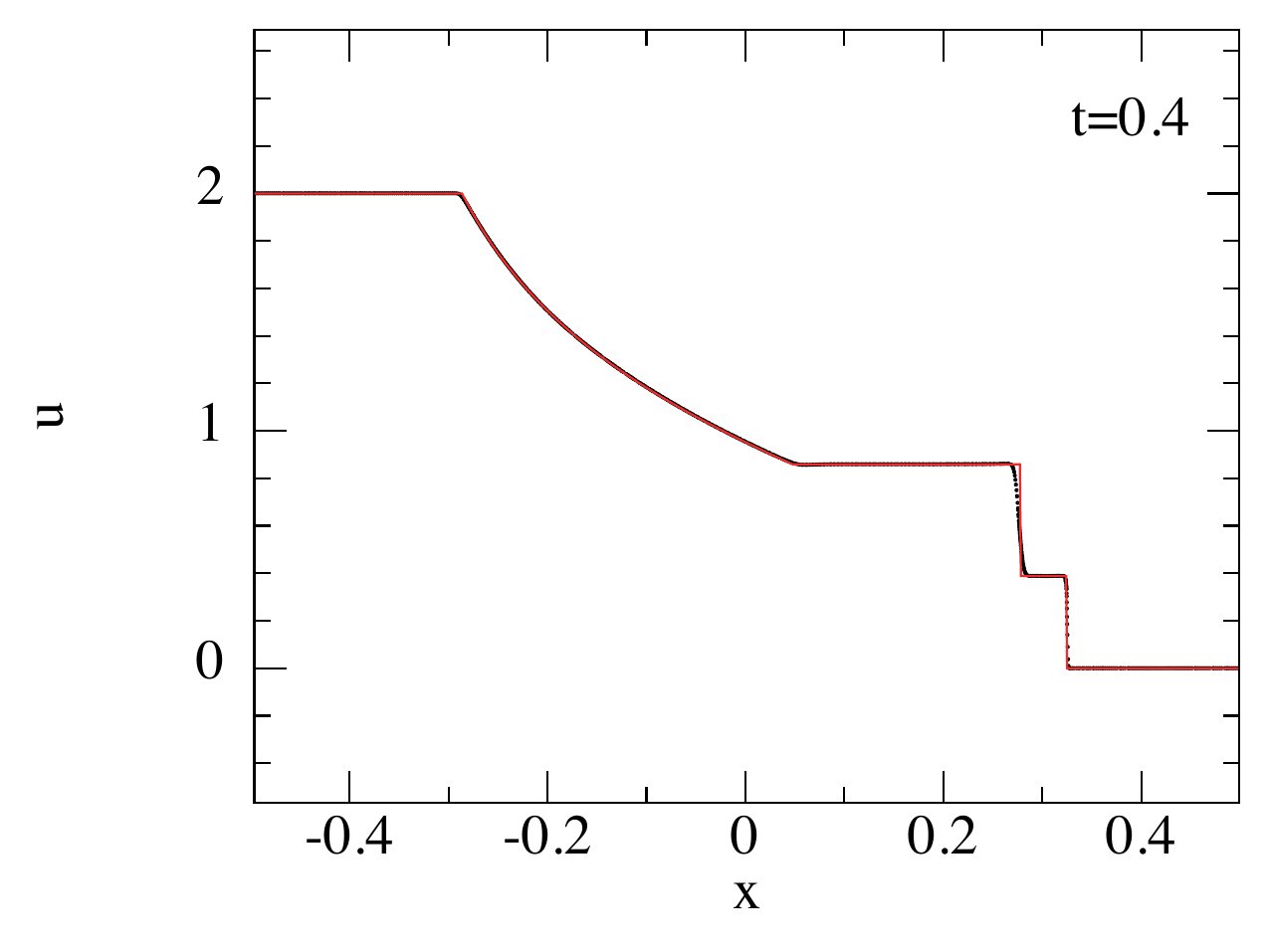} &
        \includegraphics[width=0.40\textwidth]{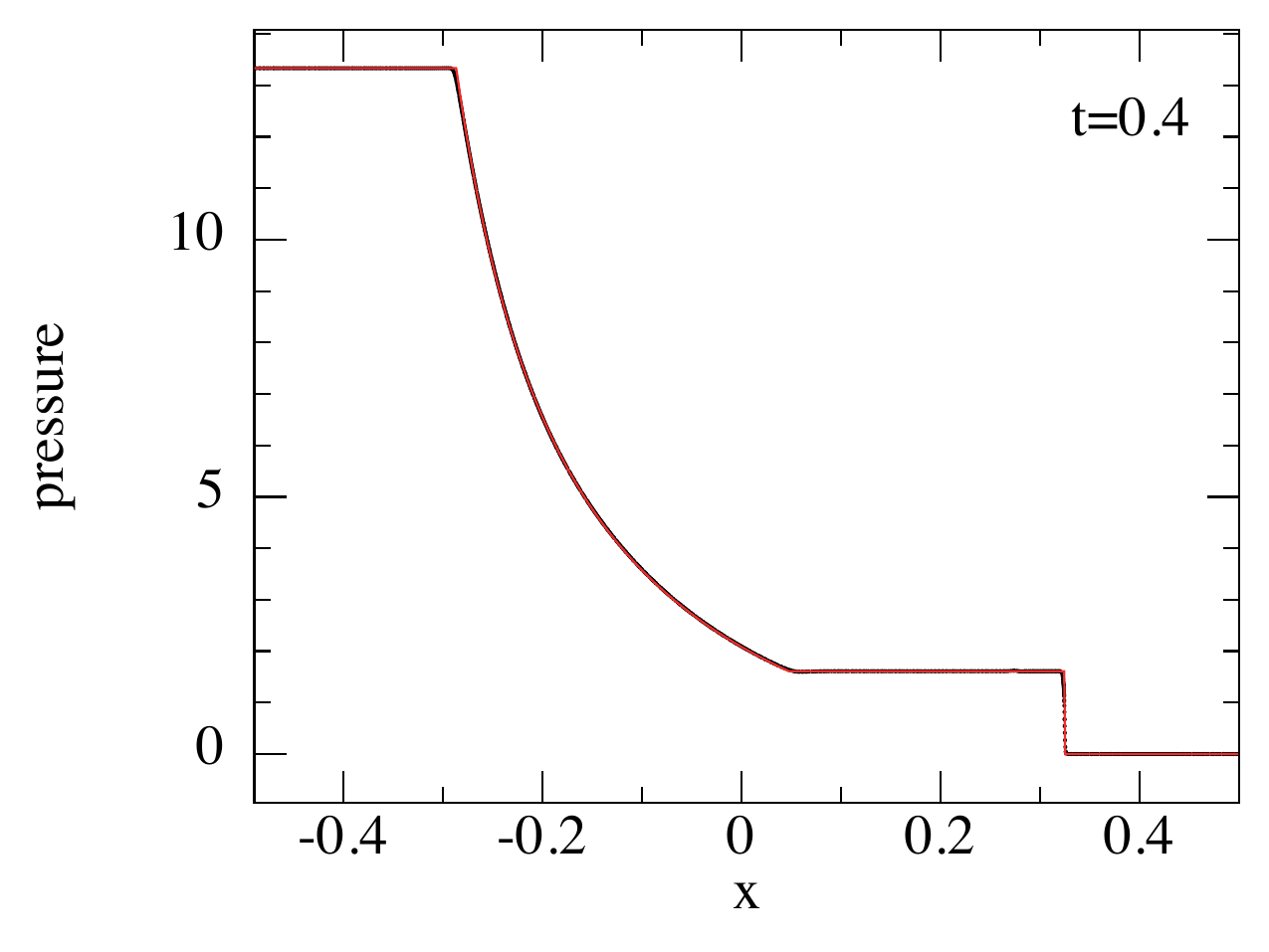}
    \end{tabular}

    \caption{3D shock tube test. Black circles show SPH particles, while the red line represents the analytical solution.}
    \label{fig:3d_shock_tube}
\end{figure}

\subsubsection{1D blast wave test}
The  Blast-wave test has been performed in 1D. The particle separation is 0.0001 on both sides of the interface. This test achieved a maximum Lorentz factor of $3.6$. For the initial configuration, we use the following set up:
\begin{equation}
    [\rho,P] = \begin{cases} [1, 1000] &\mbox{for } x < 0 \\
[1,0.001] & \mbox{for } x > 0 \\
\end{cases}
\end{equation}


\begin{figure}[H]
    \centering
    \setlength{\tabcolsep}{2pt}

    \begin{tabular}{cc}
        \includegraphics[width=0.40\textwidth]{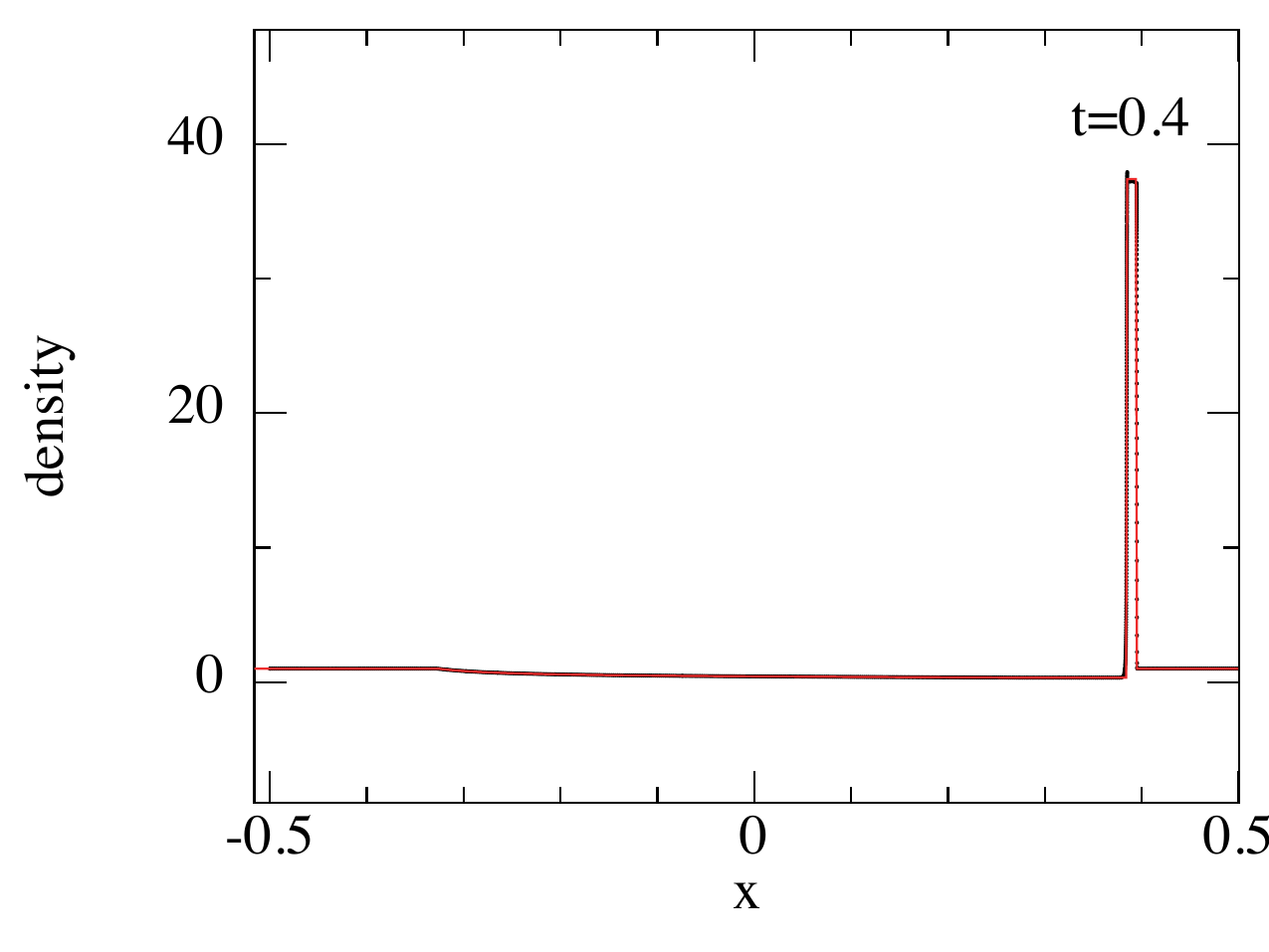} &
        \includegraphics[width=0.40\textwidth]{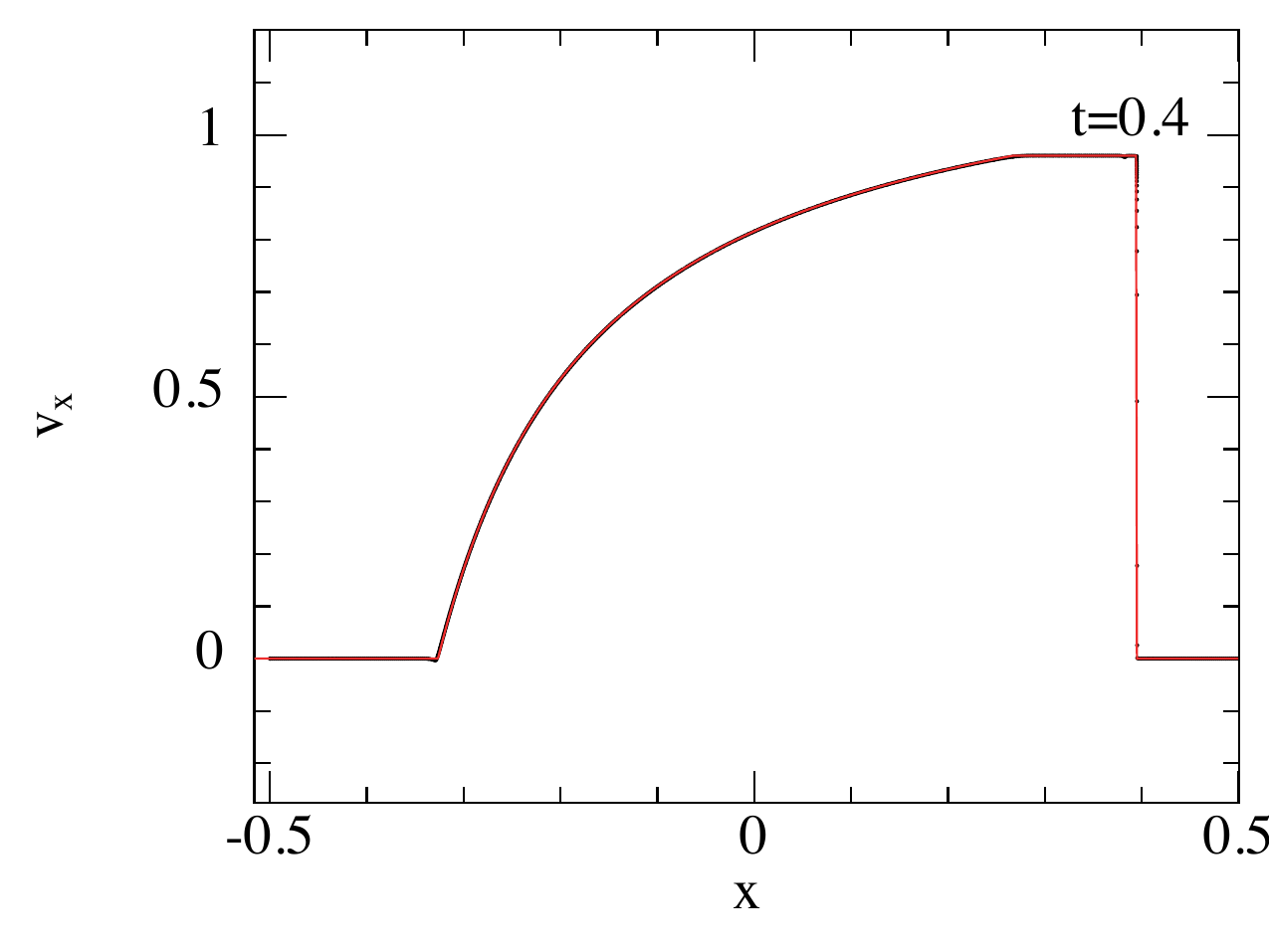} \\

        \includegraphics[width=0.40\textwidth]{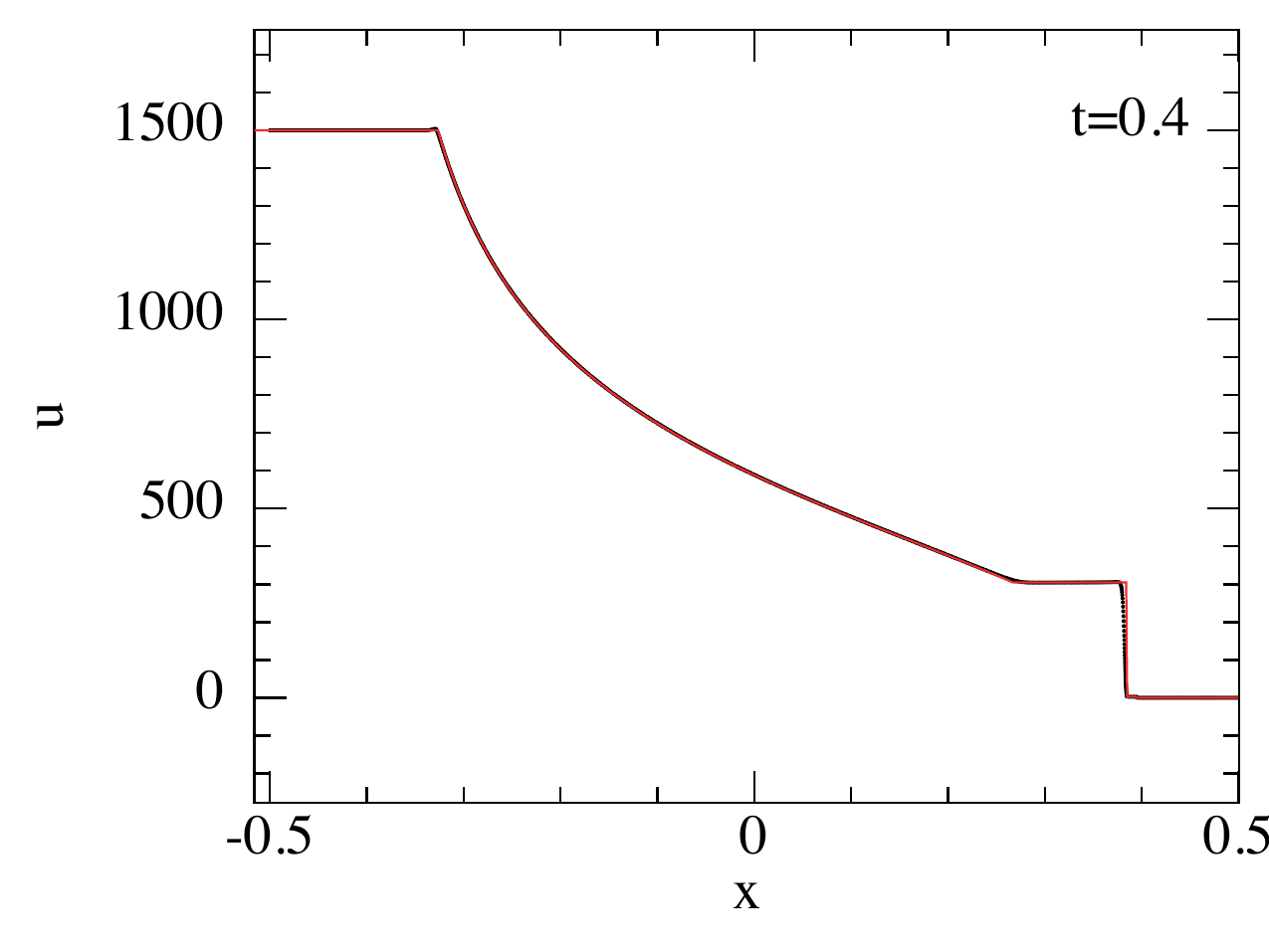} &
        \includegraphics[width=0.40\textwidth]{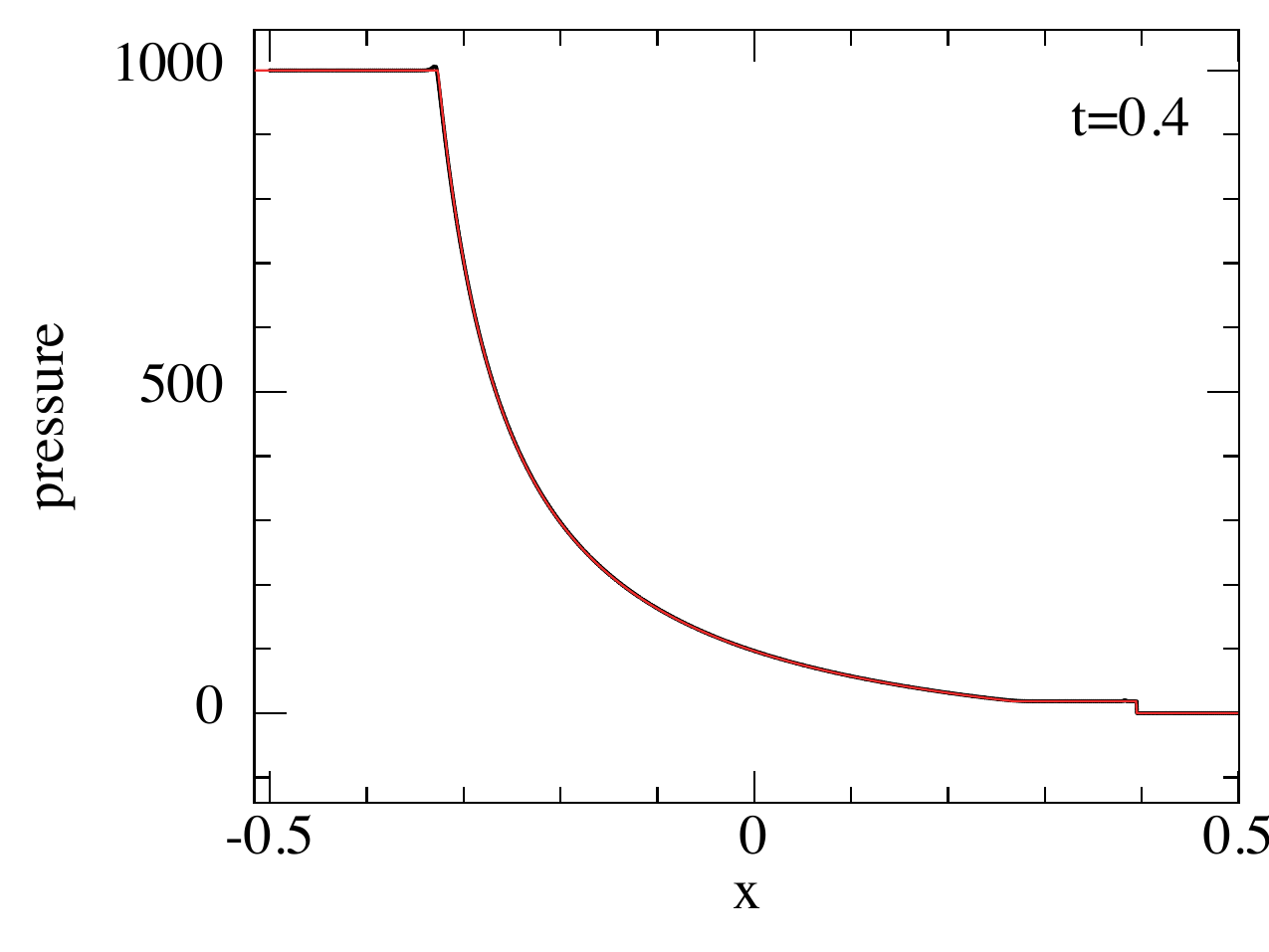}
    \end{tabular}

    \caption{1D blast wave test. Black circles denote SPH particles, while the red line shows the analytical solution from Mart\'i and M\"uller \cite{Marti1994}.}
    \label{fig:1d_blast_wave}
\end{figure}

\subsubsection{1D sinusoidal shock tube test}
Here, we perform a 1D shock tube with a sinusoidally perturbed density distribution in the right-hand side $x > 0$. The maximum value of the Lorentz factor reached is $\Gamma=1.15$. The initial condition we used for this setup is
\begin{equation}
    [\rho,P] = \begin{cases} [5, 50] &\mbox{for } x < 0 \\
[2+0.3\sin{(50x)},5] & \mbox{for } x > 0 \\
\end{cases}
\end{equation}
The test has been performed with 7001 particles. From the simulation results, we see that the sine-wave perturbation is accurately preserved across the shock front. The peaks of the propagated sine wave are consistent with the limiting analytical solutions.

\begin{figure}[t]
    \centering
    \setlength{\tabcolsep}{2pt}

    \begin{tabular}{cc}
        \includegraphics[width=0.40\textwidth]{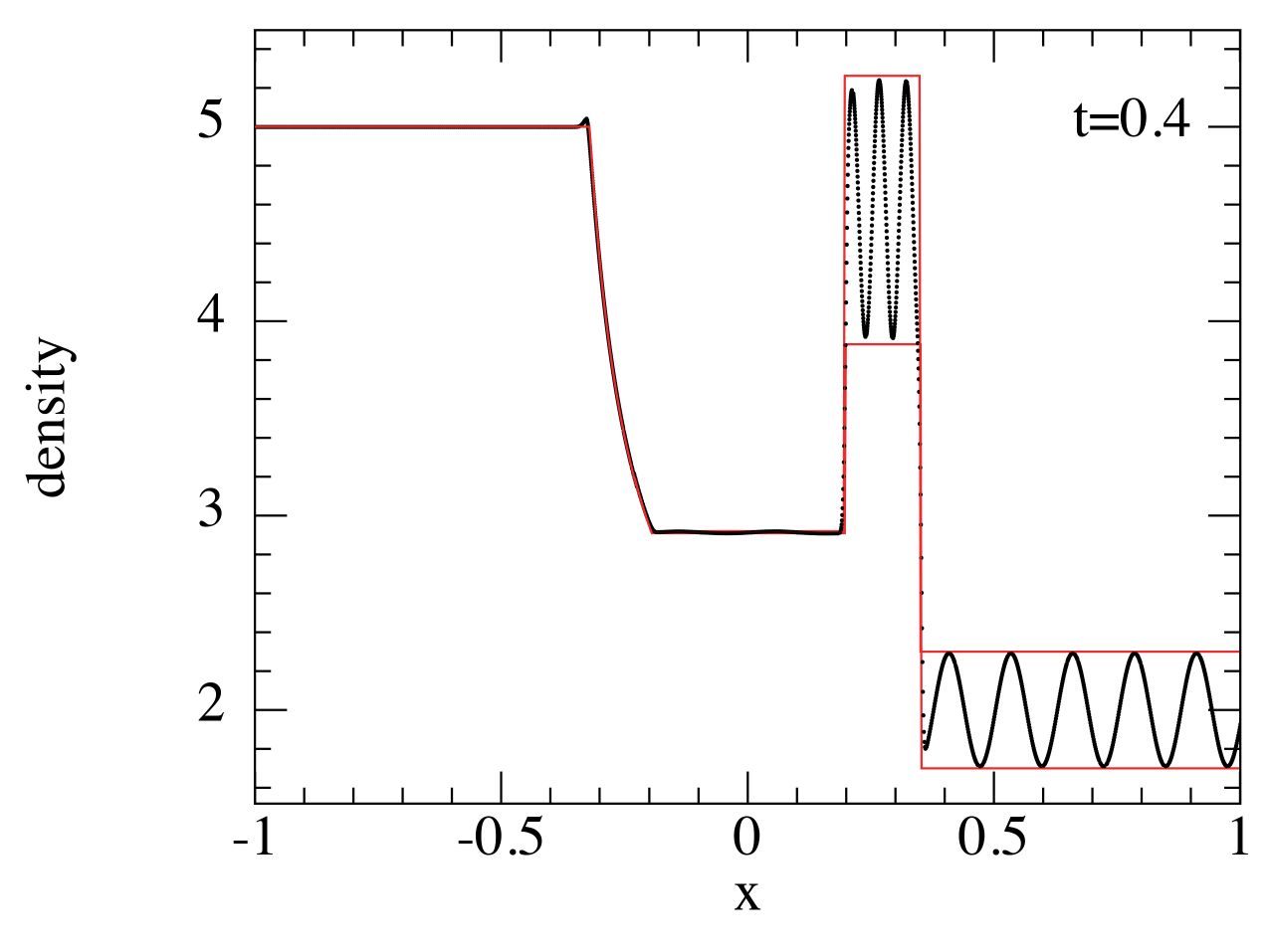} &
        \includegraphics[width=0.40\textwidth]{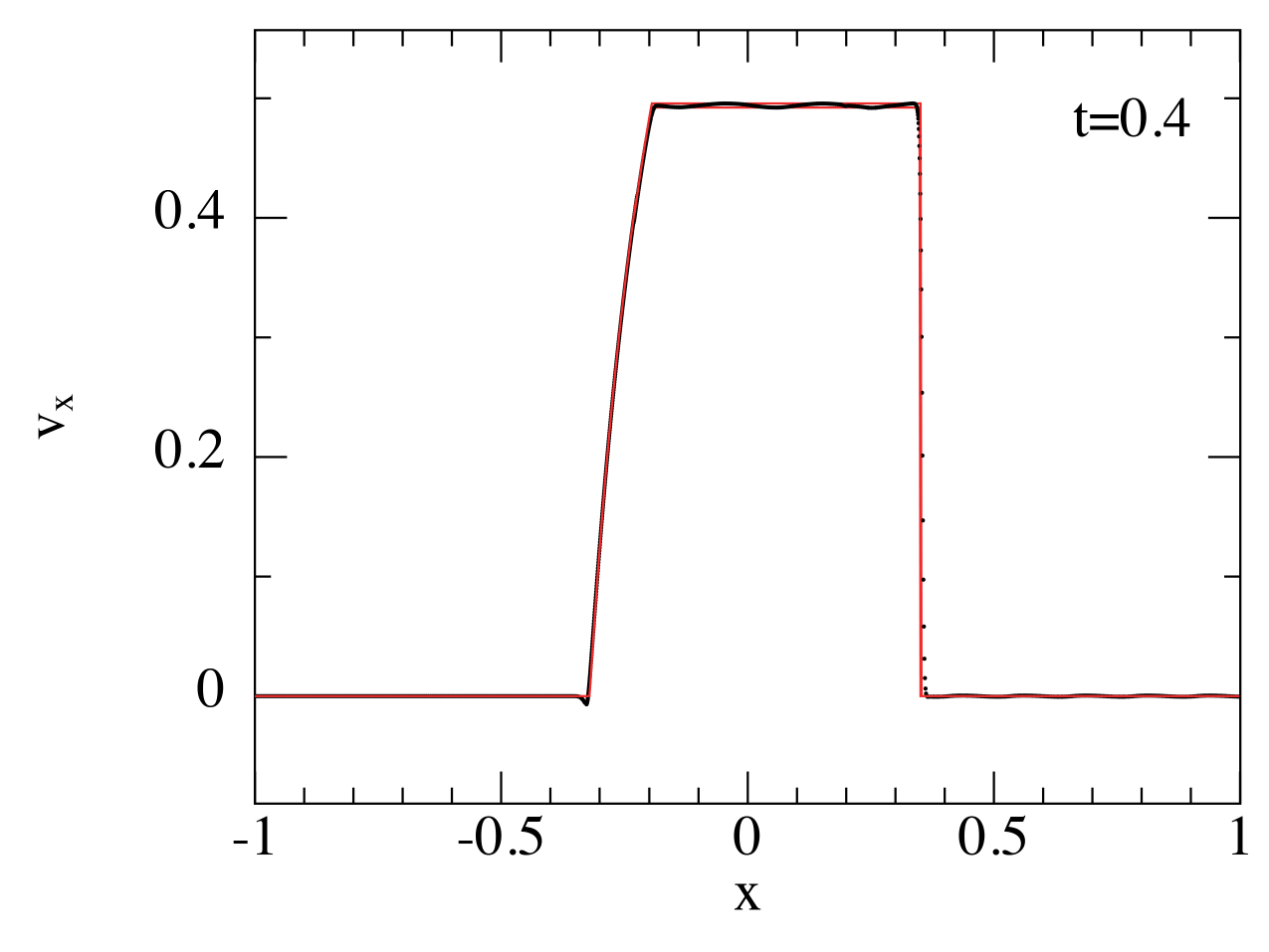} \\

        \includegraphics[width=0.40\textwidth]{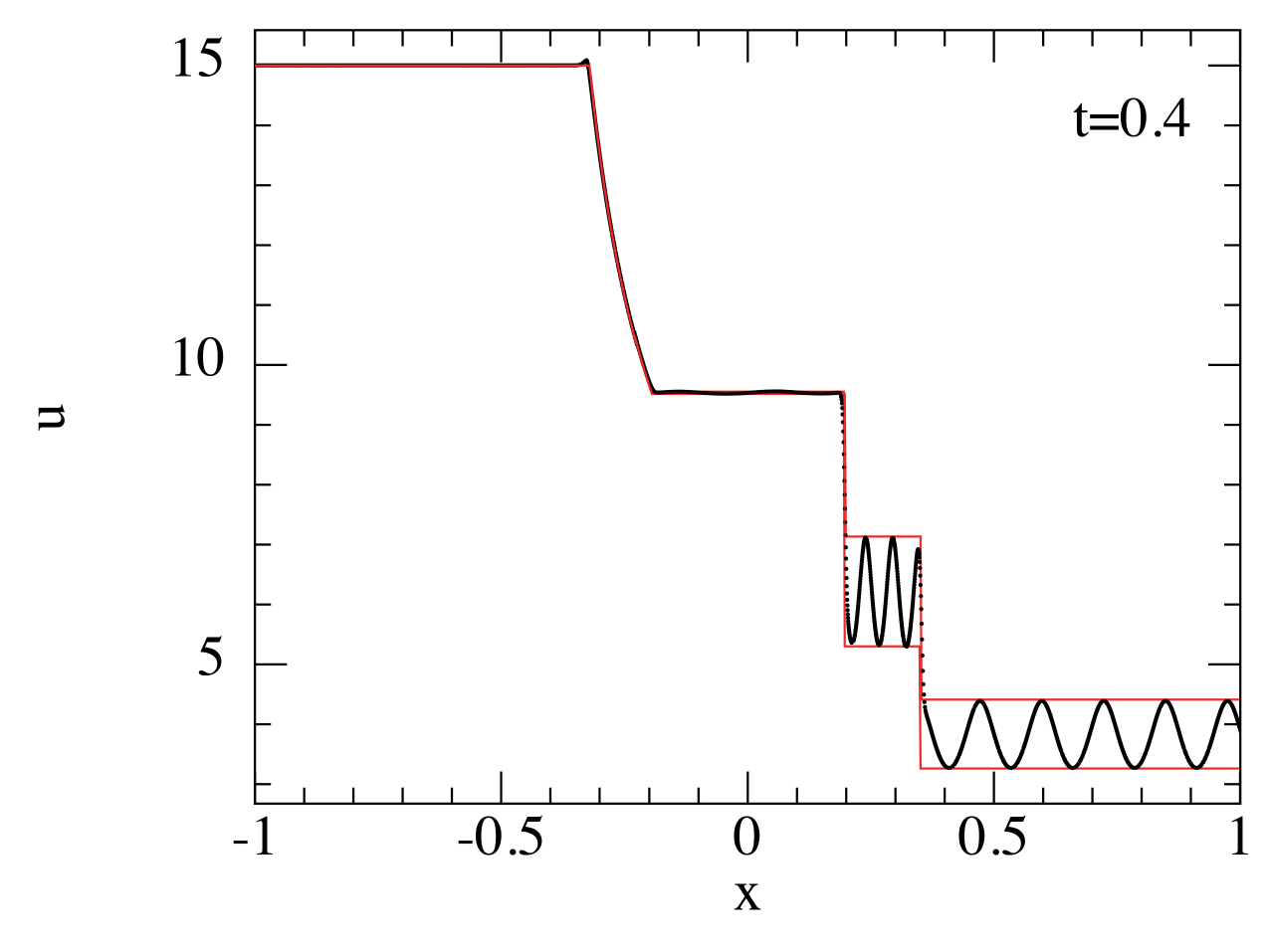} &
        \includegraphics[width=0.40\textwidth]{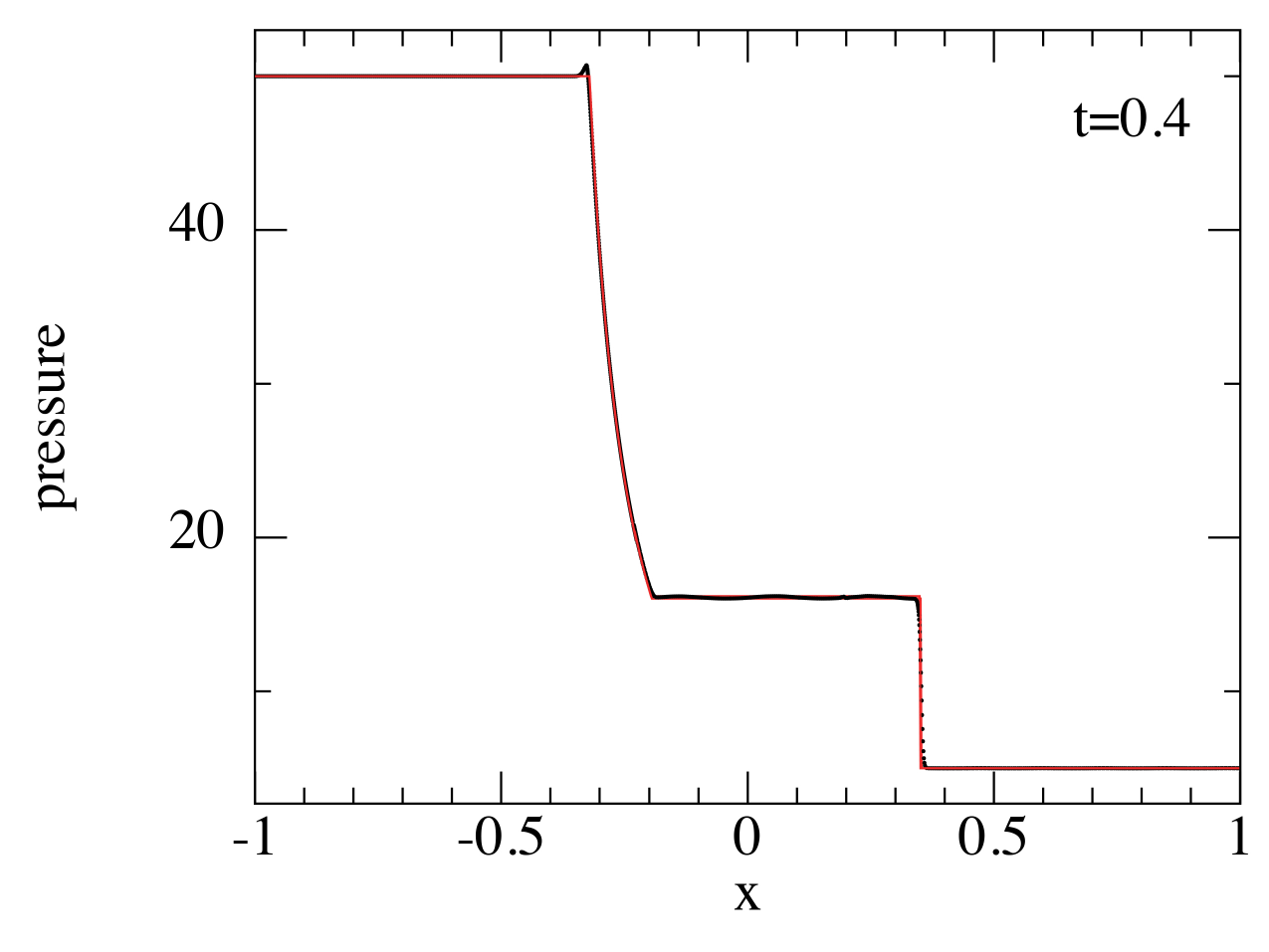}
    \end{tabular}

    \caption{1D sinusoidally perturbed shock tube test. Black circles denote SPH particles, while the red curves represent analytical solutions corresponding to the upper and lower density extremes of the sine perturbation.}
    \label{fig:1d_sine_wave}
\end{figure}

\subsubsection{3D sinusoidal shock tube test}
Here, we perform a 3D shock tube test with a sinusoidally perturbed density state on the right hand side. The maximum value of the Lorentz factor reached by the fluid is $\Gamma=1.15$. The initial conditions we used for this setup are
\begin{equation}
    [\rho,P] = \begin{cases} [8, 50] &\mbox{for } x < 0 \\
[1+0.3\sin{(50x)},5] & \mbox{for } x > 0 \\
\end{cases}
\end{equation}
This test has been performed with a total 76097 particles.
From the simulation results, we observe a consistent pattern with the 1D sine-wave shock tube test.

\begin{figure}[t]
    \centering
    \setlength{\tabcolsep}{2pt}

    \begin{tabular}{cc}
        \includegraphics[width=0.40\textwidth]{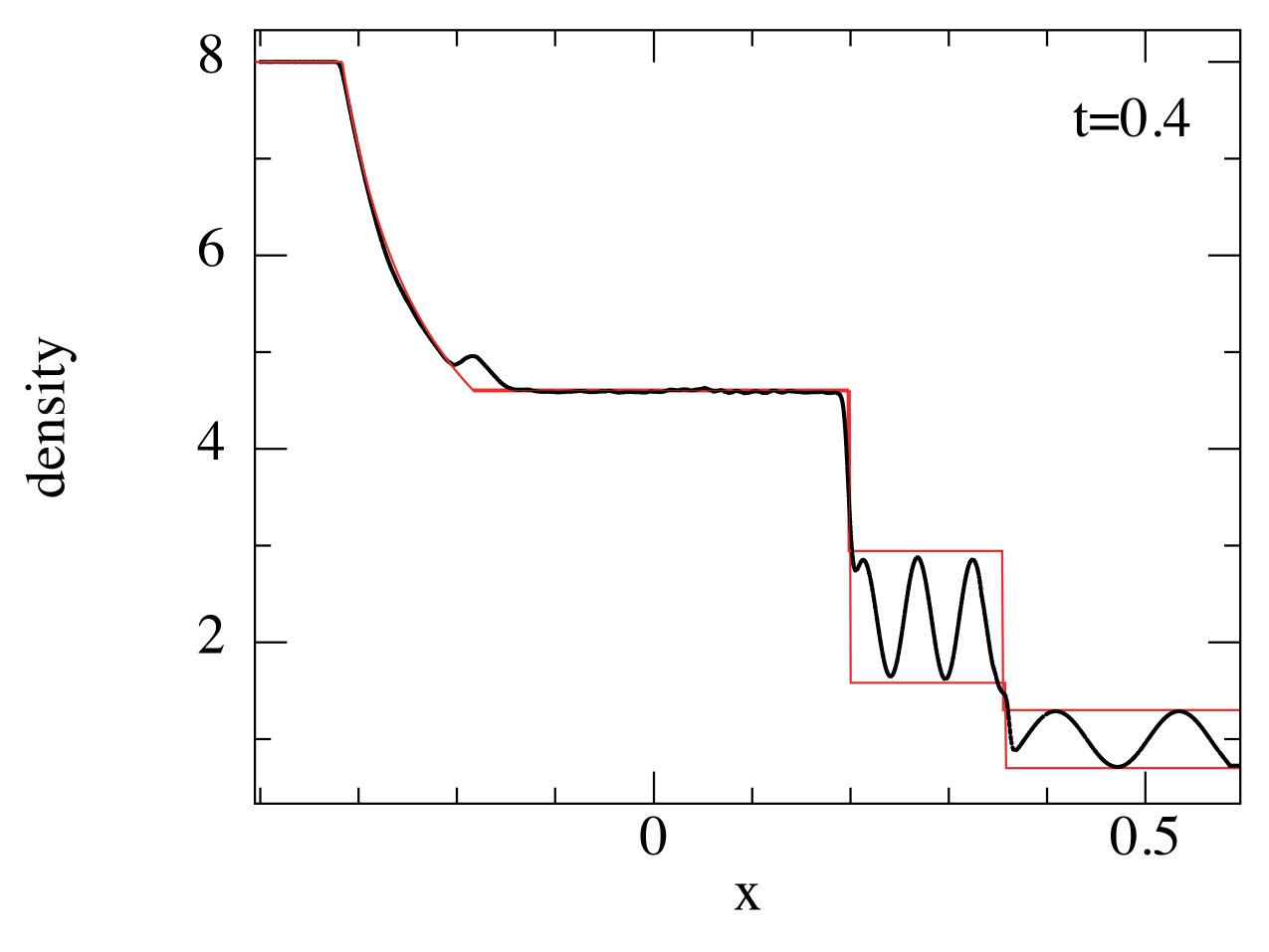} &
        \includegraphics[width=0.40\textwidth]{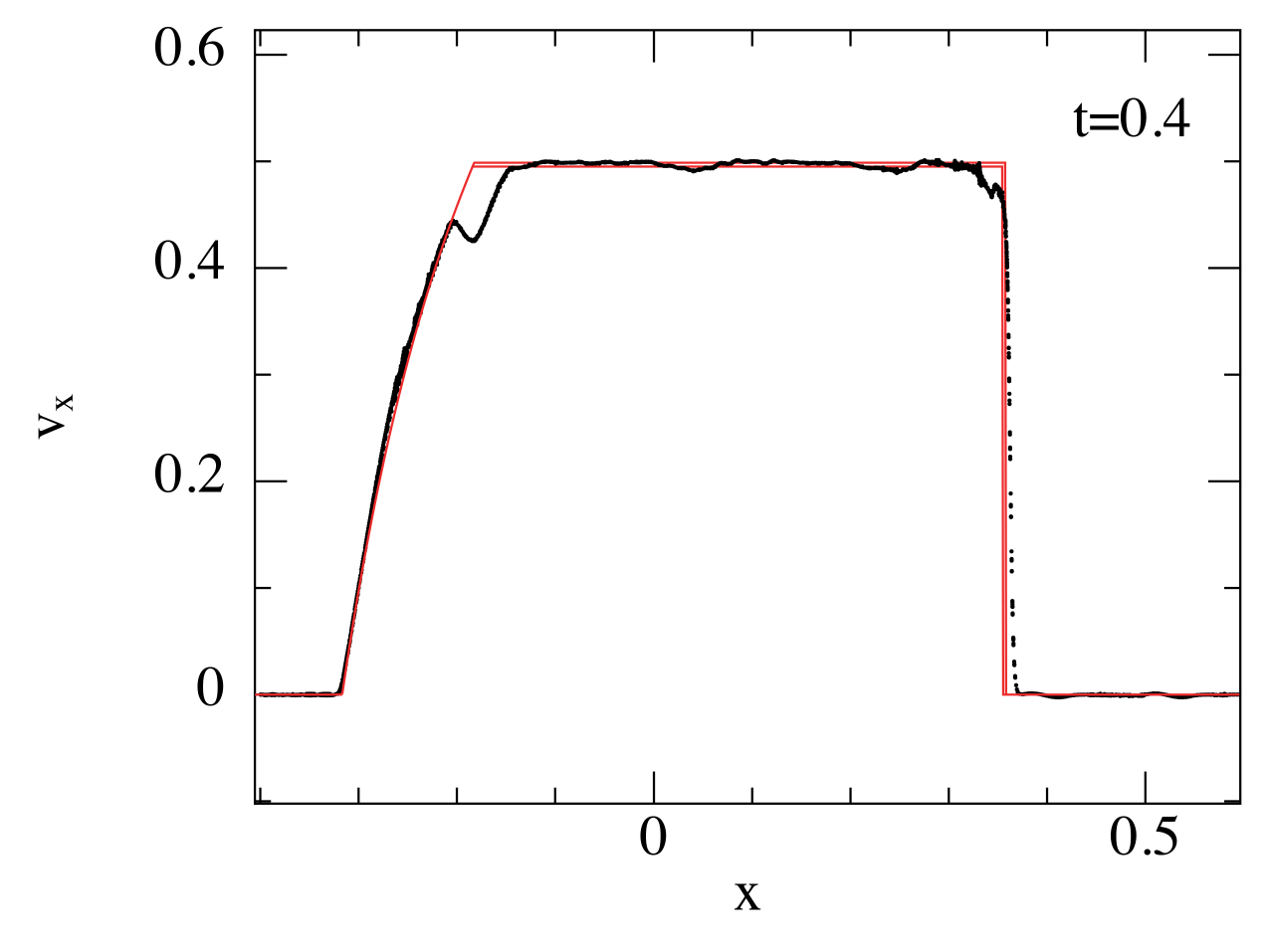} \\

        \includegraphics[width=0.40\textwidth]{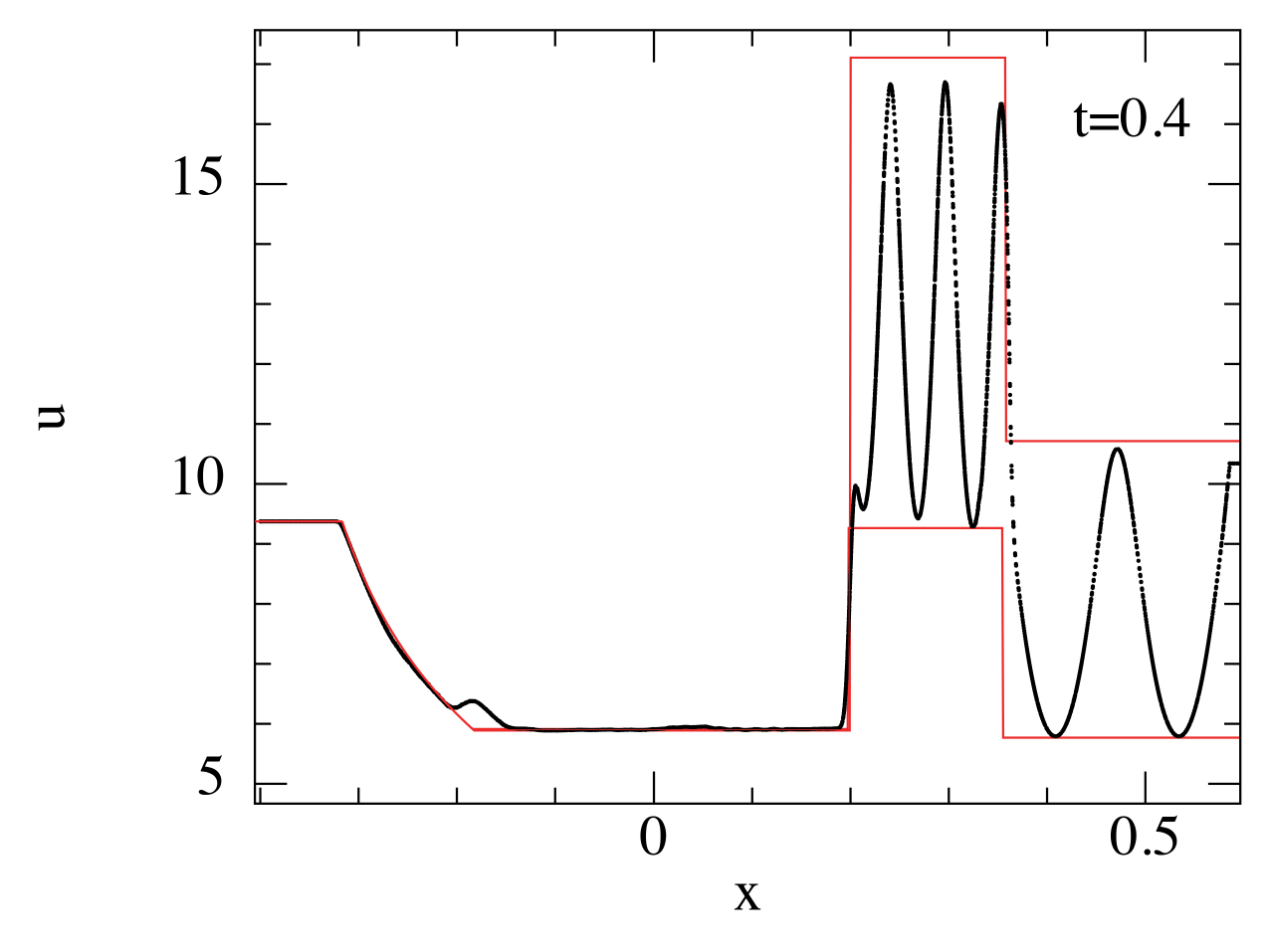} &
        \includegraphics[width=0.40\textwidth]{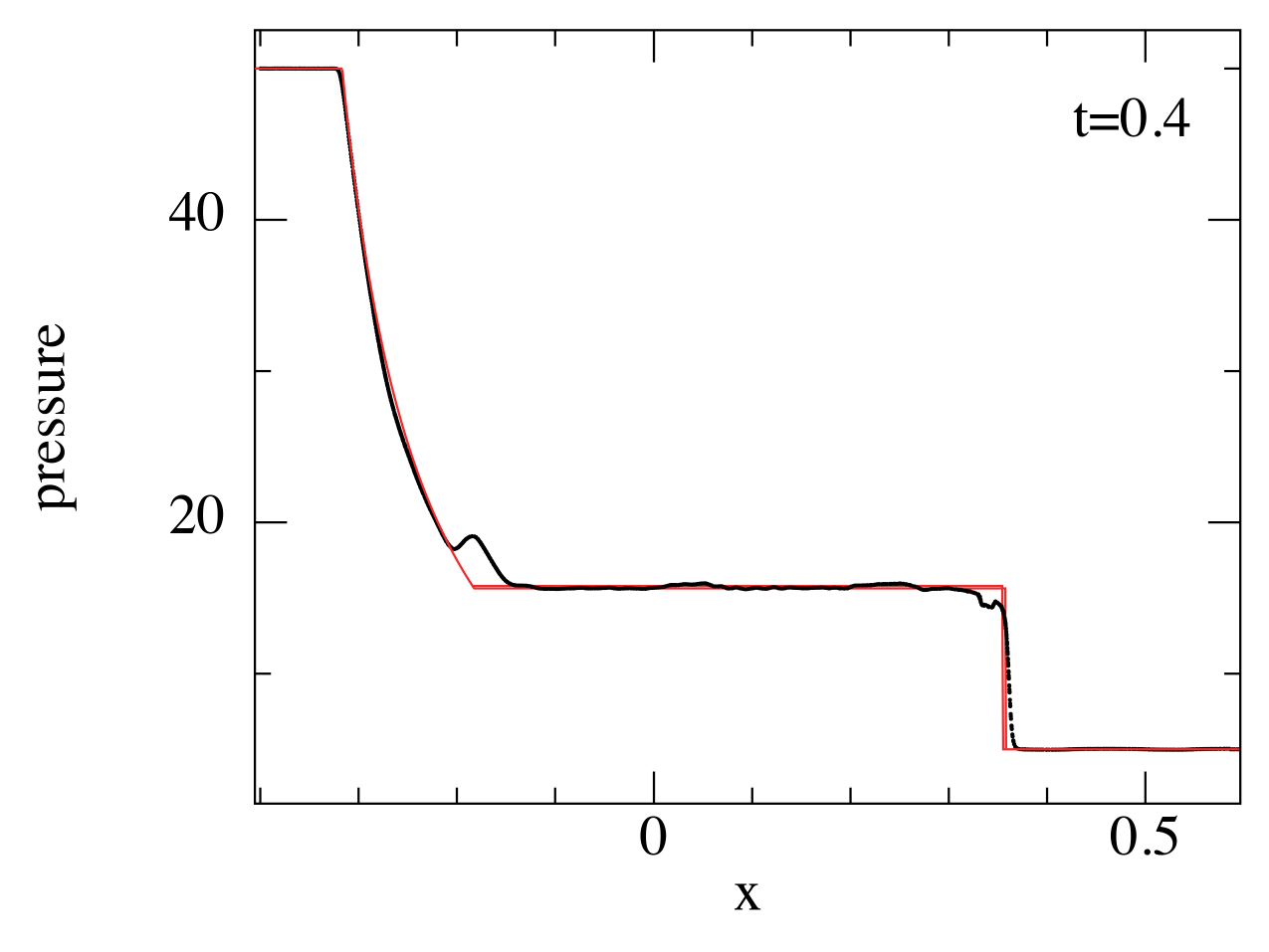}
    \end{tabular}

    \caption{3D sinusoidally perturbed shock tube test. Black circles denote SPH particles, while the red curves represent analytical solutions corresponding to the upper and lower density extremes of the sine perturbation.}
    \label{fig:3d_sine_wave}
\end{figure}

\subsubsection{3D transverse shock tube test}
In the transverse shock tube test, we consider that the fluid in the $x>0$ region has a relativistic transverse velocity. The magnitude of flow speed is constrained in relativistic flow by the speed of light. Also, transverse velocity is coupled to the evolution of the other hydrodynamic quantities through the Lorentz factor. Using the following initial conditions, we perform this test using 199996 particles. The fluid achieves a maximum Lorentz factor $\Gamma \sim 9.0$.
\begin{equation}
    [\rho,P,{v_x},{v_y}] = \begin{cases} [1,1000,0,0] &\mbox{for } x < 0 \\
[1,0.01,0,0.9938] & \mbox{for } x > 0 \\
\end{cases}
\end{equation}
\begin{figure}[t]
    \centering
    \setlength{\tabcolsep}{1pt} 
    \begin{tabular}{ccc}
        \includegraphics[width=0.32\textwidth]{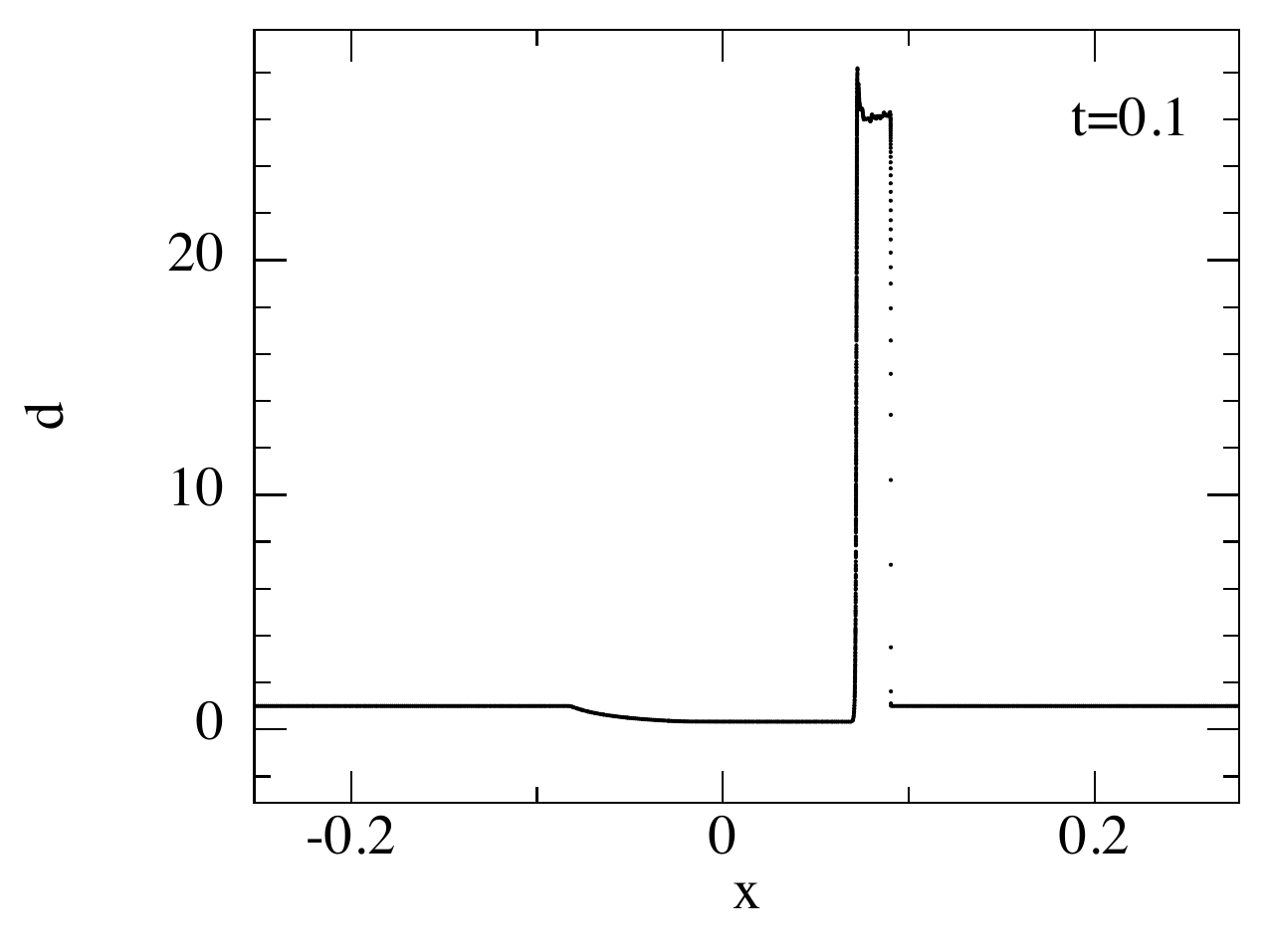} &
        \includegraphics[width=0.32\textwidth]{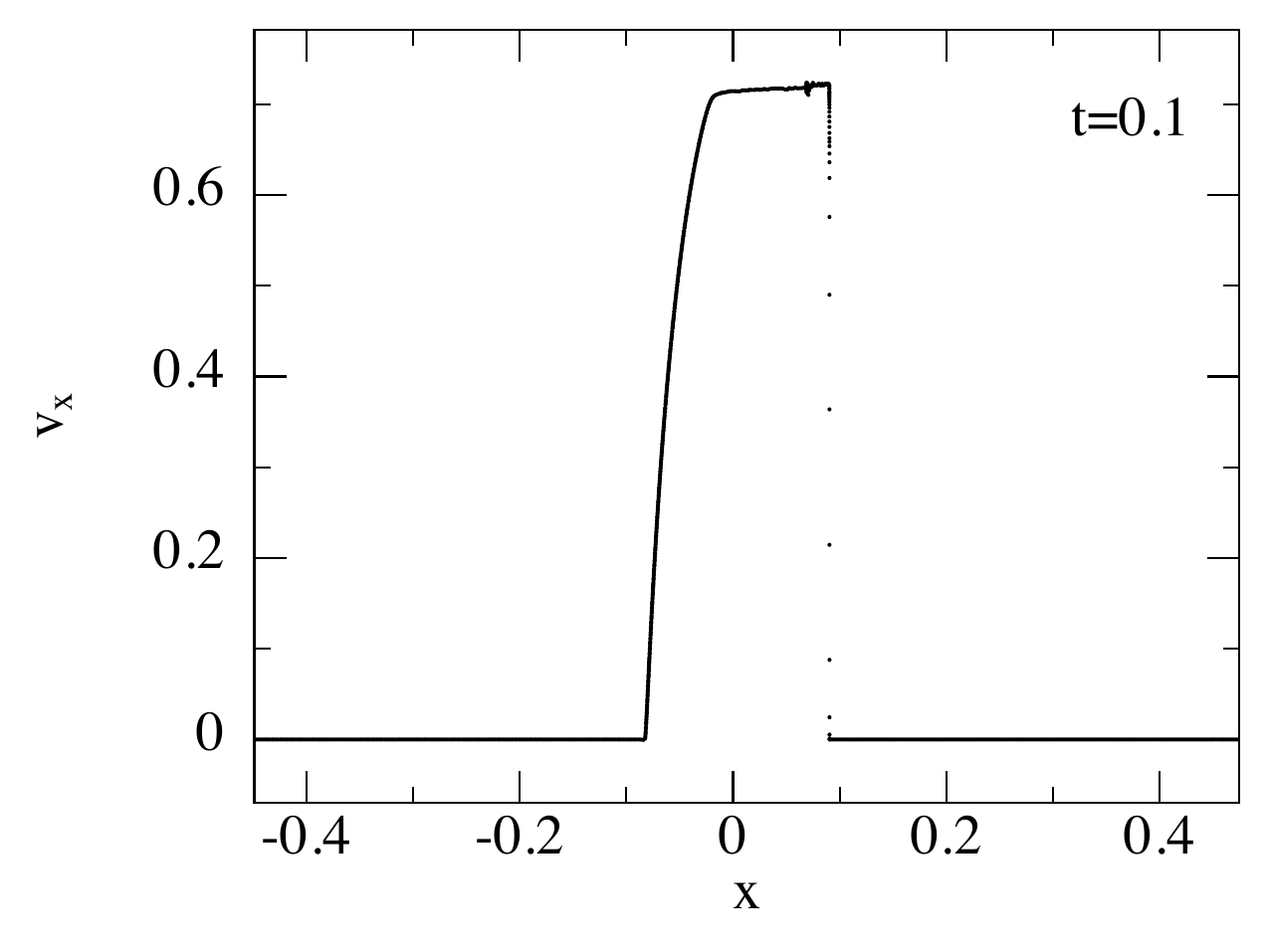} &
        \includegraphics[width=0.32\textwidth]{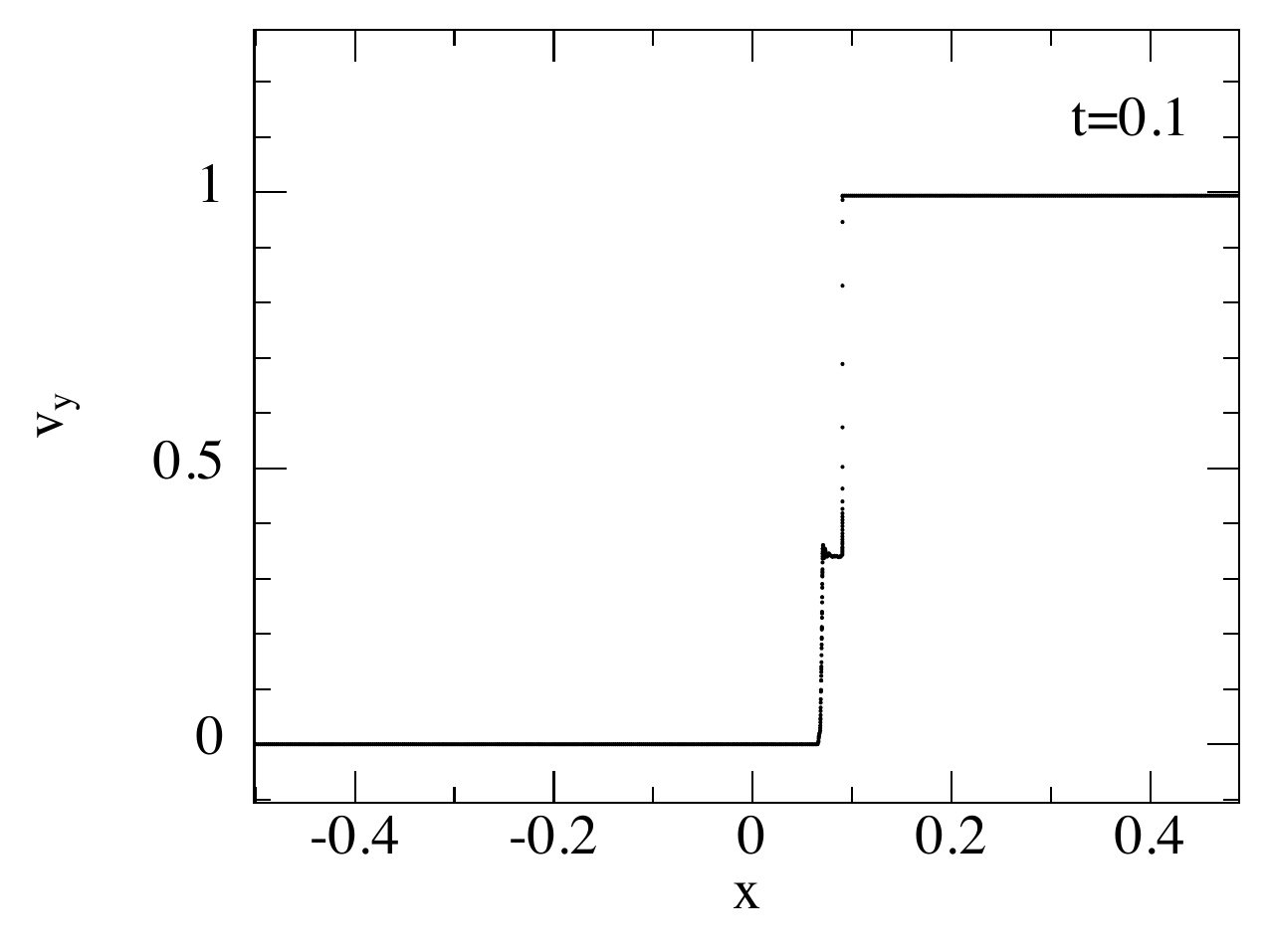} \\
        \multicolumn{3}{c}{%
            \includegraphics[width=0.32\textwidth]{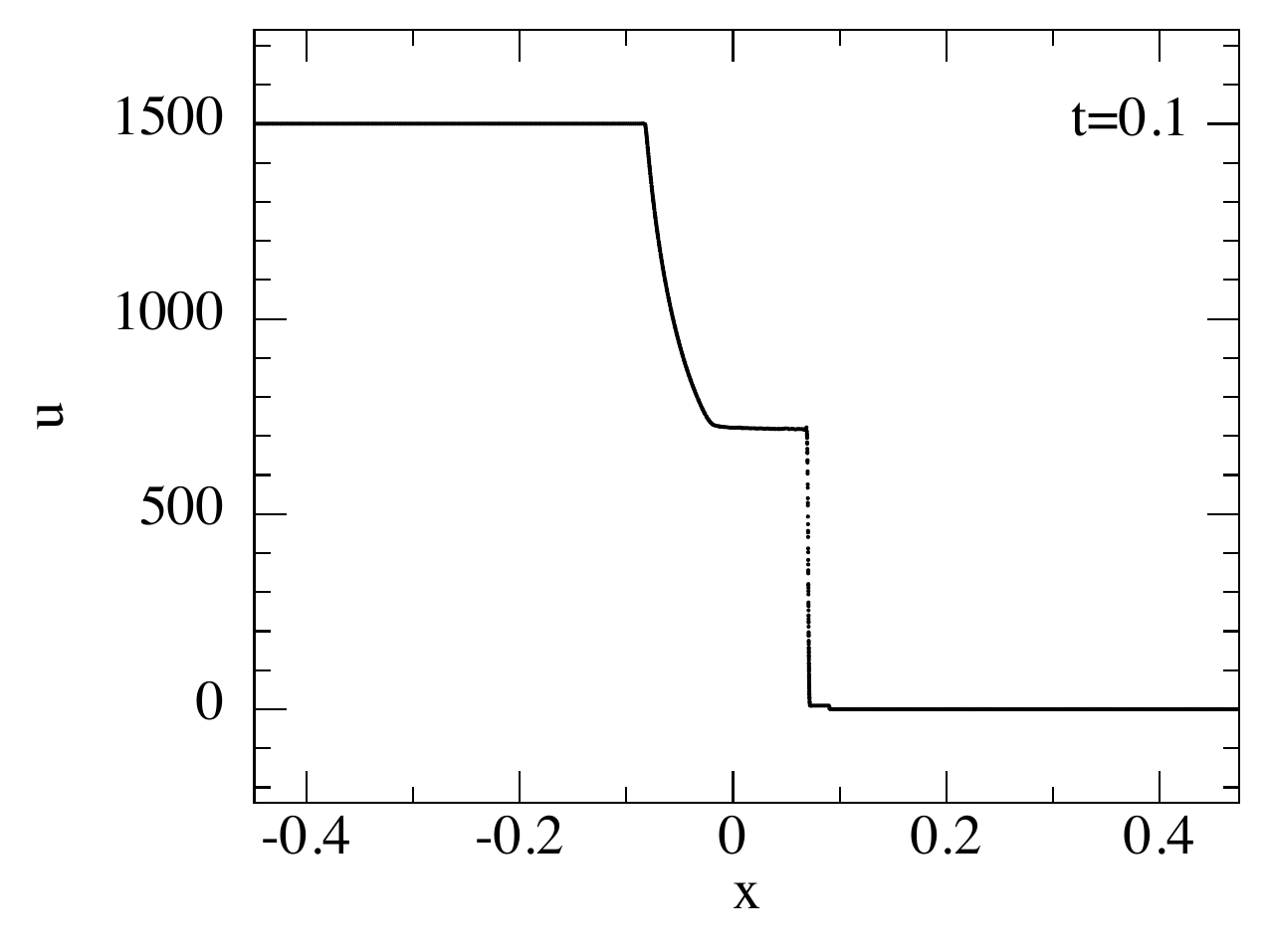}\hspace{4pt}%
            \includegraphics[width=0.32\textwidth]{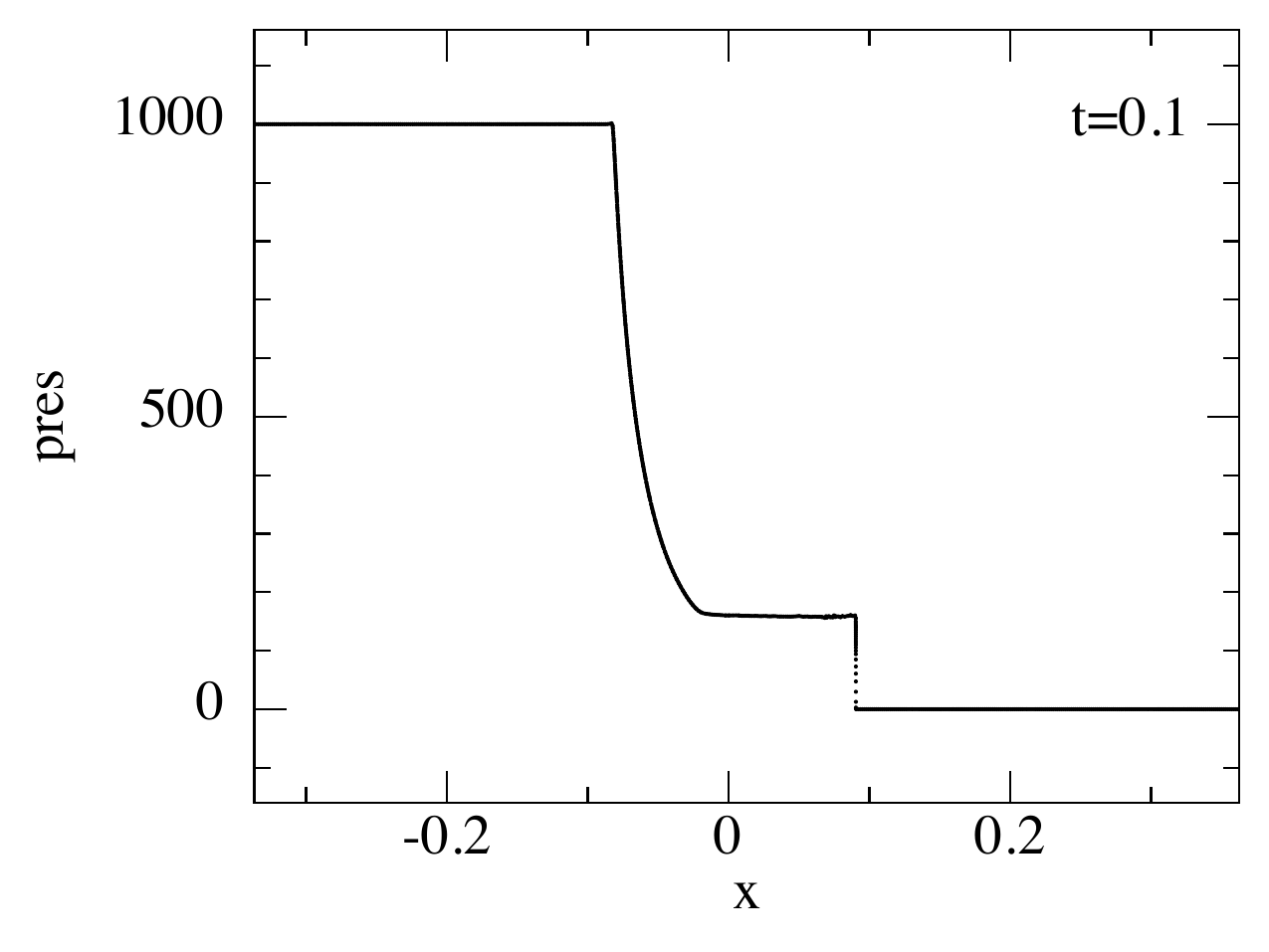}%
        }
    \end{tabular}
    \caption{Transverse shock tube test has been performed in 3D, The black circles show the SPH particles.}
    \label{fig:transverse_shocktube}
\end{figure}


\subsection{Tests of General Relativity}

\subsubsection{Geodesic test}
In order to study the ability of our code to simulate orbital dynamics, we test it for the Schwarzschild and Morris-Thorne exponential WH metric in Cartesian coordinates. To match geodesics, we have taken a test star of 1 $M_\odot$, 0.1 $R_\odot$ and $\gamma = 2$. The eccentricity and the impact parameter are $0.85$ and $0.6$ respectively, for both cases.
\begin{figure}[h]
    \centering
    \setlength{\tabcolsep}{1pt} 
    \begin{tabular}{cc}
        \includegraphics[height=0.45\textwidth]{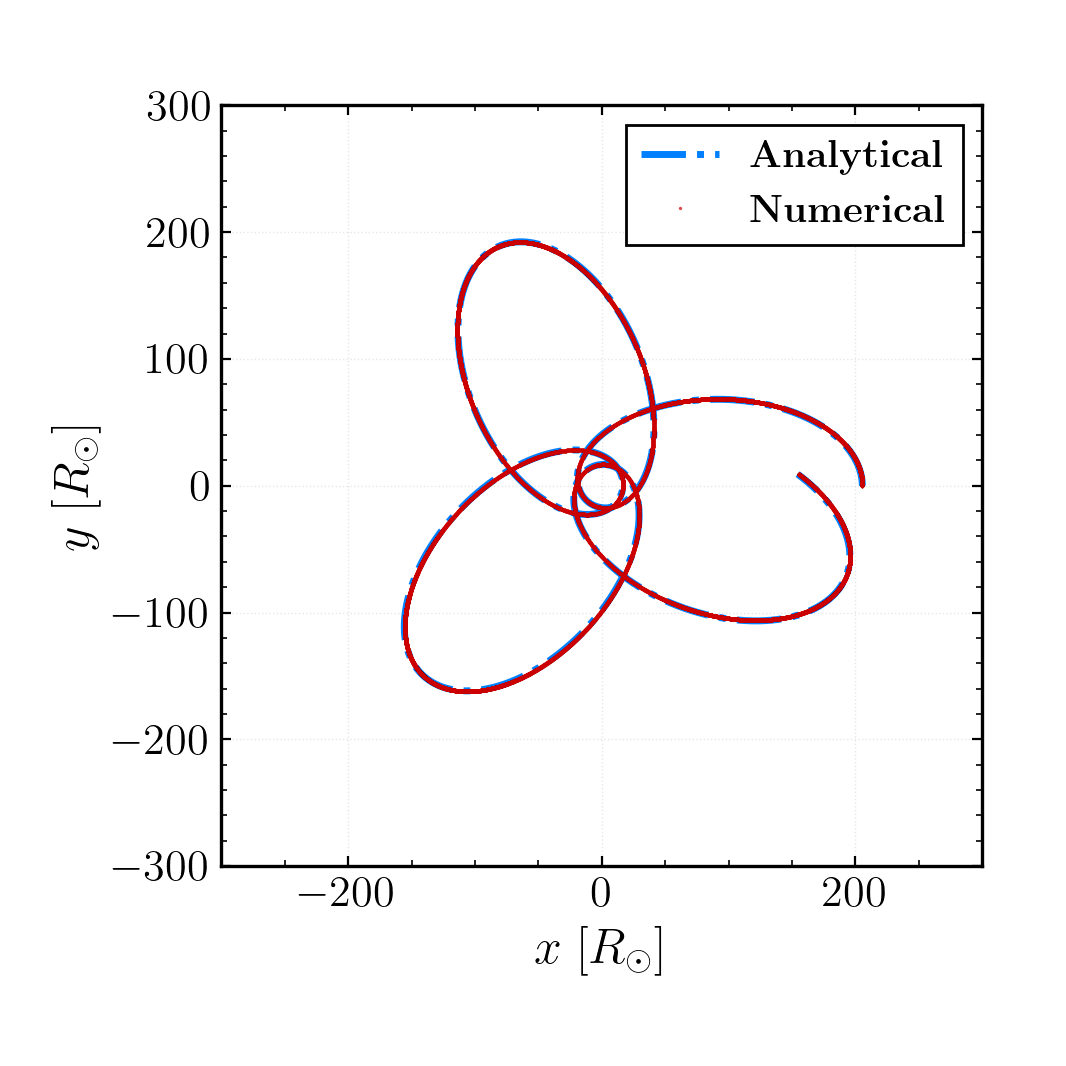} &
        \includegraphics[height=0.45\textwidth]{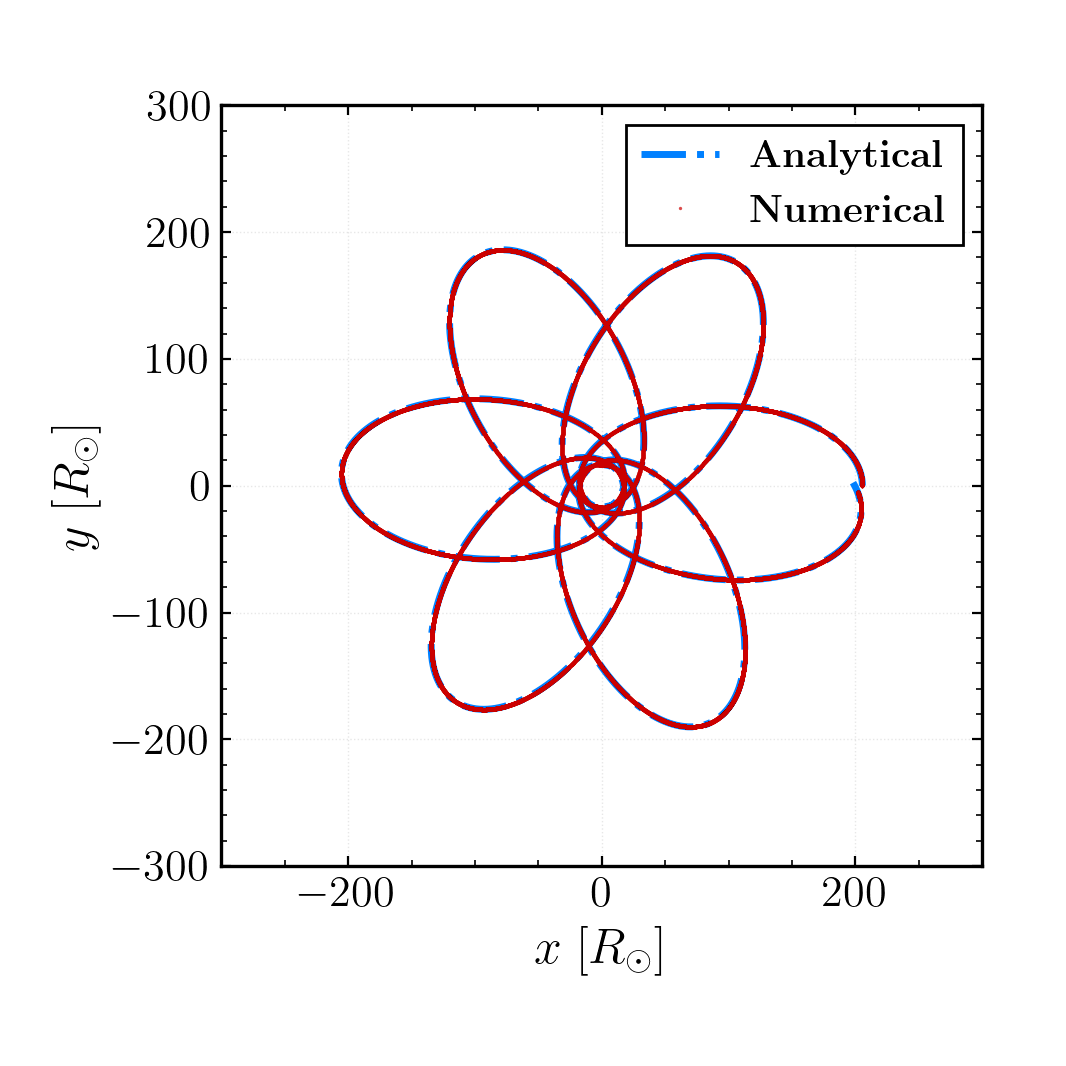}
    \end{tabular}
    \caption{On left panel: orbit around BH. On right panel: orbit around WH }
    \label{fig:orbit_test}
\end{figure}

\subsubsection{Radial Velocity}
\begin{figure}[h]
    \centering
    \setlength{\tabcolsep}{1pt} 
    \begin{tabular}{cc}
        \includegraphics[width=0.45\textwidth]{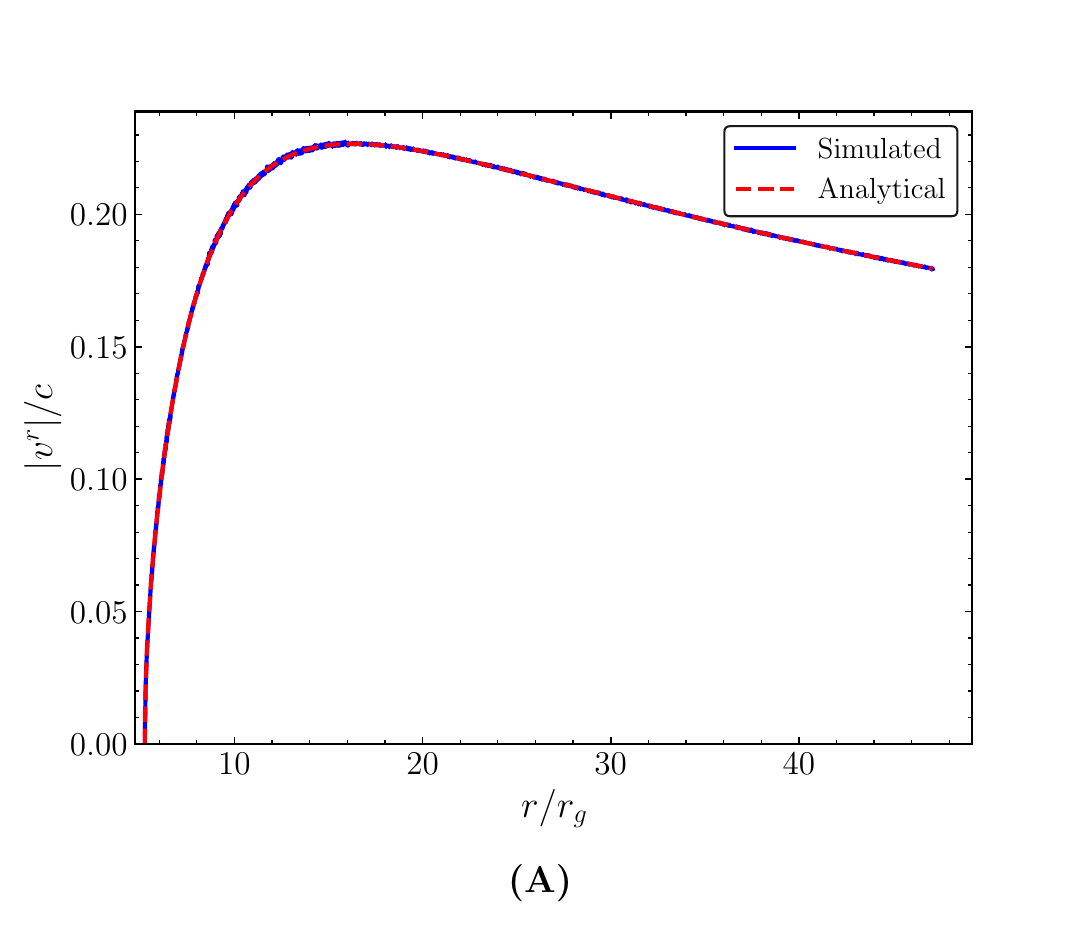} &
        \includegraphics[width=0.45\textwidth]{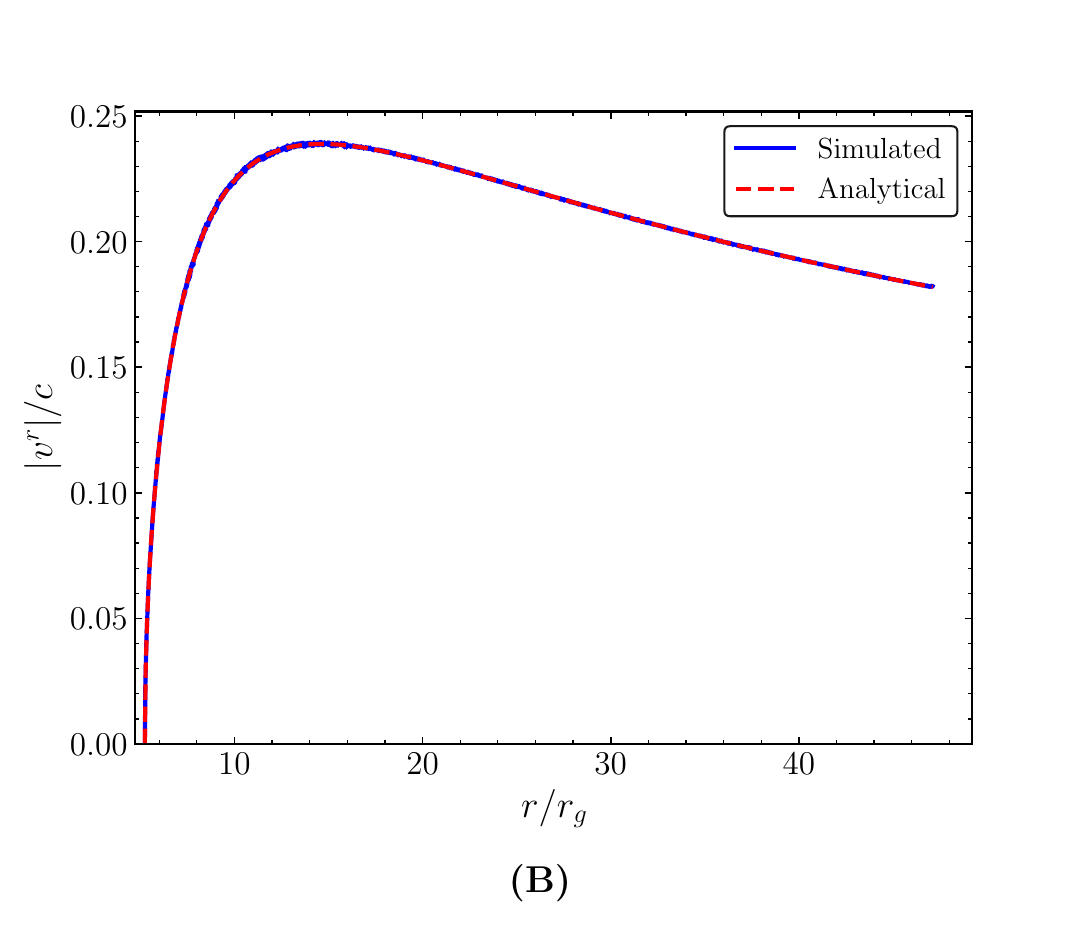}
    \end{tabular}
    \caption{Radial velocity test in the Schwarzschild BH background (A) and the exponential WH background (B).}
    \label{radial_vel}
\end{figure}

To test the behaviour of the radial velocity, we consider a compact star of mass $1\, M_\odot$ and radius $0.1\, R_\odot$ on a parabolic orbit with impact parameter $\beta = 0.9$ in the gravitational field of a $10^{6}\, M_\odot$ BH. Figure~\ref{radial_vel}A and a WH spacetime
Figure~\ref{radial_vel}B. For these parameters, tidal disruption is negligible,
allowing the star to be treated as a test object even in the strong gravity regime.

\subsubsection{Bondi accretion for a pressure-less gas}

\begin{figure}[h!]
    \centering
    \setlength{\tabcolsep}{1pt} 
    \begin{tabular}{ccc}
        \includegraphics[width=0.32\textwidth]{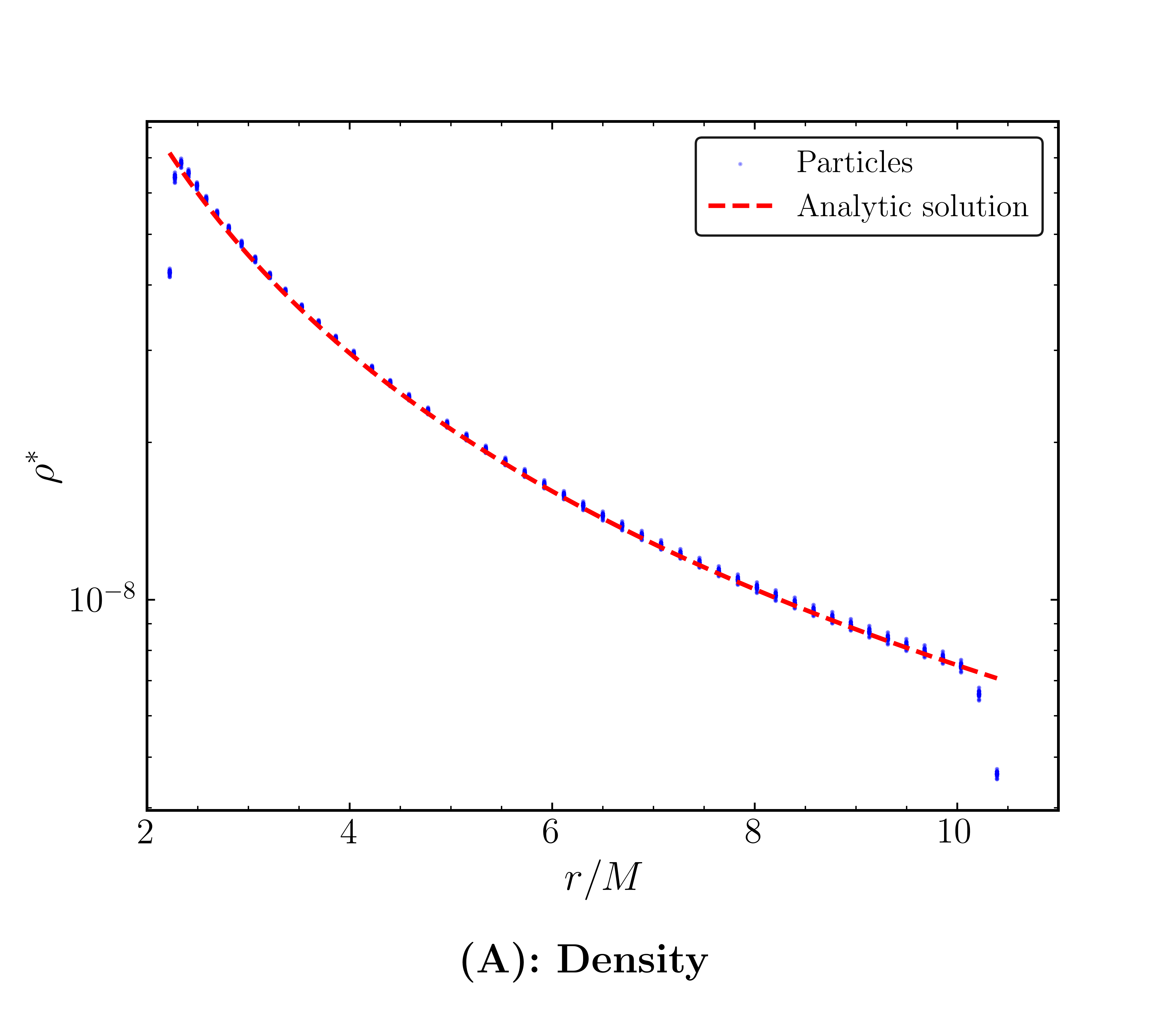} &
        \includegraphics[width=0.32\textwidth]{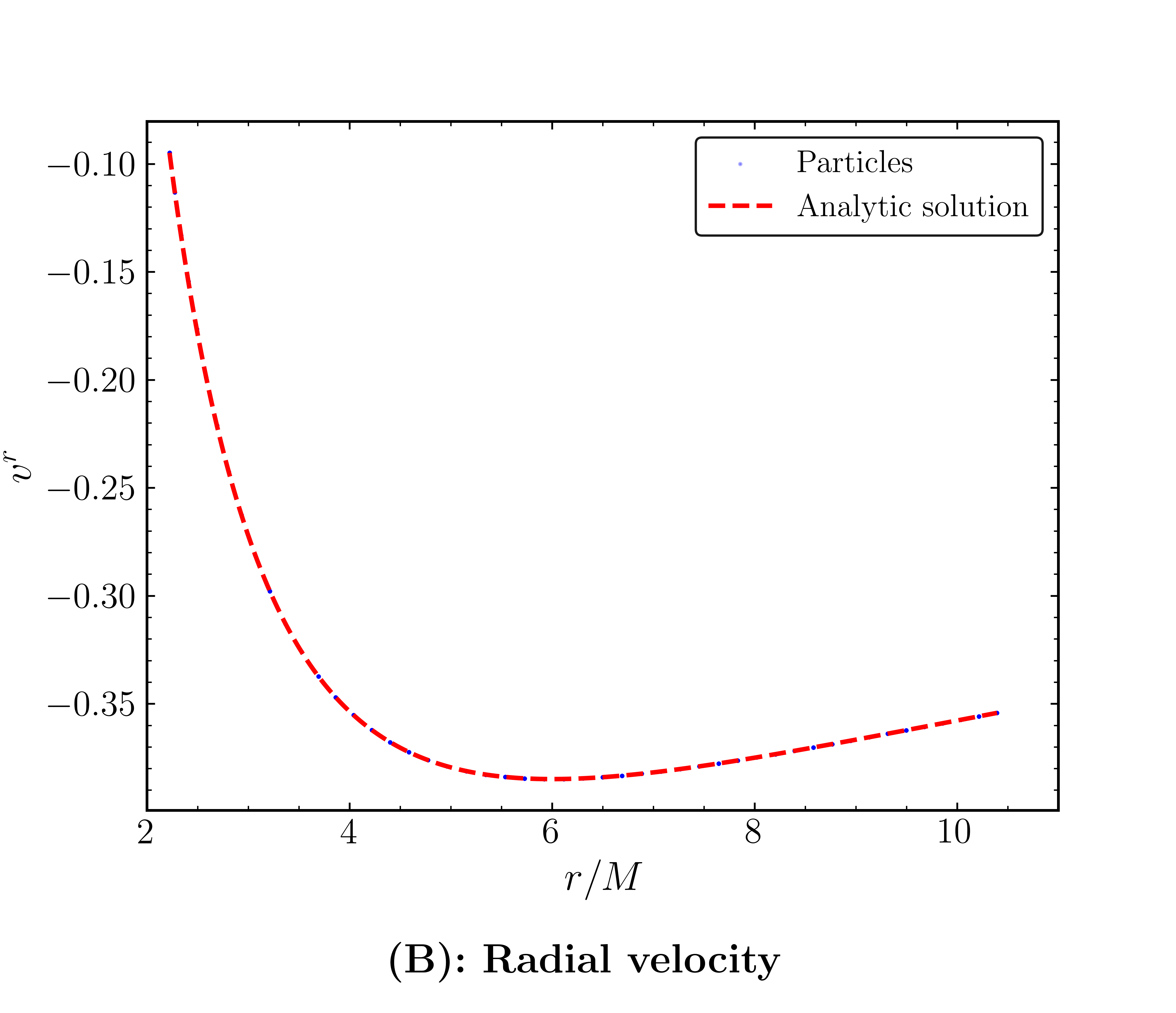} &
        \includegraphics[width=0.32\textwidth]{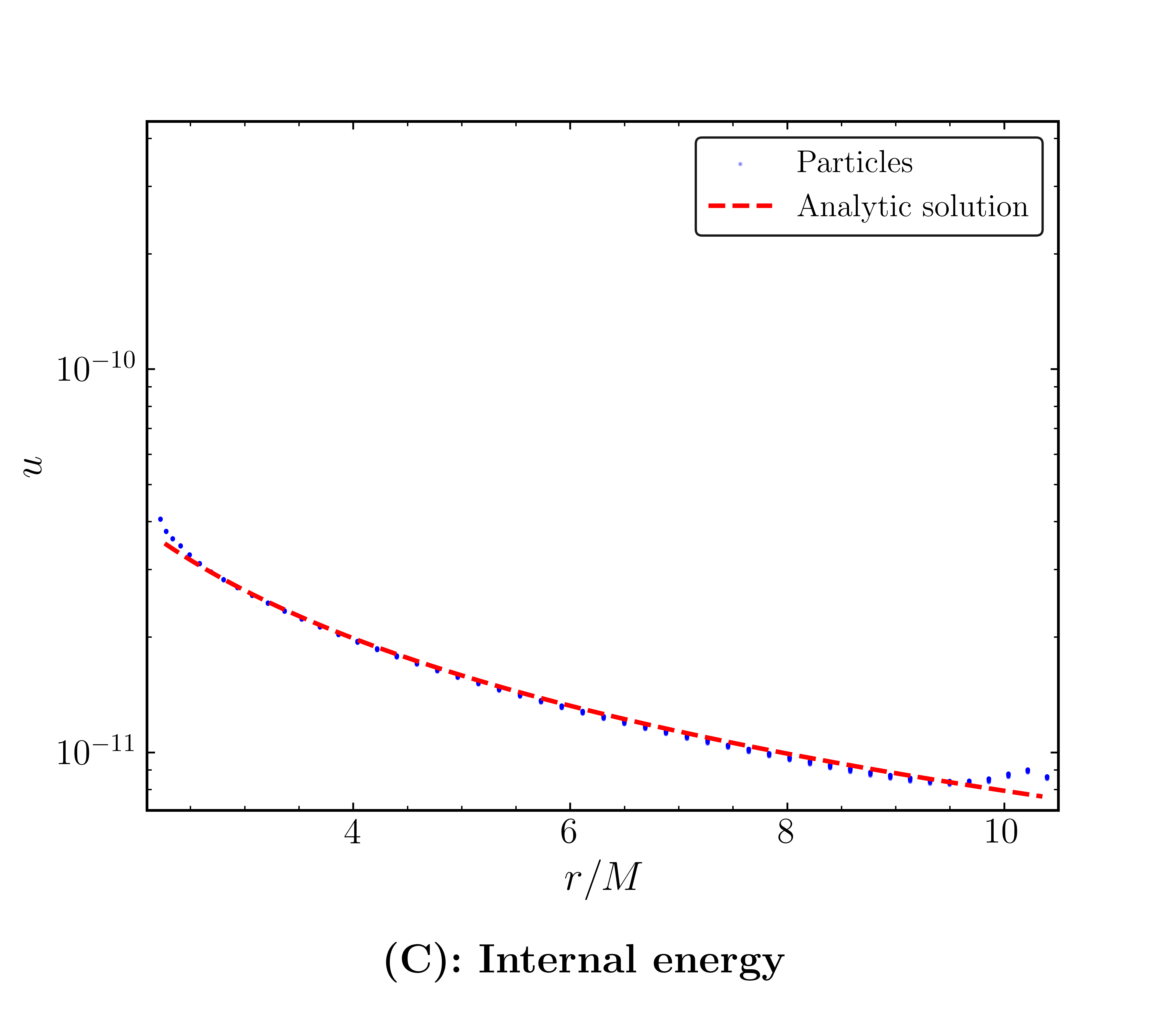}
    \end{tabular}
    \caption{Bondi geodesic test with $\alpha=0$ at $t=55$ code units.}
    \label{fig:Bondi_geodesic}
\end{figure}

To test Bondi accretion  for a pressure-less gas, we have taken the solution given by \cite{Hawley1984a}
\begin{align}
v^{r}(r) &= \sqrt{\frac{2M_\bullet}{r}}\left(1 - \frac{2M_\bullet}{r}\right)\\
\rho(r) &= \frac{\rho_0}{r^{2}} \sqrt{\frac{r}{2M_\bullet}}\\
u(r) &= u_0 \left( r^{2} \sqrt{\frac{2M_\bullet}{r}} \right)^{1-\gamma_{\mathrm{ad}}}.
\end{align}
where $\rho_0$ and $u_0$ are normalization constants. We have taken $\rho_0 = 3.35\times10^{-7}$, $M_\bullet = 1$ and $u_0 = 10^{-10}$. We keep 50 shells of 14400 particles with mass $1.4\times10^{-10}M_\bullet$ outside  $r_{in} = 20M_\bullet$ distance, each shell is kept at $0.5 M_\bullet$ temporal distance. Keeping the $\alpha_{AV} = 0$  we let the system evolve. We have shown the density and radial velocity profile in Figure \ref{fig:Bondi_geodesic}. We have used the $G=c=1$ units for this test.
\subsubsection{Bondi sonic flow test}
\begin{figure}[t]
    \centering
    \setlength{\tabcolsep}{1pt} 
    \begin{tabular}{ccc}
        \includegraphics[width=0.32\textwidth]{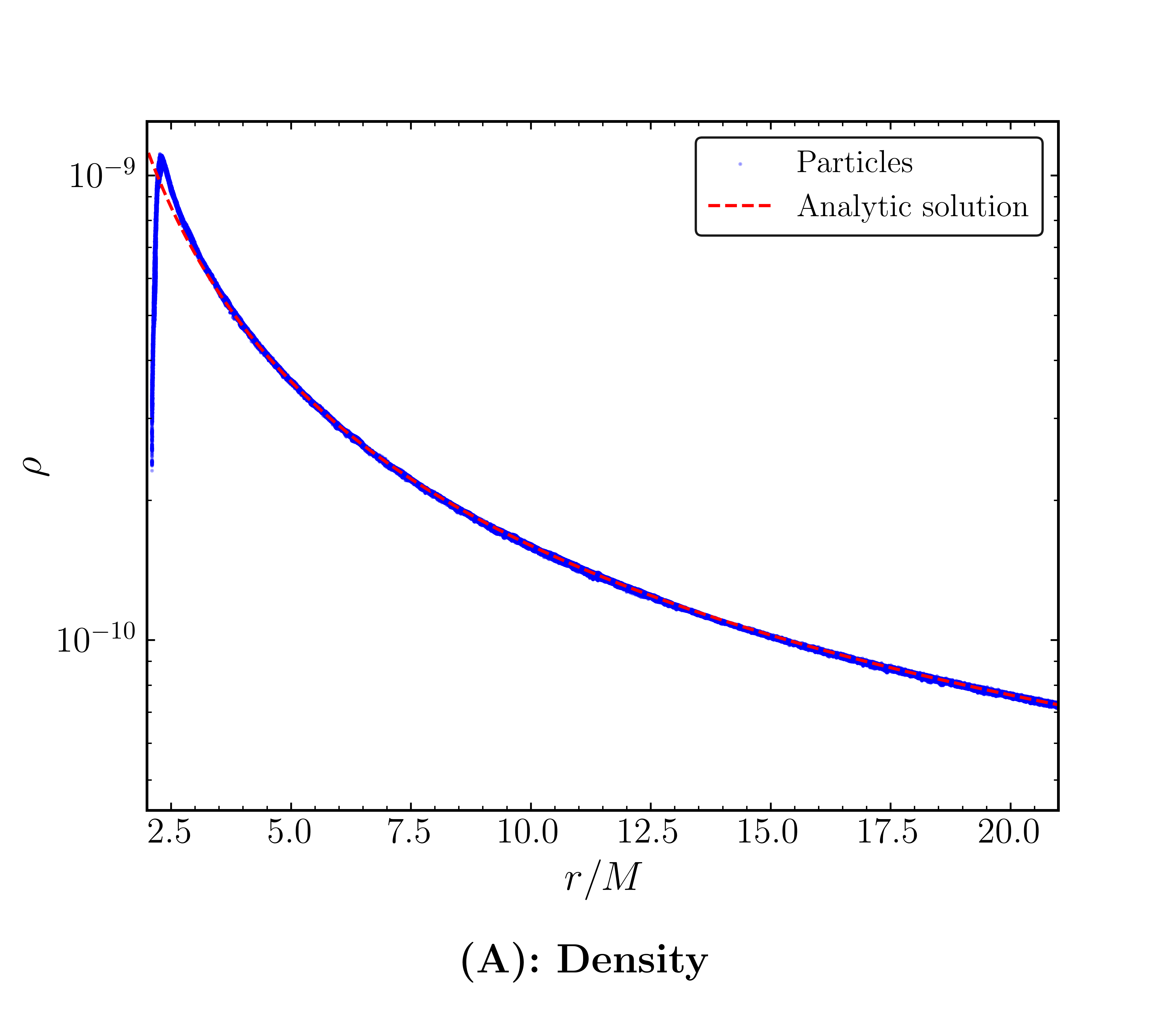} &
        \includegraphics[width=0.32\textwidth]{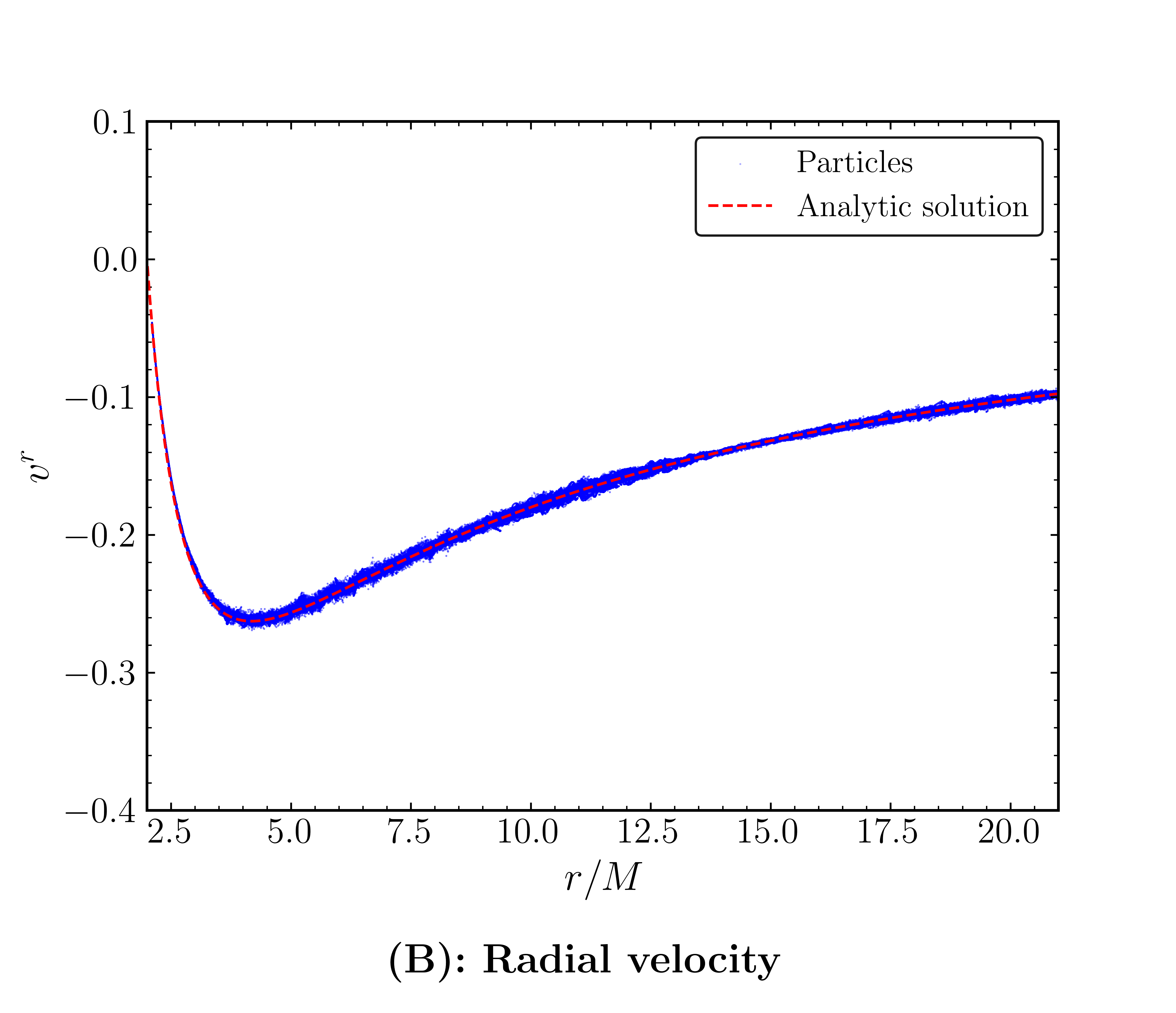} &
        \includegraphics[width=0.32\textwidth]{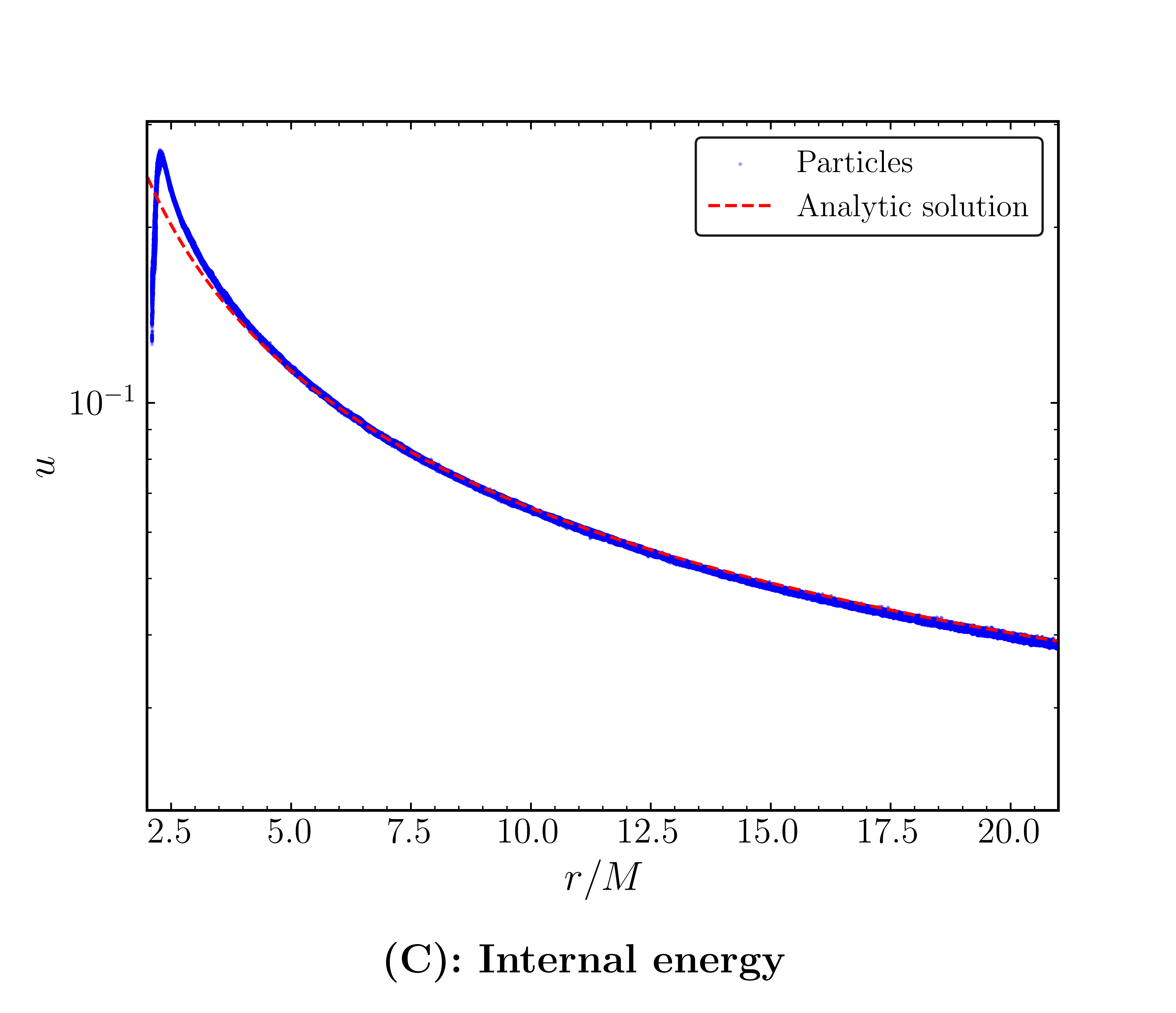}
    \end{tabular}
    \caption{Sonic flow test with $\alpha = 0.3$ at $t=577$ code units.}
    \label{fig:Sonic_0p1}
\end{figure}

To perform the sonic flow test, we follow the solution outlined in \cite{Michel1972}.
In our simulation, we filled the region from $r=2.2 M_\bullet$ to $r = 22 M_\bullet$ with spherical shells, with $5400$ particles per shell, with each particle mass $1.0\times 10^{-11} M_\bullet$ with $M_\bullet = 1$. Each shell is kept at a time interval $dt$ from the previous one. We choose $dt = 1.3 M_\bullet$, critical point $r_c=8M_\bullet$ and $\gamma = 5/3$. We continuously inject shells at $dt$ time intervals. For the simulation, we have taken $\alpha=0.3$, $\alpha_{u}=0.1$ and $G=c=1$. The simulation result at $t=577$ code units is shown in Figure \ref{fig:Sonic_0p1}.

\providecommand{\href}[2]{#2}\begingroup\raggedright

\endgroup
\end{document}